\documentclass[12pt]{article}
\usepackage{latexsym}
\usepackage{epsf}

\thicklines                         
\newcommand{\dmn}[3]{\Delta^{V,(0)}_{#1 #2} \left( #3 \right)}
\newcommand{\mma}[2]{#1_#2}
\newcommand{\mmb}[2]{\left( #1 \right)_ #2}
\newcommand{\cma}[2]{\cos{#1_#2\over 2}}
\newcommand{\cmb}[2]{\cos{\left( #1\right)_#2\over 2}}
\def\hp{\widehat p}
\def\hq{\widehat q}
\def\hkp{\widehat{p+k}}
\def\hkq{\widehat{q+k}}

\newcommand{\bibi}{\bibitem}

\def\a{\alpha}

\def\d{\delta}
\def\f{\varphi}    

\def\j{\psi}
\def\k{\kappa}     
\def\l{\lambda}
\def\m{\mu}

\def\n{\nu}

\def\p{\pi}       
\def\r{\rho}      
\def\s{\sigma}    

\def\x{\xi}

\def\D{\Delta}

\def\S{\Sigma}

\def\tr{\widetilde{r}}

\newcommand{\del}{\partial}

\newcommand{\half}{\mbox{{\normalsize $\frac{1}{2}$}} }

\newcommand{\quart}{\mbox{{\small $\frac{1}{4}$}} }
\newcommand{\eighth}{\mbox{{\small $\frac{1}{8}$}} }
\newcommand{\sixteenth}{\mbox{{\small $\frac{1}{16}$}} }

\newcommand{\ra}{\rightarrow}

\newcommand{\aplt}{ \mbox{}_{\textstyle \sim}^{\textstyle < }     }
\newcommand{\apgt}{ \mbox{}_{\textstyle \sim}^{\textstyle > }     }

\newcommand{\lag}{\langle}
\newcommand{\rag}{\rangle}

\newcommand{\tk}{\widetilde{\kappa}}

\newcommand{\ph}{\phi}

\newcommand{\ps}{\psi}

\newcommand{\khat}{\widehat{k}}

\newcommand{\tsum}{\widetilde{\sum}}

\def\tlsum#1{\widetilde{\sum_{#1}}}
\def\tlsump#1{\;\widetilde{\sum_{#1}}}
\def\tlsuma#1{\widetilde{\sum_{#1}}\;}
\def\tlsumpa#1{\;\widetilde{\sum_{#1}}\;}

\newcommand{\psb}{\overline{\ps}}

\newcommand{\hmu}{\hat{\mu}}
\newcommand{\hnu}{\hat{\nu}}
\newcommand{\phat}{\widehat{p}}

\newcommand{\be}{\begin{equation}}
\newcommand{\ee}{\end{equation}}
\newcommand{\bea}{\begin{eqnarray}}
\newcommand{\eea}{\end{eqnarray}}
\newcommand{\eq}{\ref}
\newcommand{\beq}{\begin{equation}}
\newcommand{\eeq}{\end{equation}}
\newcommand{\cc}{\cite}
\newcommand{\lb}{\label}

\newcommand{\PRL}{Phys. Rev. Lett.}
\newcommand{\PLB}{Phys. Lett. B}
\newcommand{\PRD}{Phys. Rev. D}
\newcommand{\NPB}{Nucl. Phys. B}

\def \3{\ss}

\def\footnoteitem(#1)#2{
\begin{list}{#1}{\labelwidth4.0mm \leftmargin7.0mm
\labelsep2.5mm \rightmargin7.0mm \parsep0.5ex plus0.2ex minus0.1ex
\itemsep0ex plus0.2ex }
\item #2
\end{list}
}

\def\secteq#1{ \setcounter{equation}{0}
               \renewcommand{\theequation}{#1.\arabic{equation}} }

\begin{document}
%
\begin{titlepage}

\vskip 3mm
\rightline{November 1999}

\baselineskip=20pt plus 1pt
\vskip 0.5cm

\centerline{\LARGE The Phase Diagram and Spectrum of Gauge-Fixed}     
\vskip 0.5cm
\centerline{\LARGE Abelian Lattice Gauge Theory}
\vskip 1.0cm
\centerline{\large Wolfgang Bock$^\#$, Ka Chun Leung$^\%$ }
\centerline{\em Institute of Physics, University of Siegen,
57068 Siegen, Germany}
\vskip 0.5cm
\centerline{\large Maarten F.L. Golterman$^\$$}
\centerline{\em  Department of Physics, Washington University 
St. Louis, MO 63130, USA}
\vskip 0.5cm
\centerline{\large Yigal Shamir$^\& $}
\centerline{\em Beverly and Raymond Sackler Faculty of Exact Sciences,
Tel-Aviv University,}      
\centerline{\em Ramat Aviv 69978, Israel}
\vskip 1.0cm
\baselineskip=12pt plus 1pt
\parindent 20pt
\centerline{\bf Abstract}
\textwidth=6.0truecm
\medskip

\frenchspacing
We consider a lattice discretization of a covariantly gauge-fixed
abelian gauge theory.  The gauge fixing is part of the action
defining the theory, and we study the phase diagram in detail.
As there is no BRST symmetry on the lattice, counterterms are
needed, and we construct those explicitly.
We show that the proper adjustment of these counterterms drives the theory
to a new type of phase transition, at which we recover a continuum
theory of (free) photons.  We present both numerical and (one-loop)
perturbative results, and show that they are in good agreement near this
phase transition. Since perturbation theory plays an important role,
it is important to choose
a discretization of the gauge-fixing action such that 
lattice perturbation theory is valid. Indeed, we find numerical
evidence that lattice actions not satisfying this requirement do 
not lead to the desired continuum limit.

While we do not consider fermions here, we argue that our
results, in combination with previous work, provide very strong
evidence that this new phase transition can be used to define
abelian lattice chiral gauge theories. 

\nonfrenchspacing

\vskip 0.8cm
\noindent $^\#$ e-mail: {\em bock@physik.uni-siegen.de}  \\
\noindent $^\%$ e-mail: {\em leung@physik.uni-siegen.de}  \\
\noindent $^\$$ \hspace*{0.0cm} e-mail: {\em maarten@aapje.wustl.edu }  \\
\noindent $^\&$ e-mail: {\em shamir@post.tau.ac.il}  \\
\end{titlepage}
\section{Introduction} 
\secteq{1}
\lb{INTRO}
In this paper we continue our investigation of the gauge-fixing
approach to the construction of lattice chiral gauge theories.
In this approach, gauge invariance is broken both through the
gauge-fixing terms and through the fermions.  This requires adding
a complete set of counterterms to the theory, in addition to the
gauge-fixing terms, and these counterterms will need to be tuned.
Showing that this can be done corresponds to 
demonstrating that the phase diagram contains a
continuous phase transition which can be employed to construct the
desired continuum chiral gauge theory \cc{rome,yigal}.

Here, we will restrict ourselves to the abelian case.  This avoids
many of the subtle questions concerning Gribov copies
which arise in the non-abelian case.  In particular, it makes it
possible to drop the ghost sector from consideration \cc{wmy_hn},
while still testing many of the key elements in this approach to
lattice chiral gauge theories.

We will employ (a generalization of) the lattice gauge-fixing action
proposed in Ref.~\cc{my_plb}.  Since we wish to maintain close
contact with standard weak-coupling perturbation theory, we consider
a lattice version of the Lorentz gauge.  (Other gauges may work as
well, but we believe that it is important to restrict oneself to
a renormalizable gauge.) On the lattice, {\it i.e.} in the
regulated theory, there is no gauge or BRST symmetry, and both the
transverse gauge fields and the fermions will couple to the 
gauge degrees  of freedom, represented by the longitudinal part of
the gauge field.  

This leads to the following simple questions,
both of which can be addressed without simulating the full theory
including all its dynamical degrees of freedom: 1) when we turn off
the transverse part of the gauge field, do we obtain a theory of
free (chiral) fermions in the correct representation of the gauge group
(in the abelian case, with the correct charges), decoupled from the
longitudinal modes? and 2) when we turn off the fermions, do we obtain
a theory of free photons, again decoupled from the longitudinal 
degrees of freedom?  

It is well known that (most) small
perturbations of the gauge-invariant compact lattice formulation 
of a U(1) gauge theory do not change the nature of its (weak-coupling)
continuum limit (they correspond to irrelevant directions) \cc{long}.
However, previous work has shown that it is generically not possible
to construct chiral gauge couplings near such a continuum limit
(for reviews, see Refs.~\cc{don_rev,yigal_rev}).  In our approach,
gauge fixing plays an essential role in the construction of the theory.
This means that the coupling in front of the gauge-fixing action has
to be large enough to bring us to a new type of continuous phase
transition in the phase diagram, at which both questions above
can be answered affirmatively.

We have addressed question 1) in previous work 
\cc{wmy_pd,wmy_pert,wmy_prl,wmy_edin,wmy_buk}.  We showed that, in the
``reduced" model, in which only the fermions and the longitudinal gauge
fields are kept, indeed a new type of phase transition occurs.  At this
phase transition, a continuum limit can be defined in which free
chiral fermions with the correct charges emerge, decoupled from these
longitudinal degrees of freedom.  Here, we address the second question.
We turn off the fermions, and we ask whether this new phase transition
survives when the transverse degrees of freedom are present, 
and in particular,
whether it allows us to construct a theory of free photons at the
transition.  We would like to emphasize that, even though, without
fermions, this is not a new continuum theory, our critical point 
corresponds to a new type of universality class.  This is the key
element that allows us to couple fermions chirally to the gauge fields,
without running into the problems which made many previous attempts
unsuccessful.

In addition to this fundamental question, we also address a more
technical issue.  It was argued in Ref.~\cc{yigal} that one has to be very
careful with the precise definition of the lattice version of the
gauge-fixing action.  A naive ``standard" discretization of the Lorentz
gauge-fixing action $(1/2\x)\int d^4x\; (\partial\cdot A)^2$ 
will lead to the occurrence
of a dense set of lattice Gribov copies (with no continuum counterpart).
This corresponds to a large class of uncontrolled, rough fluctuations
in the lattice theory, and may well spoil the existence of the 
critical point we are after.  The lattice gauge-fixing action proposed
in Ref.~\cc{my_plb} does not suffer from this problem, and it is
this action that we have used in our previous work.  Here, we introduce
a one-parameter class of lattice gauge-fixing actions, which
interpolates between the naive discretization and the one of
Ref.~\cc{my_plb}.  This corresponds to adding a direction to the phase
diagram, and we explore the dependence of the phase structure on this
new direction.

The organization of this paper is as follows.  In Sect. 2 we give the
full action for a gauge-fixed U(1) gauge theory, including a complete
set of counterterms.  In particular, we
introduce the parameter $\tr$, which interpolates between the naive 
gauge-fxing action and that of Ref.~\cc{my_plb}, and we discuss
the above mentioned lattice Gribov copies.  We argue that standard
lattice perturbation theory should be valid as long as $\tr$ is
large enough.  In Sect. 3 we present
analytic results in preparation of a high-statistics numerical 
study of this model.  We first explain the nature of the new phase
transition (which we will denote as the ``FM-FMD transition")
from the classical potential, and then
provide a simple-minded mean-field analysis of the model.  
Since there is good qualitative agreement between mean field and
our numerical results, we also give an overview of the structure of
the phase diagram at this stage.  
We end this section with a calculation of the
one-loop lattice photon propagator, and use it to determine some of
the counterterms at one loop.  We also consider a (composite)
scalar two-point function.  Then, in Sect. 4, we present our
numerical results, which constitute the main part.  
First we discuss in detail how we determined
the phase diagram.  After that, we compute vector and scalar 
two-point functions numerically, and compare them with perturbation theory
in order to determine whether we do indeed obtain a theory of
free photons at the FM-FMD transition.  Finally, we summarize and
discuss our findings in the last section.  There are two appendices,
containing various technical details.

We would like to end this section with mentioning that,
recently, a gauge-invariant construction of (anomaly-free) abelian
chiral gauge theories on the lattice has been proposed, based on a
Dirac operator satisfying the Ginsparg--Wilson relation 
\cc{luescher_ab} (for the non-abelian case, 
see Ref.~\cc{luescher_nonab}).  The problems
with the violation of gauge invariance discussed above 
do not apply in this case (they might if a gauge non-invariant
approximation of the Dirac operator is used, however).  
An essential ingredient is that
the fermion measure includes a gauge-field dependent phase factor
which is determined by requiring the theory to be gauge invariant and local.
Sofar, however, no explicit expression of the fermion measure was 
given, making this approach as yet 
not suitable for a numerical investigation. 
\section{The Model} 
\secteq{2}
\lb{MOD}
The central idea of the gauge-fixing approach is to control the gauge  
degrees of freedom by a gauge-fixing procedure. 
The starting point is the gauge-fixed action 
in the continuum, the ``target theory." Correlation functions of the 
target theory in the continuum satisfy Slavnov--Taylor identities,
as a consequence of the gauge symmetry.
For an abelian gauge group, which will be the subject of this paper,
the target action in the continuum is of the form, 
\be
S_{\rm c}=
S_{{\rm c, G}}(A_\m)+S_{\rm c, F}(A_\m; \j_L, \j_R) + S_{{\rm c,
g.f.}}(A_\m) \;,
\lb{CACTION}
\ee
where $S_{{\rm c, G}}(A_\m)= {1\over 4} \int d^4x\, F_{\m\n}^2$ 
designates the gauge action, 
$S_{\rm c, F}(A_\m; \j_L, \j_R)$ is the chiral fermion action,
and $S_{{\rm c, g.f.}}(A_\m)$
the gauge-fixing action.  Here, we will consider 
the Lorentz gauge, which is renormalizable, and therefore allows
us to study the (relevant part of the) phase diagram in perturbation
theory.  No ghosts are needed in the abelian case,
and henceforth we will not introduce any ghosts on the lattice either
\cc{wmy_hn}.  We then have 
\bea
S_{{\rm c, g. f.}}(A_\m )    \!\!&=&\!\!
\frac{1}{2 \x} \int d^4x \left( \sum_\m \del_\m A_\m \right)^2\;, \lb{CSGF} 
\eea
where $\x$ is the gauge-fixing parameter. 
The goal is now to transcribe 
the target theory defined by the action (\eq{CACTION}) 
to the lattice using compact lattice 
link variables and the Haar measure as integration measure 
in the path integral. 

The fermion action leads to violations of the Slavnov--Taylor
identities if we use 
any one of the standard lattice fermion formulations, such as 
Wilson, staggered or domain-wall fermions, which are in conflict 
with chiral gauge invariance.  Moreover, we will consider a class
of gauge-fixing terms, dependent on a continuous parameter $\tr$.
They will lead to additional violations of the
Slavnov--Taylor identities. The Slavnov--Taylor identities are restored 
in the continuum limit by adding a finite number 
of counterterms to the action, and appropriately tuning their
coefficients \cc{rome} in the continuum limit.  

\begin{figure}[t]
\centerline{
\epsfxsize=4.0cm
\vspace*{0.5cm}
\epsfbox{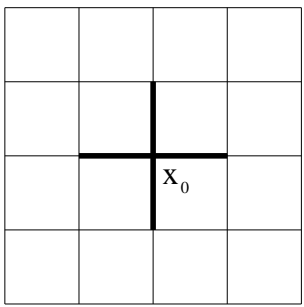}
}
\vspace*{0.0cm}
\caption{ \noindent {\em Example of a lattice Gribov copy. 
Shown is only a two-dimensional projection of the four-dimensional 
lattice. The 
$U_{\m x}$ fields on all the links attached to $x=x_0$ 
(fat lines) are equal to $-1$ and all 
other $U_{\m x}$ fields are equal to one.
}}
\label{GRIBOV}
\end{figure}

In this paper, we will drop   
the fermion sector and, as mentioned in Sect.~\ref{INTRO}, 
focus on the question whether the lattice discretization 
of the target action with U(1) symmetry provides a valid 
formulation of a theory of free photons on the lattice. 
The lattice action is formulated in terms of 
the compact link variables                                
$U_{\m x}=\exp(i a g  A_{\m x})$, with  
$g$ the gauge coupling and 
$a$ the lattice spacing (which we will set to one throughout
this paper). 
The lattice action is then given by the expression   
\be
S=S_{\rm G}(U)+S_{{\rm g. f.}}(U)
+ S_{{\rm c.t.}}(U)\;.
\lb{FULL_ACTION}
\ee
where $S_{\rm G}(U)$ is the gauge action, $S_{{\rm g. f.}}(U)$ the 
gauge fixing, and $S_{{\rm c.t.}}(U)$ the counterterm action on the lattice.
For the lattice transcription of the 
gauge action $S_{{\rm c, G}}(A_\m )$ 
we employed the standard plaquette action,               
\be
S_{{\rm G}}(U) = \frac{1}{g^2} \; \sum_{x}\sum_{\m <\n}
\left\{ 1-\mbox{Re } U_{\m \n x} \right\} \;,  \lb{SG} 
\ee
where $U_{\m \n x}=U_{\m x} U_{\n x+\hmu} U^{\dagger}_{\m x+\hnu}
U^{\dagger}_{\n x}$ is the usual lattice plaquette variable.
The lattice transcription of the gauge-fixing action 
is more subtle. A naive lattice discretization of 
Eq.~(\eq{CSGF}) leads to 
\be
S_{{\rm g. f.}}^{\rm naive}(U) = \tk \;  
\sum_x  \left( \sum_\m \D^{-}_\m V_{\m x}  \right)^2 \;,  
\lb{SNAIVE}
\ee
where 
\be
V_{\m x}=\mbox{Im }U_{\m x} \;,
\lb{VDEF}
\ee
\be
\tk=\frac{1}{2 g^2 \xi} \;,  \lb{TKGXI}
\ee
and $\D_\m^- V_{\m x}=V_{\m x}- V_{\m x-\hmu}$ ($\D_\m^-$
is the backward nearest-neighbor lattice derivative).
The problem with the naive lattice discretization 
is that the classical vacuum of the action 
$S_{{\rm G}}(U)+S_{{\rm g. f.}}^{\rm naive}(U)$
is not unique \cc{yigal}. It is easy to see that 
$S_{{\rm G}}(U)+S_{{\rm g. f.}}^{\rm naive}(U)$
has absolute minima for a  
dense set of lattice Gribov copies 
$U_{\m x}=g_x 
\; 1 \; g_{x+\hmu}^{\dagger}$, of the classical vacuum $U_{\m x}=1$
for particular sets of $g_x$. 
An example of such lattice 
Gribov copies is displayed in Fig.~\ref{GRIBOV},
%
where we set $g_{x}=-1$ at site $x=x_0$ 
and $g_x=+1$ at all sites $x \neq x_0$. 
The $U_{\m x}$ fields on the links attached to the site $x_0$ (fat lines) are 
equal to $-1$ whereas the $U_{\m x}$ fields on all other links 
(thin lines) are equal to $+1$.
These lattice Gribov copies are a high momentum 
lattice artifact with no counterpart in the continuum. (They 
should not be confused with continuum Gribov 
copies, which are a long-distance phenomenon.)   It is clear that
perturbation theory around only one of the absolute minima
of $S_{\rm G}(U)+S^{\rm naive}_{\rm g.f.}(U)$, in particular
the classical vacuum $U_{\m x}=1$, may not
give a valid description of the theory, and the phase diagram could
(and will) turn out very different from what one would expect from naive
perturbation theory.  

It is however possible to remove the unwanted lattice Gribov copies 
of the classical vacuum (and therefore, by continuity, of field
configurations perturbatively close to the vacuum) by adding 
a higher dimensional operator to the gauge-fixing action 
in Eq.~(\eq{SNAIVE}). This procedure is similar to Wilson's
idea of removing the species doublers of the naive lattice fermion action 
by adding an operator of dimension larger than four (the Wilson term). 
Such higher dimensional operators 
do not affect the small-momentum behavior of the theory, but 
can be used to change the behavior at large momenta.

The gauge-fixing action we will use in this paper is given by the 
expression 
\be
S_{{\rm g. f.}}(U) =S_{{\rm g. f.}}^{\rm naive}(U)
+ \tr \; \tk \; \sum_x 
\left\{ \quart\;\left(C_x +C_x^{\dagger} \right)^2 -B_x^2 \right\}
\;,  \lb{SGF}
\ee
where 
\be
C_x=\sum_y \Box_{x y} (U)\;,
\lb{CC}
\ee
\be
B_x=\sum_\m \left( \frac{V_{\m x-\hmu} + V_{\m x} }{2} \right)^2 \;, \lb{BX}
\ee
and 
\be
\Box (U)_{xy} = \sum_\m \left\{ U_{\m x}\; \d_{x+\hmu,y}
                               + U_{\m x-\hmu}^{\dagger}\; \d_{x-\hmu,y}
                               -2 \; \d_{x,y} \right\} \lb{LL}
\ee
is the covariant lattice Laplacian. In Eq.~(\eq{SGF}) we have 
multiplied the higher dimensional operator by a new parameter $\tr$ which  
can be viewed as the analog 
of the Wilson parameter $r$ that multiplies the 
Wilson term. It can be shown that the action
$S_{{\rm G}}(U)+S_{{\rm g. f.}}(U)$
has, for $\tr> 0$,  a unique absolute
minimum at $U_{\m x}=1$ \cc{my_plb} so that, for
$\tr>0$, standard perturbation theory in $g$ is valid.
The gauge-fixing action provides 
a continuous interpolation between 
the naive gauge-fixing action (\eq{SNAIVE}) ($\tr=0$) and 
the gauge-fixing action at $\tr=1$ which was 
introduced previously in Ref.~\cc{my_plb} and was used in 
Refs.~\cc{wmy_pd,wmy_pert,wmy_prl,wmy_edin,wmy_buk,asit}. 

We will study the $\tr$ dependence of the phase diagram of the 
purely bosonic theory, and explore the effects of the lattice Gribov 
copies on the phase structure. Obviously, lattice Gribov copies 
introduce rough gauge degrees of freedom. It is 
therefore conceivable that  
lattice Gribov copies at small $\tr$ 
give rise to a disordered or symmetric phase. It has been argued earlier 
that a chiral gauge theory cannot be obtained in such a phase \cc{yigal_nogo}.
This would teach us that, in the gauge-fixing approach to lattice
chiral gauge theories, one needs to choose $\tr$ of order one.

Finally, we have to specify the counterterm action. 
It contains all relevant and marginal operators
which are allowed by the exact lattice symmetries~\cite{rome}.
In the case of the U(1) gauge theory there are six such terms:
\bea
S_{{\rm c.t.}} (U)\!\!&=&\!\! -\k
\sum_{\m x} \left\{ U_{\m x}+ U_{\m x}^{\dagger}  \right\}
- \; \l_1 \; \sum_x \sum_{\m \n} 
\left( \D_\n^- \; \mbox{Im }U_{\m x} \right)^2 \nonumber \\
\!\!&-&\!\!\l_2 \; \sum_x \sum_\m \left( \D_\m^- \mbox{Im }U_{\m x} \right)^2 
-\l_3\; \sum_x \left( \sum_\m \D_\m^- \mbox{Im }U_{\m x} \right)^2 
\nonumber \\
\!\!&-&\!\!\l_4\;\sum_x \left( \sum_\m (\mbox{Im }U_{\m x})^2\right)^2
-\l_5 \; \sum_x \sum_\m \left(\mbox{Im }U_{\m x} \right)^4
\;. \lb{SC}
\eea
The term proportional to $\k$ is a mass counterterm for the gauge 
field. It is the only dimension-two counterterm. All other counterterms 
are of dimension four. An expansion 
in $g$ shows that the counterterms with coefficients $\l_1$, 
$\l_2$ and $\l_3$ are wave-function renormalization counterterms.
The terms with the coefficients $\l_4$ and $\l_5$ are, to leading order
in $g$, quartic in the gauge potential, and are needed 
to eliminate photon self-interactions. The coefficients of the 
counterterms have to be tuned such that 
the Slavnov--Taylor identities are satisfied in the continuum limit
({\it cf.} Sect. \ref{LAMBDAS}).  Note that terms such as 
$\sum_x \sum_\m (\D_\m^- \mbox{Im }U_{\m x})\sum_\n (\mbox{Im }U_{\n x})^2$
do not occur because of the symmetry $U_{\m x}\to U^\dagger_{\m x}$.
However, such counterterms will have to be considered once the theory
is coupled to fermions.

The path integral is defined by  
\bea
Z\!\!&=&\!\!\int D U \; \exp( -S(U)) \lb{PATHIV} \\
\!\!&=&\!\!
\int D U D \phi \;
   \exp( -S(\phi_x^{\dagger} U_{\m x} \phi_{x+\hmu} ))  \; , \lb{PATHIH}
\eea  
where the action $S(U)$ is given in Eq.~(\eq{FULL_ACTION}), and 
the integration over the link variables is performed with 
the usual Haar measure. Only the gauge action in Eq.~(\eq{FULL_ACTION}) 
is invariant under the local gauge transformation 
\be
U_{\m x} \ra U_{\m x}^g=g_x U_{\m x} g_{x +\hmu}^{\dagger} \;. \lb{GAUGET}
\ee

In Eq.~(\eq{PATHIH}) 
we made the integration over the longitudinal 
gauge degrees of freedom $\phi_x$
explicit. It is easy to see that 
Eq.~(\eq{PATHIV}) follows from Eq.~(\eq{PATHIH}) 
by performing the gauge transformation (\eq{GAUGET}) with 
$g_x=\phi^\dagger_x$ and using the fact that $\int D \phi =1$ \cc{long}.
The longitudinal modes can be viewed as
group-valued Higgs fields $\phi_x$. We will therefore denote in the following 
$S(\phi_x^{\dagger} U_{\m x} \phi_{x+\hmu} )$ as the action in the ``Higgs 
picture'' and $S(U)$ as the action 
in the ``vector picture.'' Both pictures are equivalent and 
every observable in the vector picture has, according to Eq.~(\eq{GAUGET}), 
a counterpart in the Higgs picture. 
The path integral in the Higgs picture is invariant under the 
local transformation 
\be
\phi_x \ra h_x \phi_x\;, \;\;\;\; 
U_{\m x} \ra h_x U_{\m x} h_{x+\hmu}^{\dagger} \;.
\ee

Finally, we mention here             
that in $d$ space-time dimensions the path integrals 
(\eq{PATHIV}) and (\eq{PATHIH}) 
are invariant under the additional discrete symmetry 
(with all $\l_i=0$)
\be
U_{\m x} \ra -U_{\m x}  \;, 
\;\;\;\;\; \k \ra -\k -8 \; d\; \tk \; \tr
\;, \;\;\;\;\; \tk \ra \tk \;, \;\;\;\;\; \tr \ra \tr
\;, \;\;\;\;\; g \ra g \;, \
\lb{SPTRAF}
\ee
which implies that we can restrict ourselves to the 
$\k > -4d \tk \tr$ region of the $(\k,\tk, \tr)$ phase diagram. 
In the following we will refer to 
\be
\k=-4 d \tk \; \tr \lb{SP}
\ee
as the symmetry surface, or symmetry line (point) 
if $\tk$ or (and) $\tr$ are kept fixed.  

At $\tk=0$ and $\l_i=0$, $i=1,\ldots,5$, the model reduces to the 
U(1) gauge-Higgs model with group-valued Higgs fields. The 
resulting $(\k,g)$ phase diagram corresponds 
to the $\tk=0$ hyperplane of the four-dimensional $(\k, g, \tk,
\tr)$ phase diagram (for $\tk=0$ the model is independent of $\tr$). 
In the limit $g \ra 0$ the phase diagram of the gauge-Higgs model 
reduces to that of the four-dimensional XY model. This phase diagram 
of the XY model contains three phases: 
a ferromagnetic (FM) or broken phase at  
$\k > \k_{\rm FM-PM} \approx 0.15$, a paramagnetic (PM) or symmetric phase 
at $-\k_{\rm FM-PM} \leq  \k \leq  \k_{\rm FM-PM}$, and an antiferromagnetic 
(AM) phase at $\k < -\k_{\rm FM-PM}$. The $\k > 0$ region is mapped 
by the symmetry (\eq{SPTRAF}) into the $\k < 0$ region. This implies that 
the FM phase is mapped into the AM phase and vice versa. 
The FM-PM transition, and, because of the symmetry (\eq{SPTRAF}), also the 
PM-AM phase transition are second-order phase transitions. 
At $g>0$ the FM phase turns into a Higgs and the 
PM phase into a Coulomb phase, which we will still 
denote as FM and PM phases, respectively.
There is strong evidence that also 
the FM-PM phase transition at $g>0$ is of second order \cc{order}.
The spectrum in the FM phase contains a massive vector boson 
with quantum numbers $J^{\rm PC}=1^{- -}$ and a massive 
Higgs boson 
with quantum  numbers $J^{\rm PC}=0^{++}$. The masses of both particles
are expected to scale when $\k$ is tuned towards the 
second-order FM-PM phase transition. 
The spectrum in the Coulomb phase contains only a massless 
photon if one keeps away from the FM-PM phase transition \cc{spectrum}. 
The model in the Coulomb phase provides a valid formulation of free
photons on the lattice but, as we explained 
before, because of the strongly fluctuating longitudinal gauge modes, 
it is not possible to formulate a chiral gauge theory with a gauge 
non-invariant fermion action in that phase. 
The spectrum near the FM-PM phase transition contains, apart from the 
massless photon, also positronium-like bound states of scalars \cc{spectrum}. 
At large $g$ 
and small $\k$ the phase diagram contains a confining (CF) phase, 
which is separated from the PM phase by a phase transition, 
located at $g \approx 1$.

In Refs.~\cc{wmy_pd,wmy_pert,wmy_prl,wmy_edin,wmy_buk}, 
we have studied the U(1) model 
with $\l_i=0$, $i=1,\ldots,5$ in the reduced limit where 
the gauge coupling is tuned to zero, while keeping 
$\tk=1/(2 g^2 \xi)$ fixed (reduced model). In this limit the transverse 
gauge degrees of freedom are gone and only the dynamics 
of the longitudinal gauge degrees of freedom remains. 
The path integral of this reduced model is given by 
Eq.~(\eq{PATHIH}) with all link variables set equal to one.  
The phase diagram of the reduced model at $\tr=1$ 
contains, apart from the FM and PM phases, 
also a new type ferromagnetic 
directional (FMD) phase where the vector field 
$\mbox{Im}\,\phi_x^{\dagger} \phi_{x+\hmu}$ condenses and  
hypercubic rotation invariance on the lattice is broken
({\it cf.} the $\tr=1$ plane of Fig.~\ref{SCHEM}, which
is very similar to the $\tr=1$ reduced-model phase diagram). 

The phase transition between the FM and FMD 
phases, which, as we will see, is also present in the full U(1) model,
plays a crucial role in the gauge-fixing approach. 
While the photon is massive in the FM phase, 
one obtains a massless photon by tuning $\k$, from within 
the FM phase, towards this FM-FMD phase transition.  Tuning all other
counterterms as well should then lead to a theory of free, relativistic
photons.  We should note here that $\k$ is the only relevant
counterterm parameter, whereas all the $\l_i$ are marginal.  We 
therefore may expect that only $\k$ needs to be tuned nonperturbatively,
while, with a given numerical precision, a one-loop, or even 
tree-level determination of the $\l_i$ will suffice.  Indeed, the
results which we will present in the following sections provide
evidence of this. 

With chiral fermions, we demonstrated that, in the reduced limit, 
the fermion spectrum 
contains only the desired chiral states \cc{wmy_prl,wmy_edin,wmy_buk}. 

The phase diagram at small $g$ 
is expected to look similar to the phase diagram of the reduced model, 
since the transverse components of the gauge fields are still 
very small.  We will show in the following sections that 
even at a relatively large value of $g$ ($g=0.6$) the phase diagram is 
qualitatively very similar to the one for the reduced model, if
$\tr\approx 1$.
\section{Analytical results}
\secteq{3}
\lb{ANA}
In this section we present our analytical results. Subsect.~\ref{CLASS}
deals with the constant-field approximation, which already gives 
some insight into the phase structure. 
In Subsect.~\ref{MEAN} we will determine the $(\k, \tk, \tr)$ phase 
diagram at fixed small $g$ in the mean-field approximation. 
The counterterm coefficients $\k$, $\l_1$, $\l_2$ and $\l_3$ 
are calculated to one-loop order in perturbation theory 
in Subsect.~\ref{LAMBDAS}. 
In Subsect.~\ref{PROP_AN} we analyze the spectrum near the 
FM-FMD phase transition in perturbation theory.
\subsection{Constant-field approximation}
\lb{CLASS}
In the constant-field approximation we set 
\be
U_{\m x}=\exp( i g A_\m) \;,  \lb{CONSTF}
\ee
where $A_\m$ is a space-time-independent vector potential. After inserting 
(\eq{CONSTF}) into Eq.~(\eq{FULL_ACTION})
all terms which contain derivatives of the gauge field vanish, and 
we obtain an expression for the classical potential density.           
Expanding the resulting expression 
in powers of $g$, we find  
\bea
V_{{\rm cl}}(A_\m)\!\!&=&\!\! \k \left\{ g^2 \sum_\m A_\m^2 + \ldots \right\}
+ \half \; g^6 \; \tk \; \tr \; \left\{ \left( \sum_\m A_\m^2 \right)
\left( \sum_\m A_\m^4 \right) +\ldots \right\} \nonumber \\
\!\!&\phantom{=}&\!\! -\l_4 \; g^4 \; \left\{ \left( \sum_\m A_\m^2 \right)^2
+ \ldots \right\}
-\l_5 \; g^4 \; \left\{  \sum_\m A_\m^4  +\ldots \right\}\; , \lb{VCL}
\eea
where ``$\ldots$'' represents terms which are of higher 
order in $g^2$. A massless non-interacting photon is 
obtained for $\k =0$, $\l_4 =0$ and $\l_5 =0$.
The relation $\k=\k_{\rm FM-FMD}=0$ defines a critical surface in the 
three-dimensional 
$(\k, \tk, \tr)$ phase diagram where the photon mass vanishes. 
We will see later that the critical coupling
$\k_{{\rm FM-FMD}}(\tk ,\tr)$ is shifted away from zero by perturbative
corrections.                              

The minimization of the classical potential 
density (\eq{VCL}) for $\l_4=\l_5=0$ shows that
\be
\begin{array}{ll}
\lag  g \; A_\m \rag = 0\;, &
 \mbox{for } \k \geq \k_{{\rm FM-FMD}} \;,  \\
\lag g \; A_\m \rag = \pm \left( |\k -\k_{{\rm FM-FMD}}| / (6 \; \tk \; \tr)
\right)^{1/4}
\;,  &  \mbox{for } \k < \k_{{\rm FM-FMD}} \;, 
\end{array}
\lb{CFMD}
\ee
for $\m=1,\ldots,4$ (see Ref.~\cc{my_plb}, which also deals with the
case $\lambda_4\ne0$, $\lambda_5\ne 0$).
This implies that $\k=\k_{{\rm FM-FMD}}=0$ corresponds to a phase transition
between the FM phase where $\lag A_\m \rag$ vanishes
and  the gauge boson has a nonzero mass, and
the FMD phase with a nonvanishing vector condensate $\lag A_\m \rag$. 
The hypercubic rotation invariance on the lattice is broken in the 
FMD phase by the nonvanishing vector condensate.  

The constant-field approximation is supposed to provide a satisfactory 
description of the model only at large $\tk\sim 1/g^2$, where
strongly fluctuating gauge configurations are suppressed by 
a small Boltzmann weight and the smoother gauge configurations 
are well approximated by a constant field.  Indeed, as we will see,
a perturbative expansion starting from the constant-field approximation
can be developed by expanding $U_{\m x}={\rm exp}(igA_{\m x})$ and 
setting $\tk=1/(2\x g^2)$, with $\x$ fixed. 
The situation is different at small $\tk$ where rough longitudinal
fields 
are not sufficiently suppressed. This picture is confirmed by our previous 
investigation of the reduced model at 
$\tr=1$, where we find indeed a FM-FMD phase transition  at 
large $\tk$ (consistent with the constant field-approximation), but an 
FM-PM phase transition at small $\tk$. 
The emergence of the disordered PM phase at small $\tk$ is 
due to the dominance of rough gauge field configurations. 
Equation~(\eq{VCL}) shows that the classical potential density 
depends only on the product $\tk \tr$. This suggests that a small 
value of $\tr$ has the same effect as a small value of $\tk$
and that the ``small $\k$''-region around $\tr=0$ might 
indeed be filled with a PM phase. 
This conjecture will be confirmed by our mean-field analysis and 
Monte Carlo simulations.
\subsection{Mean Field Calculation of the Phase diagram} 
\lb{MEAN}
In the following we set again $\l_i =0$, $i=1,\ldots ,5$ and determine
the phase boundaries in the $(\k,\tk,\tr)$ space. 
The mean-field analysis for gauge theories on the
lattice is ambiguous as it is in conflict with
local gauge invariance \cc{dz}.  For 
gauge-Higgs models the mean-field calculation in the 
vector picture leads to a wrong phase structure 
at small values of $\k$, whereas the mean-field approximation 
in the Higgs picture leads to a phase diagram which, at least qualitatively, 
complies at small $\k$ with Monte Carlo simulations.
Since we are interested in the phase diagram at small $\k$, 
we performed our mean-field analysis in the Higgs picture. 
Here, we will only describe the results, while
relegating all the technicalities to Appendix A. 

\begin{figure}
\vspace*{-1.5cm}
\begin{tabular}{ll}
 \hspace*{-1.3cm} \epsfxsize=10.00cm
 \vspace*{-0.5cm}
\epsfbox{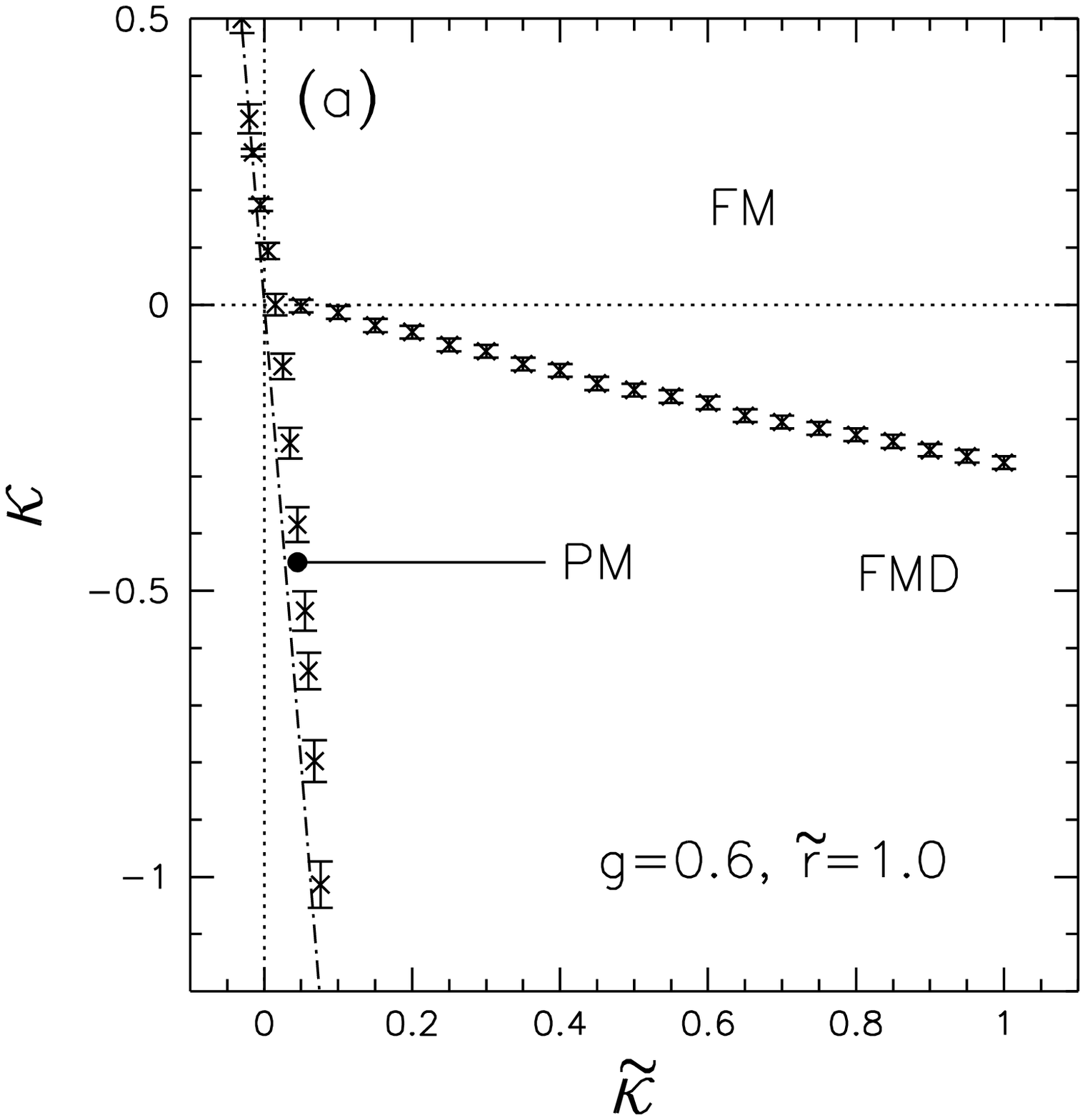}     &
 \hspace{-4.0cm} \epsfxsize=10.00cm
 \vspace*{-1.0cm}
\epsfbox{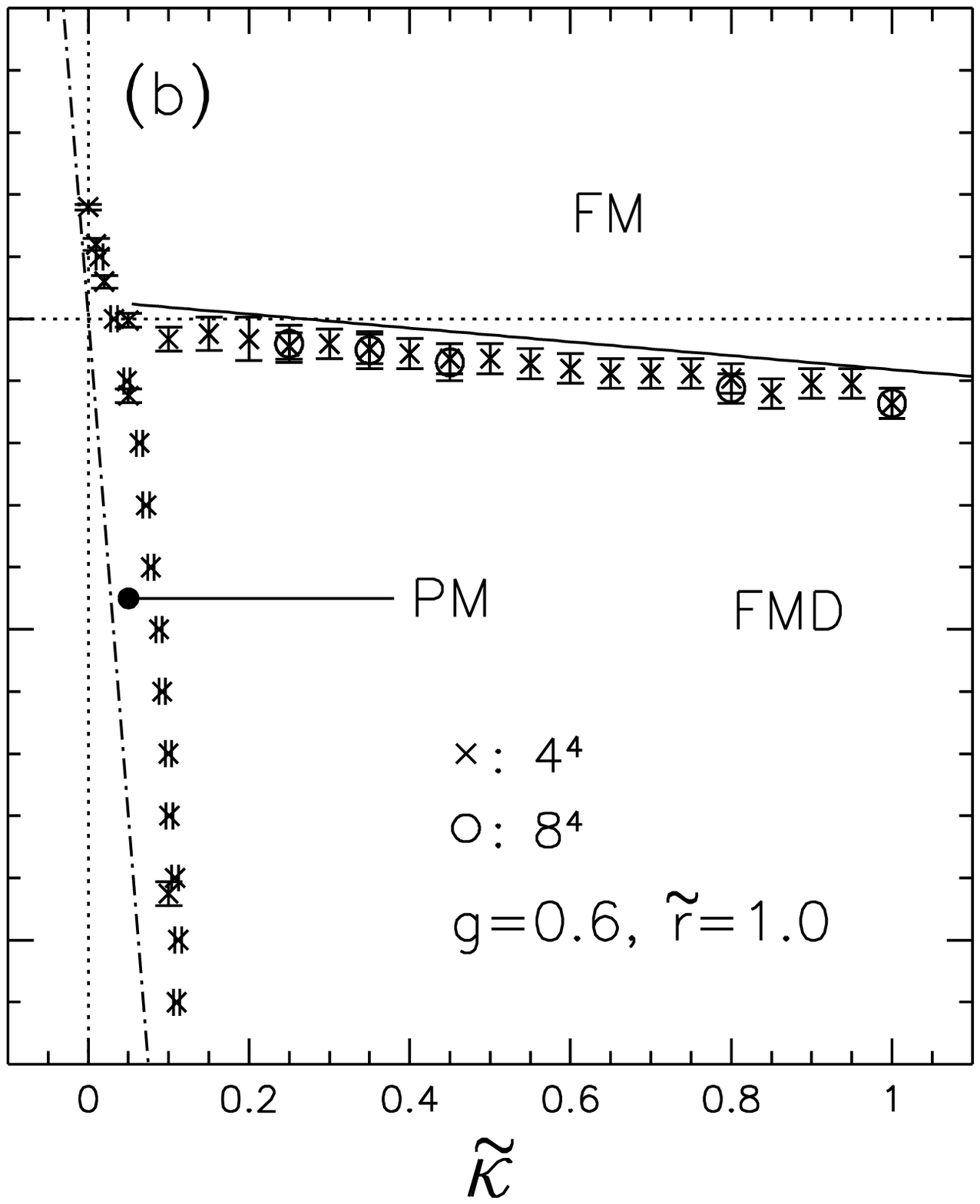}     \\
 \hspace*{-1.3cm} \epsfxsize=10.00cm
 \vspace*{-0.5cm}
\epsfbox{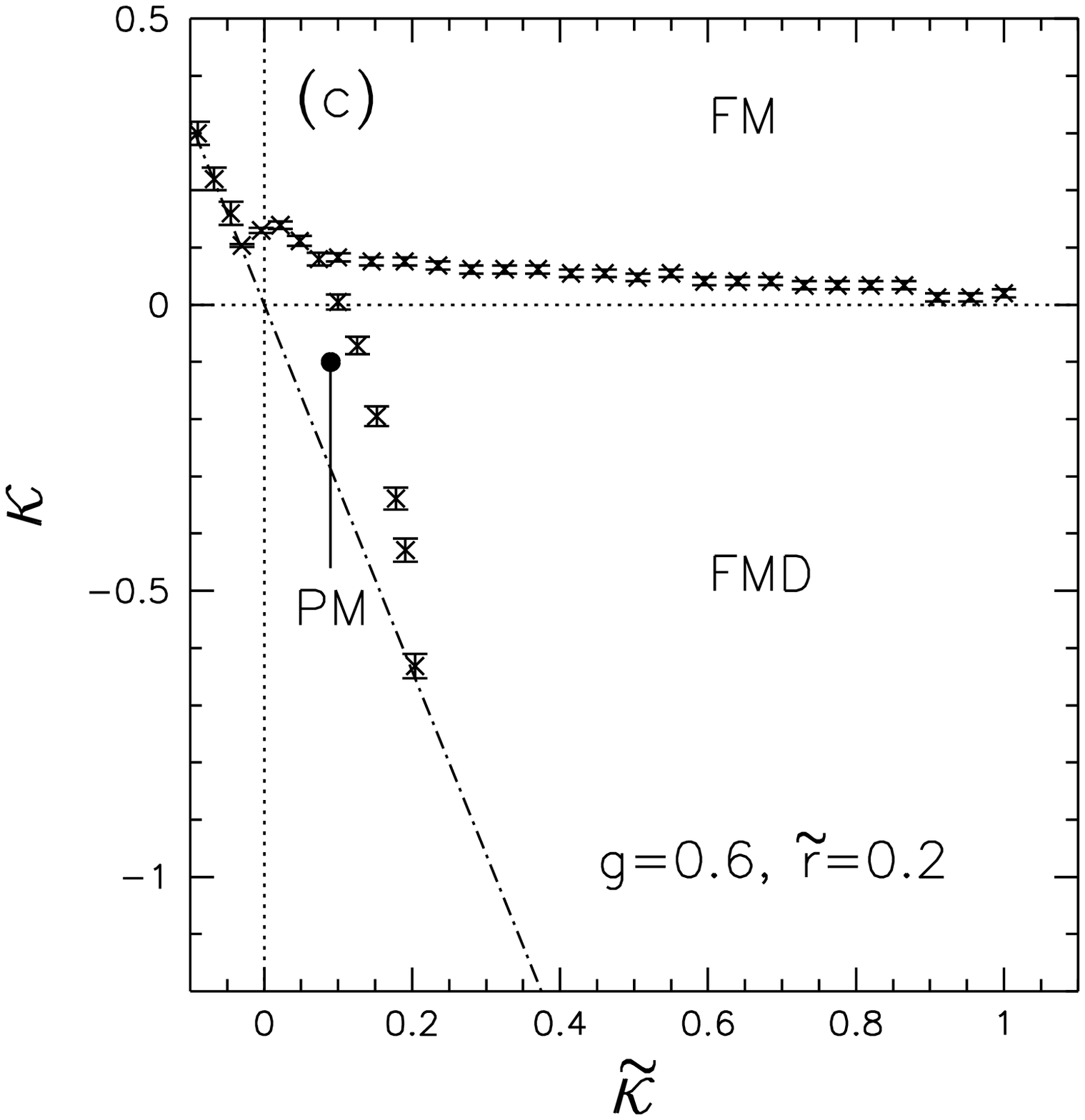}     &
 \hspace{-4.0cm} \epsfxsize=10.00cm
 \vspace*{-1.0cm}
\epsfbox{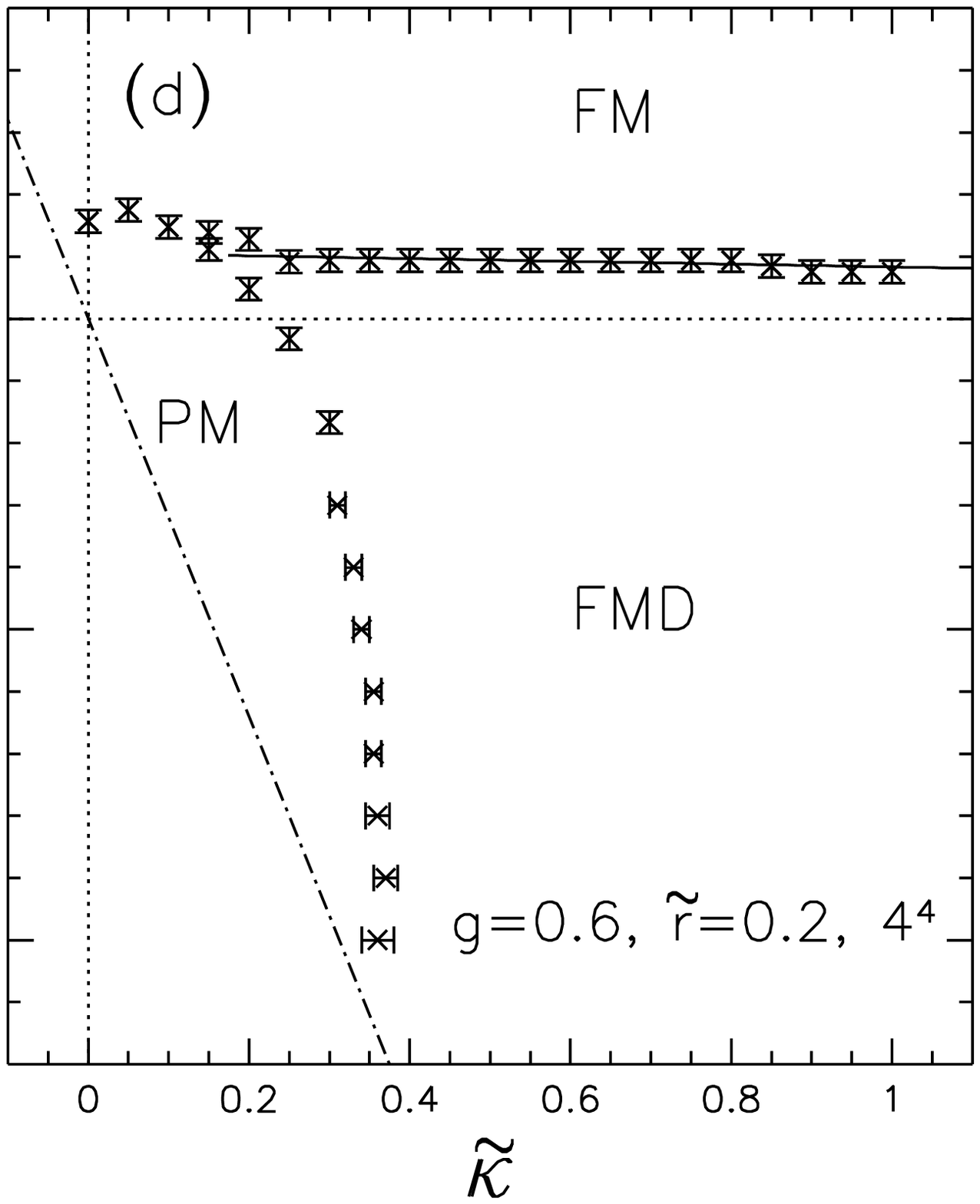}     \\
\end{tabular}

\vspace*{0.2cm}
\caption{ \noindent {\em  The $(\k, \tk)$ phase diagrams            
at $\tr=1$ (Figs. a and b) and $\tr=0.2$ (Figs. c and d).
The plots on the left were obtained by a mean-field calculation, and
those on the right by Monte Carlo simulations on a $4^4$ lattice. 
The dash-dotted 
lines mark the symmetry line ({\it cf.} Eq.~({\protect \eq{SP}})). 
The perturbative result for the FM-FMD transitions is represented in the
plots on the right by solid lines. The relation between 
the perturbative and Monte Carlo results will be discussed in 
Sect.~{\protect \ref{NUM_PHAS}}.
}}
\label{PHASED1}
\end{figure}

\begin{figure}
\vspace*{-0.5cm}
\begin{tabular}{ll}
 \hspace*{-1.3cm} \epsfxsize=10.00cm
 \vspace*{-0.5cm}
\epsfbox{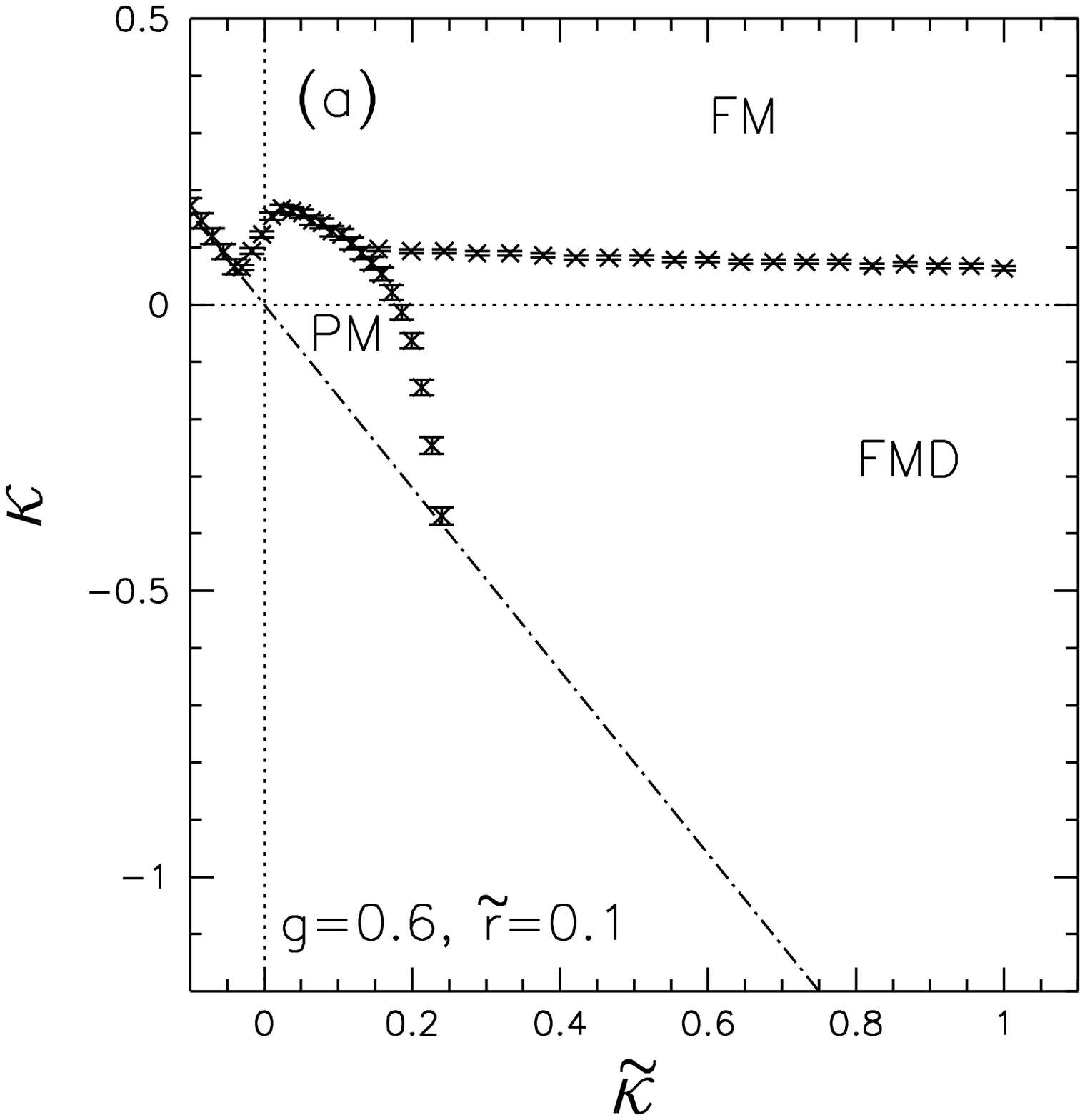}     &
 \hspace{-4.0cm} \epsfxsize=10.00cm
 \vspace*{-1.0cm}
\epsfbox{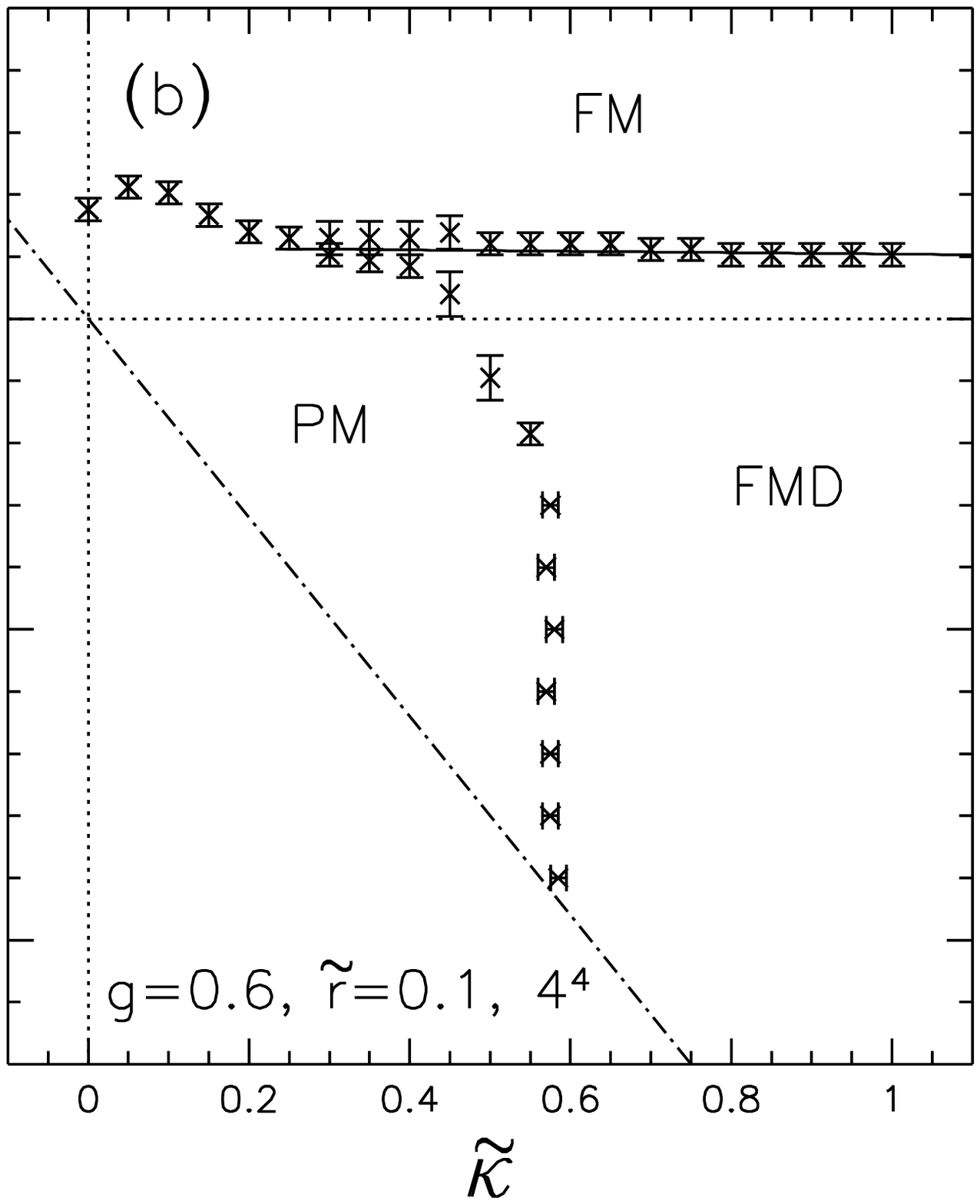}     \\
 \hspace*{-1.3cm} \epsfxsize=10.00cm
 \vspace*{-0.5cm}
\epsfbox{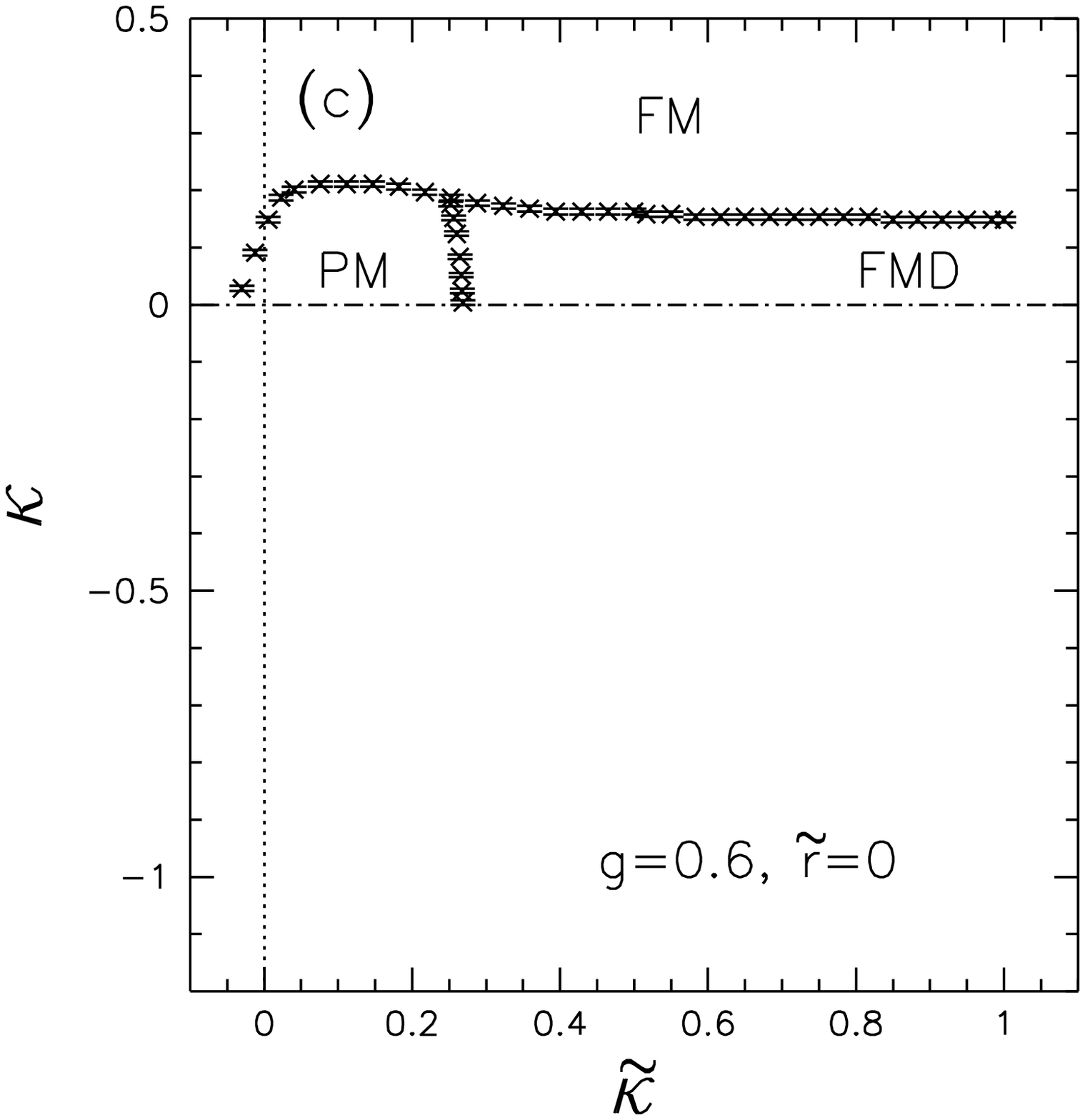}     &
 \hspace{-4.0cm} \epsfxsize=10.00cm
 \vspace*{-1.0cm}
\epsfbox{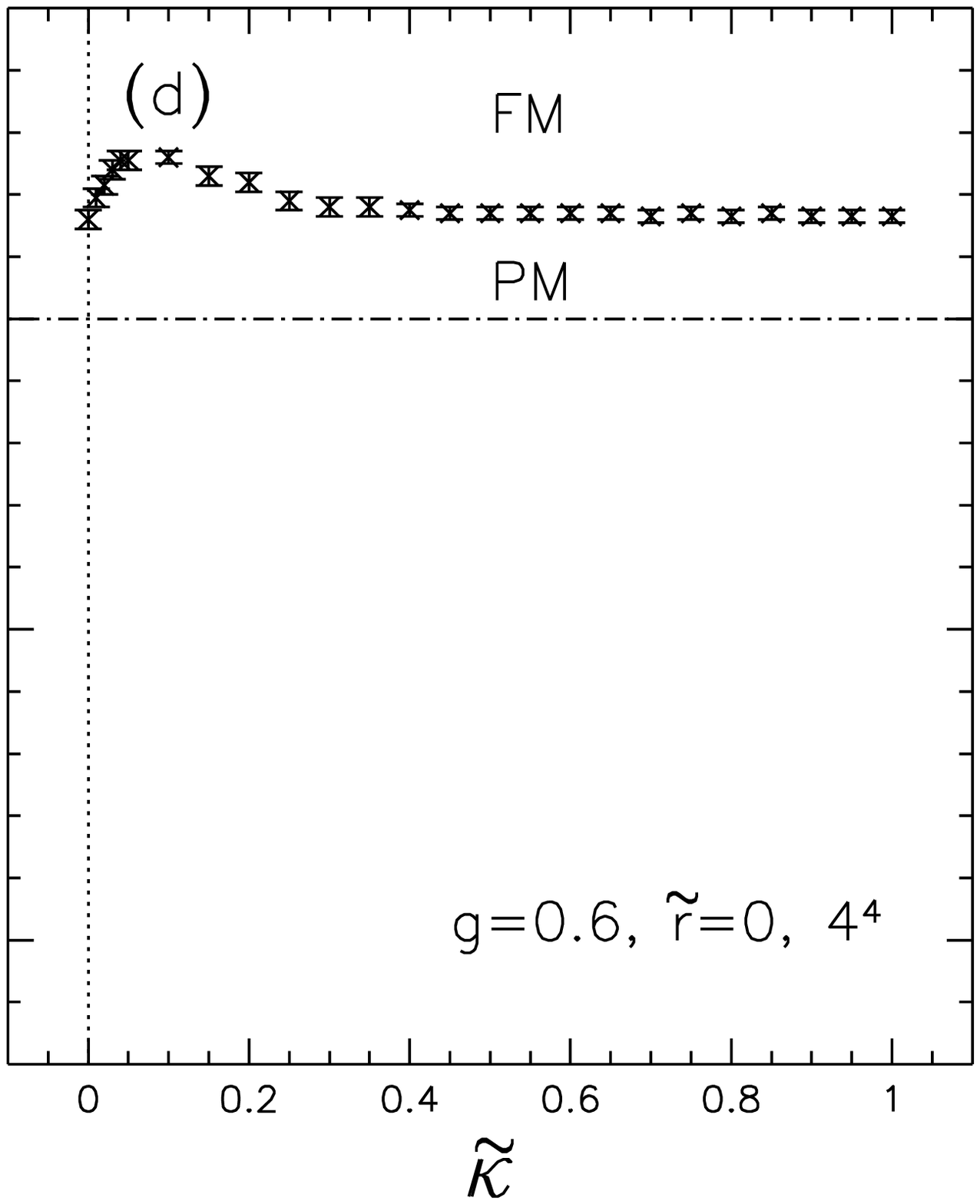}     \\
\end{tabular}
\vspace*{-0.0cm}
\caption{ \noindent {\em Same as Fig.~{\protect \ref{PHASED1}}, but for 
$\tr=0.1$ (Figs. a and b) and $0$ (Figs. c and d). 
Note that the symmetry line in Figs. c and d 
coincides with the $\k=0$ axis. 
}}
\label{PHASED2}
\end{figure}

\begin{figure}
\vspace*{-0.5cm}
\begin{tabular}{ll}
 \hspace*{-1.3cm} \epsfxsize=10.00cm
 \vspace*{-0.5cm}
\epsfbox{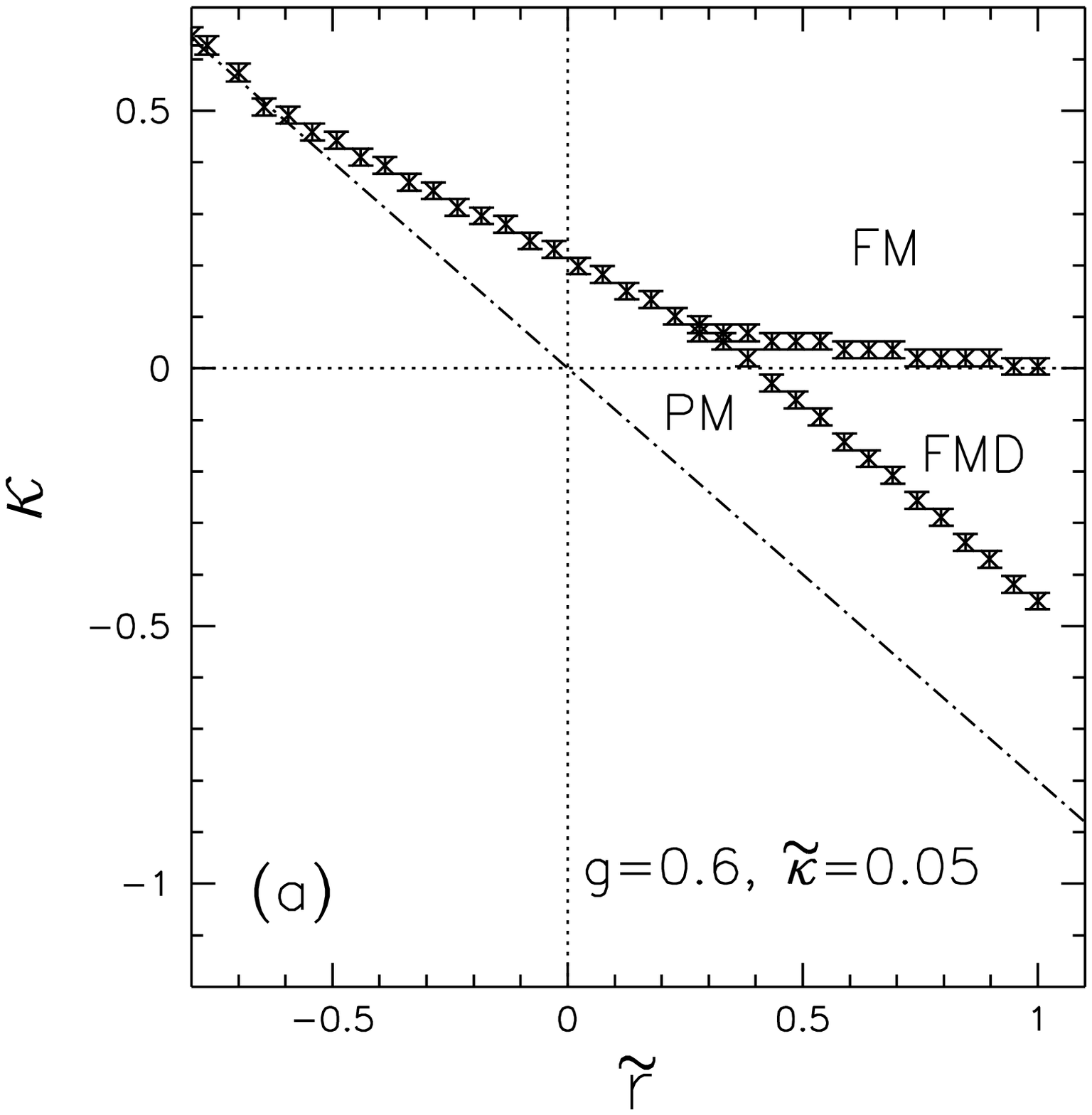}     &
 \hspace{-4.0cm} \epsfxsize=10.00cm
 \vspace*{-1.0cm}
\epsfbox{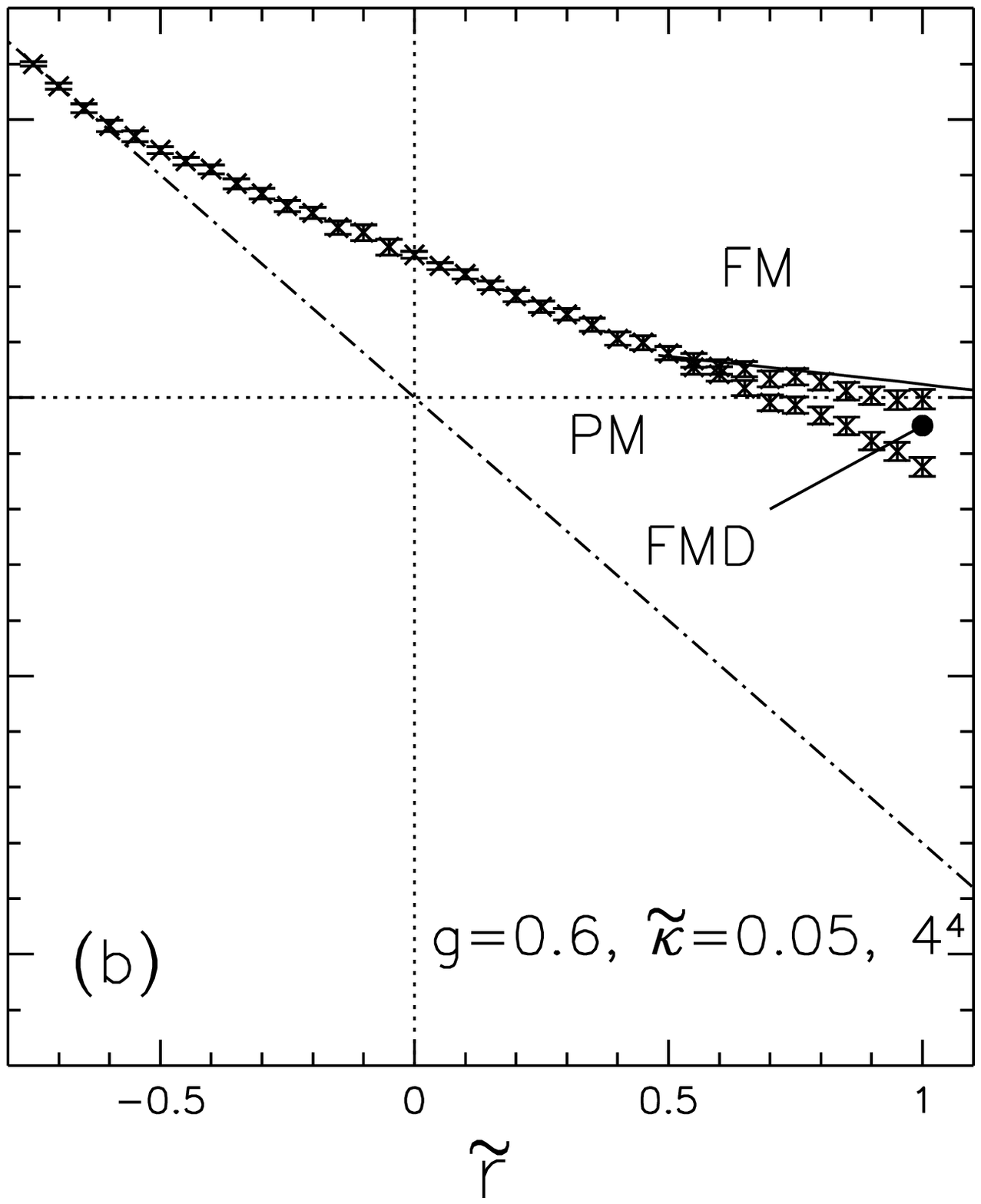}     \\
 \hspace*{-1.3cm} \epsfxsize=10.00cm
 \vspace*{-0.5cm}
\epsfbox{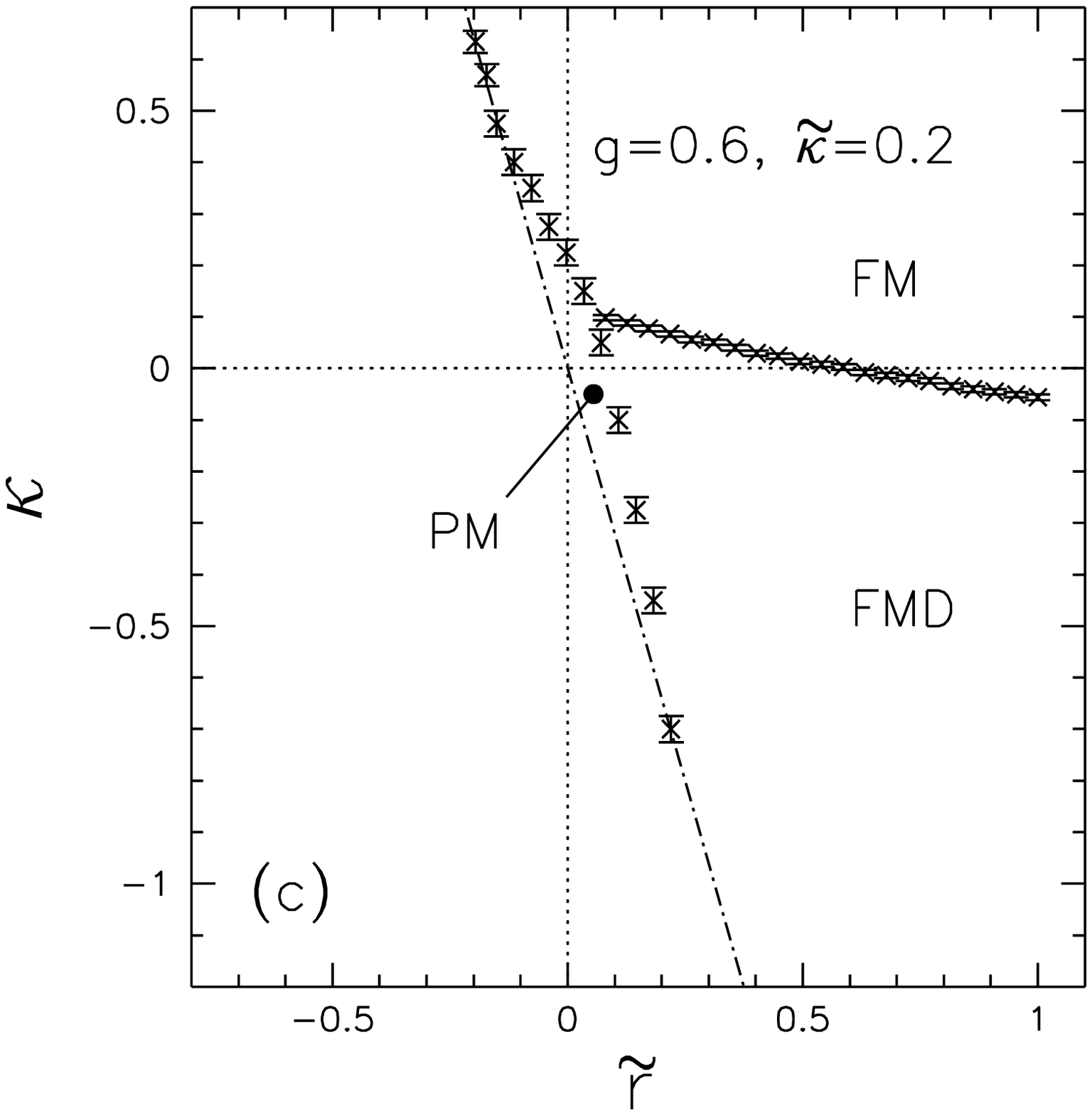}     &
 \hspace{-4.0cm} \epsfxsize=10.00cm
 \vspace*{-1.0cm}
\epsfbox{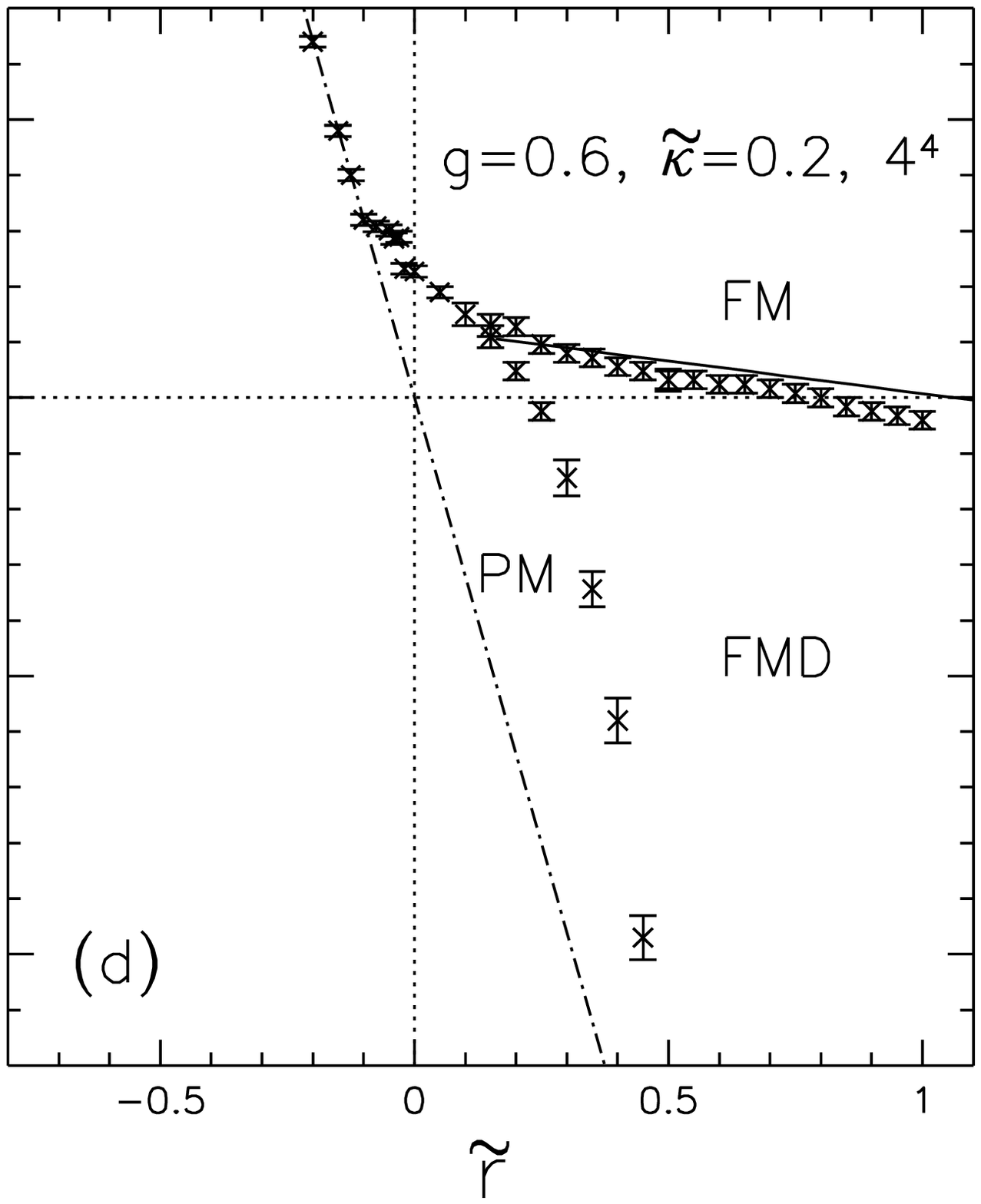}     \\
\end{tabular}
\vspace*{-0.0cm}
\caption{ \noindent {\em  The $(\k, \tr)$ phase diagrams for           
$\tk=0.05$ (Figs. a and b) and $0.2$ (Figs. c and d). The 
mean-field phase diagrams are displayed again on the 
left, and those from the Monte Carlo simulations on the 
right.  The dash-dotted line is the symmetry line.
The perturbative result for the FM-FMD transitions is represented in the
plots on the right by solid lines. The relation between 
the perturbative and Monte Carlo results will be discussed in 
Sect.~{\protect \ref{NUM_PHAS}}.
}}
\label{PHASED3}
\end{figure}

\begin{figure}
\vspace*{-0.5cm}
\begin{tabular}{ll}
 \hspace*{-1.3cm} \epsfxsize=10.00cm
 \vspace*{-0.5cm}
\epsfbox{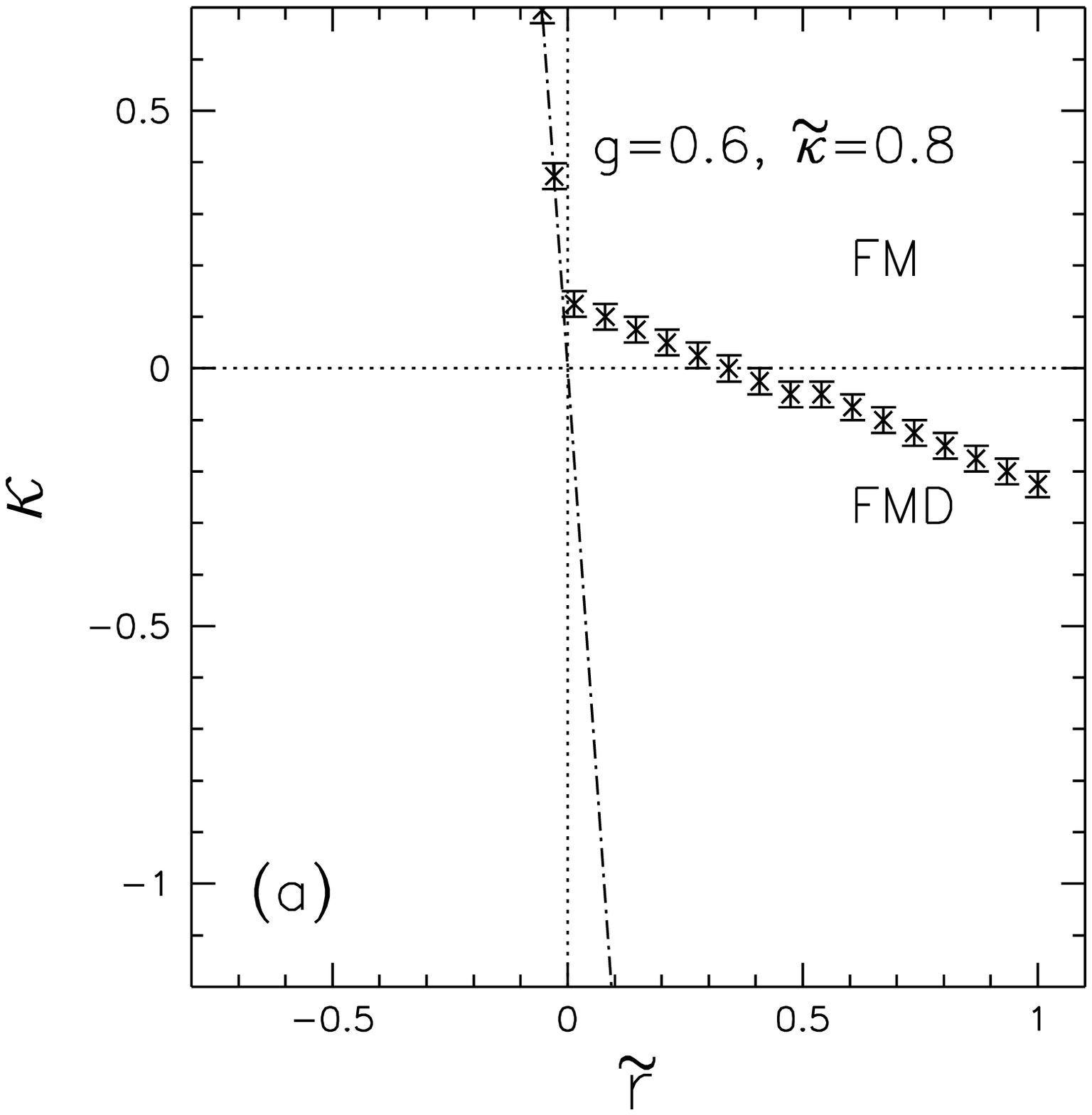}     &
 \hspace{-4.0cm} \epsfxsize=10.00cm
 \vspace*{-1.0cm}
\epsfbox{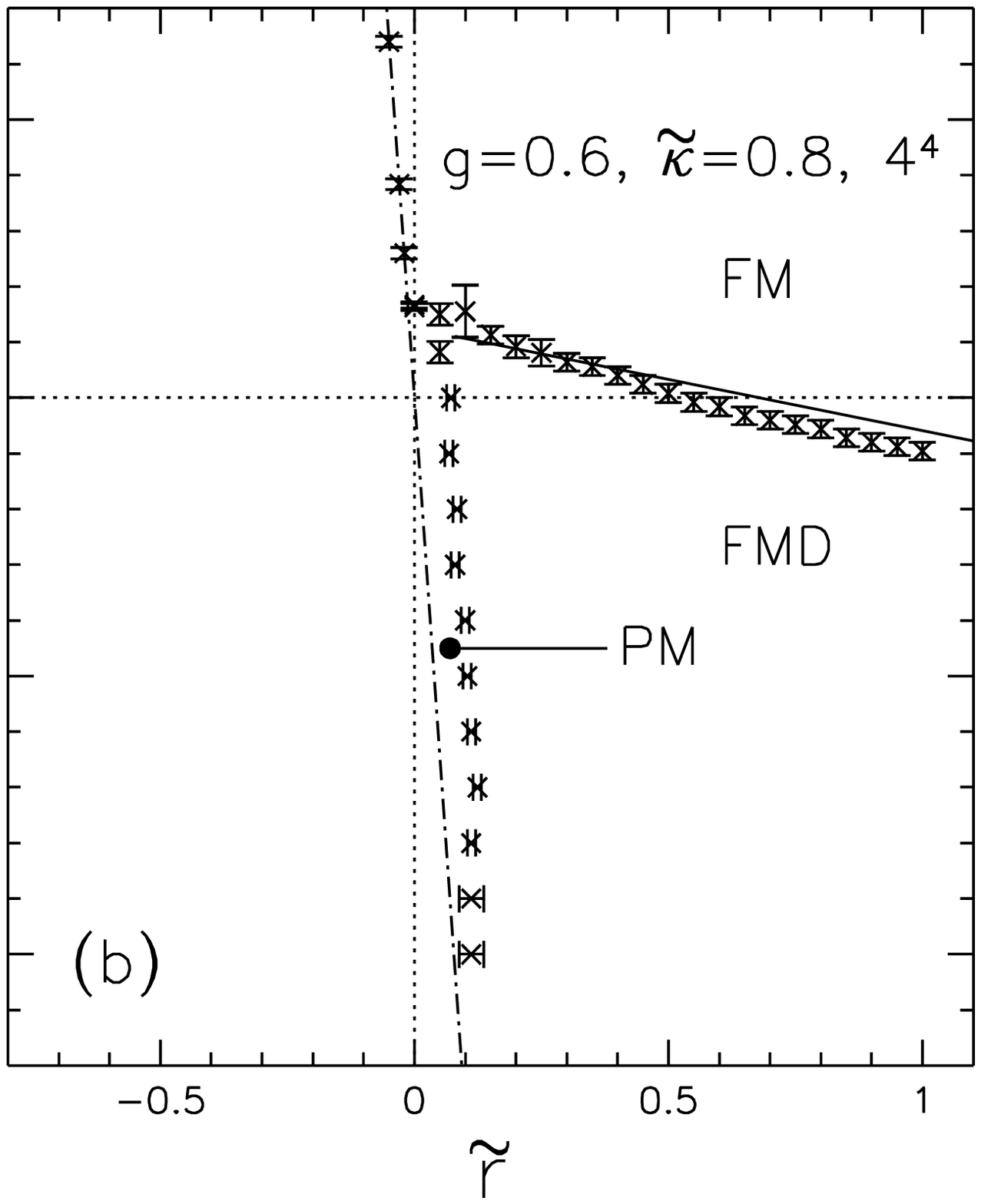}     \\
\end{tabular}
\vspace*{-0.0cm}
\caption{ \noindent {\em Same as Fig.~{\protect \ref{PHASED3}} but for 
$\tk=0.8$. 
}}
\label{PHASED4}
\end{figure}

All the results we are going to present 
in the following were obtained at $g=0.6$, scanning the
three-dimensional $(\k,\tk,\tr)$ phase diagram at $g=0.6$
by keeping either $\tk$ or $ \tr$
fixed. (We also performed a few scans 
of the four-dimensional $(g,\k,\tk,\tr)$ parameter 
space in the $g$ direction at $\tr=1$, $\k=0$ and several values of $\tk$ 
and find that the transition to the 
confining phase occurs always at values of $g$ which are 
larger than $0.6$.) 
The  resulting two-dimensional 
sections through the $(\k,\tk,\tr)$-phase diagram are displayed
in the left columns of Figs.~\ref{PHASED1}-\ref{PHASED4}.
The corresponding phase diagrams obtained from the Monte Carlo 
simulations are displayed in the right columns. They 
will be discussed in more detail in Sect.~\ref{NUM_PHAS} but 
a first glimpse shows that they are, in most cases, similar 
to the mean-field phase diagrams. 

The dash-dotted lines in 
Figs.~\ref{PHASED1}-\ref{PHASED4} mark the symmetry line 
(\eq{SP}); we only determined the phase diagram above that line. 
The error bars in the mean-field phase diagrams 
mark the distance between 
two successive points in our scans of the phase diagram ({\it cf.}
Appendix A).

Figs.~\ref{PHASED1}a and b show the 
$(\k,\tk)$ phase diagram at $\tr=1$.  There is a PM phase for
small $|\tk|$ and small $|\k|$, and for large $\tk$ there is a 
phase-transition line separating an FM and an FMD phase. 
The plots in Figs.~\ref{PHASED1}-\ref{PHASED2} show that the
situation remains qualitatively the same when $\tr$ is lowered. 
The phase diagram at $\tr=0$ contains 
at small $\tk$ and $\k$ a PM phase, as for $\tr>0$.
The mean-field calculation, however, predicts that, at $\tr=0$, 
the region at large $\tk$ and small $|\k|$ is filled by 
an FMD phase, which is in conflict with our Monte Carlo simulations 
(plot in the right column), which show that the PM phase extends 
to very large $\tk$ with no sign of an FMD phase. 
We therefore believe that the FMD phase at large $\tk$ 
is an artifact of the mean-field approximation 
(which tends to favor ordered over disordered phases).  
Figures~\ref{PHASED3}-\ref{PHASED4} show three 
$(\k, \tr)$ phase-diagram plots at fixed $\tk$ values.   
The phase diagrams  at $\tk=0.05$ and $0.2$ look very similar to those
at fixed $\tr>0$. For $\tk<0<\tr$ or $\tr<0<\tk$ we 
find an FM-AM phase transition which coincides with 
the symmetry line, Eq.~(\eq{SP}). Figure~\ref{PHASED4} shows again that 
the mean-field calculation 
at $\tk=0.8$ does not lead to a PM phase at small $\tr$.       

\begin{figure}[t]
\centerline{
\epsfxsize=11.0cm
\vspace*{-.5cm}
\epsfbox{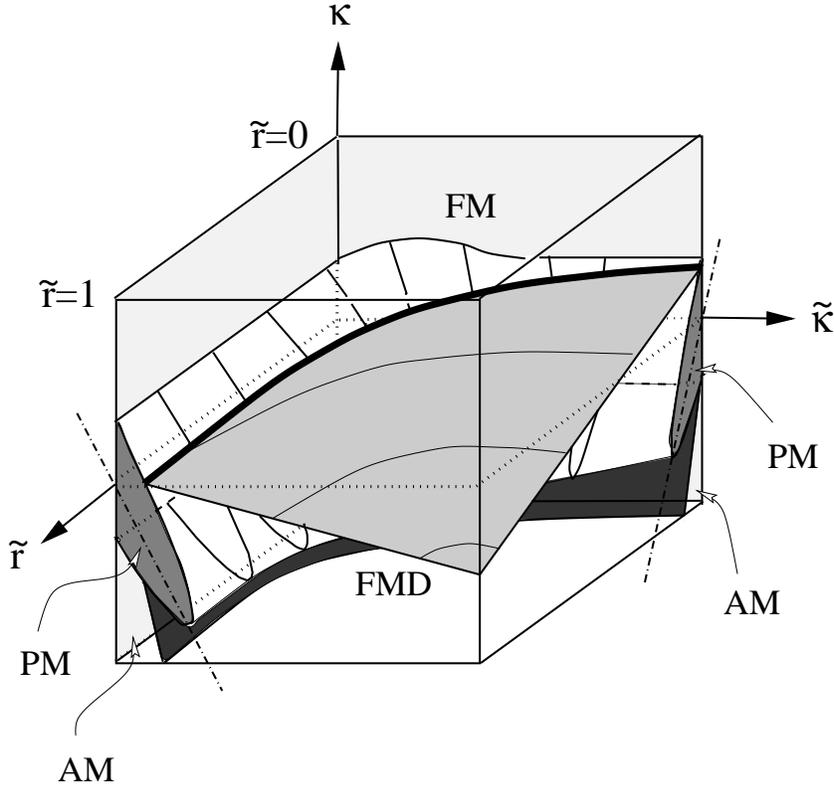}
}
\vspace*{1.2cm}
\caption{ \noindent {\em Schematic three-dimensional 
plot of the $(\k,\tk,\tr)$ phase diagram 
in the range from $\tr=0$ to $\tr=1$, $\tk=0$ to $\tk=1$. 
The two dashed lines on the front faces of the
phase diagram cube indicate the position  of the 
symmetry surface (Eq.~{\protect \eq{SP}}) on those faces. 
The thick line represents the FM-FMD-PM tricritical line.
}}
\label{SCHEM}
\end{figure}

\begin{figure}[t]
\vspace*{-0.5cm}
\begin{tabular}{ll}
 \hspace*{-1.3cm} \epsfxsize=10.00cm
 \vspace*{-0.5cm}
\epsfbox{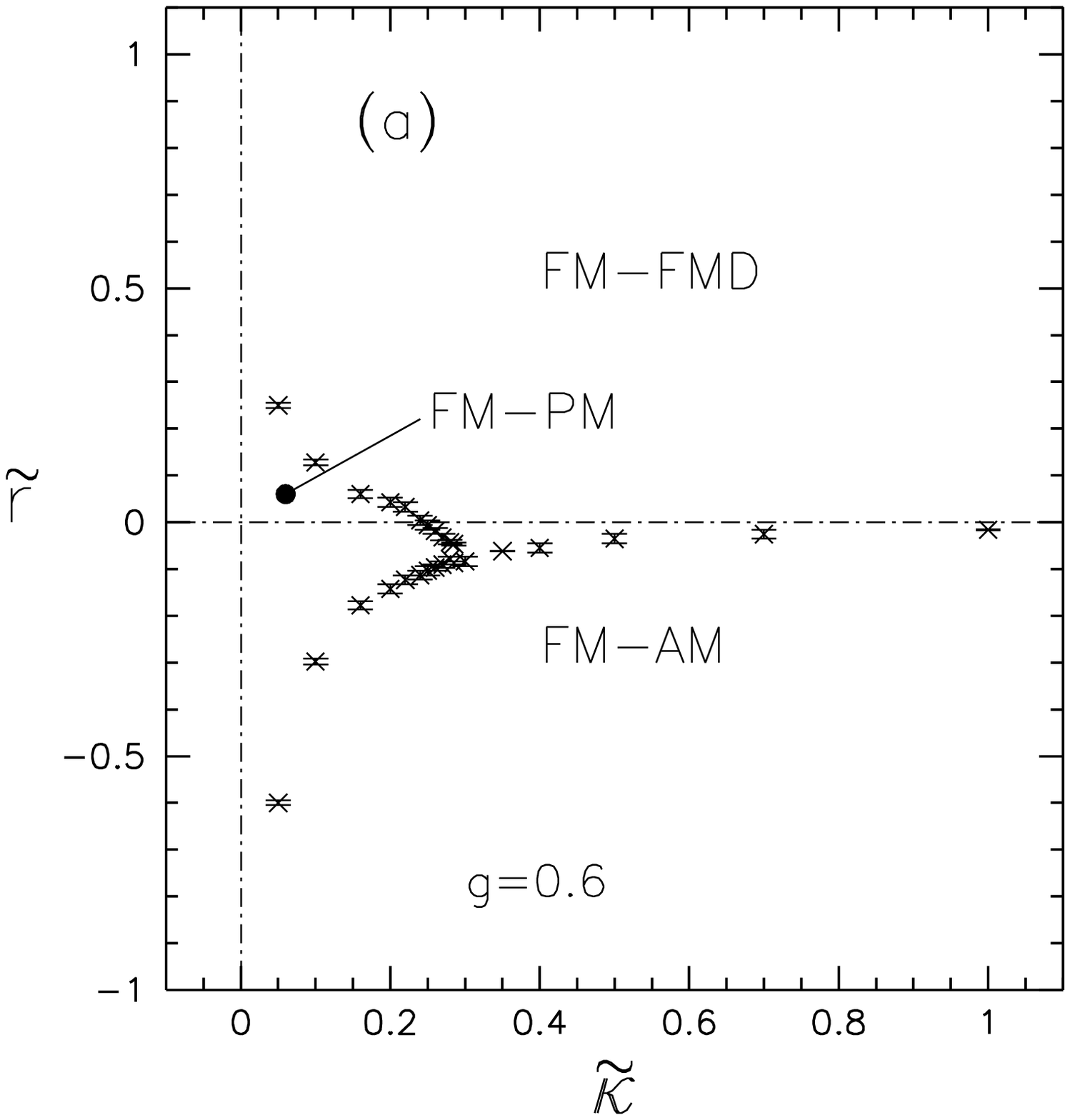}     &
 \hspace{-4.0cm} \epsfxsize=10.00cm
 \vspace*{-1.0cm}
\epsfbox{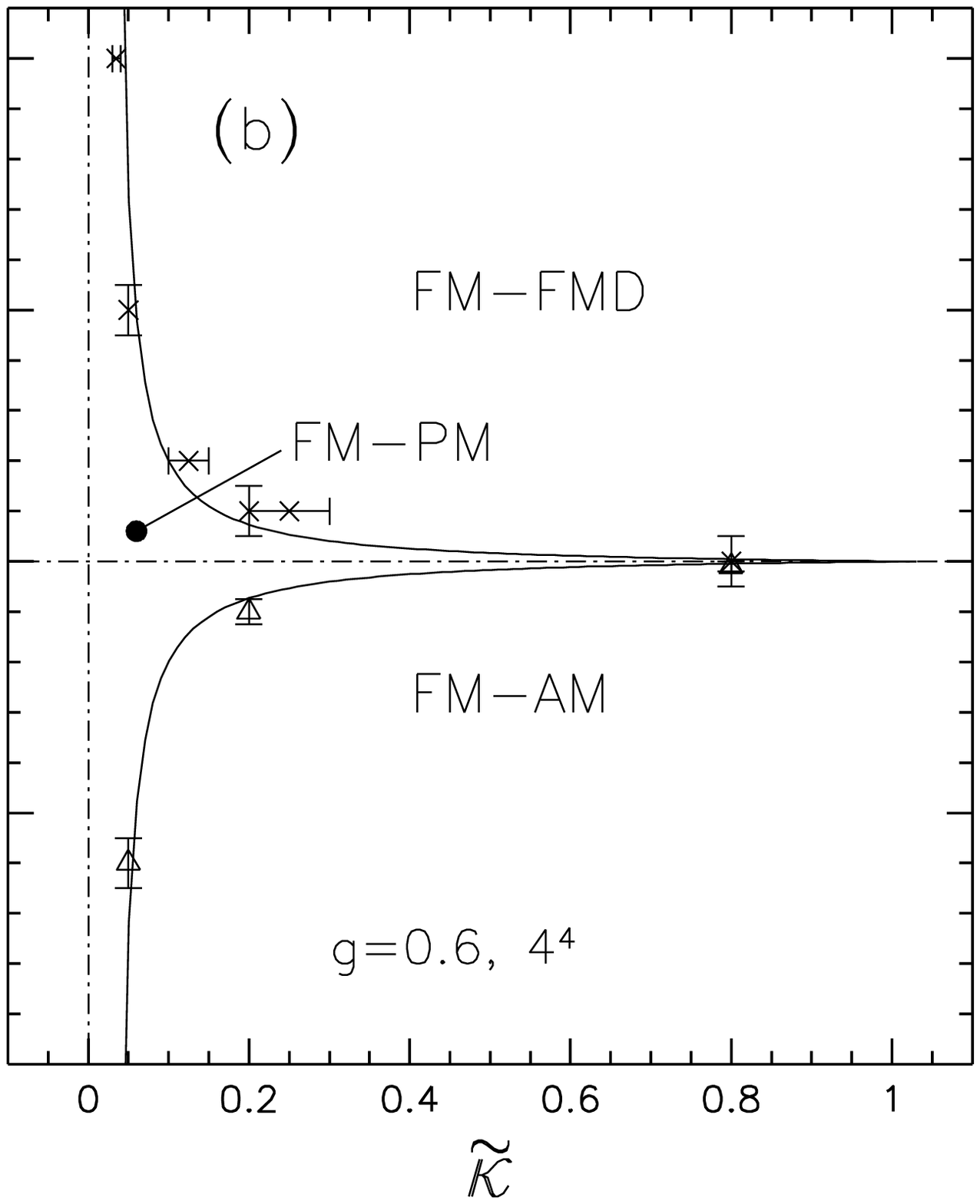}     \\
\end{tabular}
\vspace*{0.0cm}
\caption{ \noindent {\em Projection of the FM-FMD-PM 
and FM-AM-PM tricritical lines onto a $\k=\mbox{const}$ plane obtained 
by a mean-field calculation (Fig.~a) and by a Monte Carlo simulation 
(Fig.~b). The FM-FMD-PM and FM-AM-PM tricritical lines in Fig.~a 
appear to merge into an FM-FMD-AM tricritical line at large $\tk$. 
The two solid curves in Fig.~b are to guide the eye.
}}
\label{LP}
\end{figure}

We have compiled the Monte Carlo results for the $(\k,\tk,\tr)$ 
phase diagram into a schematic graph, 
shown in Fig.~\ref{SCHEM}. The FMD phase at large $\tk$ and 
$\tr=0$ is not shown 
in that graph since, as we said, there is no evidence from Monte Carlo 
simulations that the FMD extends down to $\tr=0$. Fig.~\ref{SCHEM}
shows that the FM-PM and FM-FMD phase-transition 
sheets are separated by a tricritical line where three phases 
(FM, PM and FMD) meet; we will call this the FM-FMD-PM line. 
Similarly, also the FM-PM and FM-AM phase transition sheets 
are separated by an FM-AM-PM tricritical line
(not shown in Fig.~\ref{SCHEM}). The projections of 
these tricritical lines onto a constant-$\k$ plane are shown 
in Fig.~\ref{LP}a (mean field) and in Fig.~\ref{LP}b (Monte Carlo). 
We see that the discrepancy between the mean field and Monte Carlo 
results at $\tr=0$ correlates with the fact that the mean-field
and Monte Carlo locations of these tricritical lines differ slightly.
In mean field, the PM phase ends at $\tk \approx 0.3$, and
the FM-FMD-PM and FM-AM-PM lines merge
into an FM-FMD-AM tricritical line, whose
projection approaches the $\tr=0$ axis for $\tk=\infty$.

In the mean-field approximation we find that the FM-PM transition
is second order at $\tk=0$ (this is the ``standard" gauge-Higgs model). 
It is still second order at $\tk>0$ and $\tr=1$, but 
changes into a first order phase transition when $\tr$ is lowered. 
Similarly we find that the FM-FMD transition 
is of second order at $\tr=1$, but of 
first order at small $\tr$. The FM-AM phase 
transition is always of first order. 
%
%
%
%
\subsection{Perturbative determination of the counterterm coefficients} 
\lb{LAMBDAS}
The counterterm coefficients $\k$, $\l_1+\l_3$ and $\l_2$ 
can be calculated order by order in 
perturbation theory by demanding the Slavnov--Taylor identity 
\be
\sum_{\m \n} p_{\m} \; q_{\n}\; \lag A_\m(p) \; A_\n (q) \rag =\xi \; \d(p+q)
 \lb{STCONTP}
\ee
to be satisfied in the continuum limit $a \ra 0$.  (The Slavnov--Taylor
identity does not determine the other linear combination,
$\l_1-\l_3$, because it corresponds to a gauge-invariant operator.)

First, we transcribe the left-hand side of 
eq.~(\eq{STCONTP}) to  a finite lattice, 
\be
W(p)
=\sum_{\m \n} \phat_\m \; \phat_\n \; \D_{\m \n}^V (p)\; ,
\lb{BRST_I_TREE_A}
\ee
where $\phat_\m=2 \sin (p_\m/2)$ is the lattice momentum, and 
\be
\D_{\m \n}^V (p)= 
\frac{1}{g^2\; L^3 \; T} \left\lag \sum_{x,y}
\mbox{Im} \; U_{\m x} \; \mbox{Im} \; U_{\n y} \exp(i\;  p \; (x-y)  )
\right\rag \;. \lb{PROP_V} 
\ee
is the vector two-point function in momentum space, with $L^3 T$ 
the lattice volume of a cylindrical lattice 
of spatial extent $L$ and temporal extent $T$.  
The vector propagator $\D_{\m \n}^V (p)$ 
is now computed to a given order in $g^2$. The resulting expression 
for $\D_{\m \n}^V (p)$,  and hence 
also for $W(p)$, will be a function 
of the counterterm coefficients. 
The counterterm coefficients are then determined such 
that in the continuum limit
\be
\lim_{a \ra 0} W(p)
=\xi  \lb{BRST_I_TREE}
\ee
(keeping the physical volume fixed).
Similarly, the coefficients $\l_4$ and $\l_5$ of the quartic counterterms
can be calculated by requiring the Slavnov--Taylor identity 
\bea
&&\sum_{\m \n \r \s} p_\m q_\n k_\r l_\s\;
\lag A_\m (p) A_\n (q) A_\r (k) A_\s (l) \rag \nonumber \\
&&= \xi^2 \left( \d(p+q)\; \d(k+l)+
\d(p+k)\; \d(q+l)+\d(p+l)\; \d(q+k) \right)\; 
\lb{BRST_II_TREE}
\eea
to be satisfied in the continuum limit $a \ra 0$.  

After inserting 
\be
A_{\m x} =\tlsum{p} \exp \left(i \; p\cdot x \right) \; 
\exp  \left(i \; p_\m /2 \right)  \; A_{\m} (p) \;,    \lb{ATRAF}              
\ee
with $\tsum_p=(L^3 T)^{-1} \sum_p$,
into the action (\eq{FULL_ACTION}), we obtain from the term bilinear 
in vector potential $A_\m(p)$ for the tree-level vector propagator 
\be
\D^{V,(0)}_{\m \n}(p) =\left[ \left( \phat^2 + m^2 \right)
\; \d_{\m \n} - \left( 1-\frac{1}{\xi}\right) \; \phat_\m \; \phat_\n 
\right]^{-1}\;,\lb{PROP_TREE} \\
\ee
where
\be
m^2=2\k g^2\;.\lb{MASS}
\ee
We included the mass counterterm in the tree-level propagator, since
it also functions as an infrared cutoff.
%
%
%
%

We have calculated the critical values of the coefficients $\k$, $\l_1$,
$\l_2$ and $\l_3$ to one-loop order in perturbation theory. 
To one-loop order the vector propagator is 
\bea
\D^{V,(1)}_{\m \n}(p) \!\!&=& \!\! Z_{\m \n}(g^2)\; 
\left[ \phantom{\left( \frac{1}{\xi} \right)} 
\!\!\!\!
\!\!\!\!
\!\!\!\!
\left( m^2 + (1+\l_1 \;g^2 ) \; \phat^2 + \l_2 \; g^2 \; \phat_\m^2 
\right) 
\; \d_{\m \n} \right. \nonumber \\
\!\!&-& \!\!
\left.
\left( 1-\frac{1}{\xi}-\l_3 \; g^2 \right) 
\; \phat_\m \; \phat_\n +\S_{\m \n}^{\rm bare}(p) \right]^{-1}\;,
\lb{PROP_ONE} 
\eea
with
\be
Z_{\m \n}(g^2)=1-g^2 \; \half \; \tlsum{k} 
\left( \D_{\m \m}^{V,(0)}(k)+\D_{\n \n}^{V,(0)}(k) \right)\;, \lb{WAVEF}
\ee
and
\bea
\!\!\!\!\!\! \S_{\m \n}^{\rm bare}(p) \!\!&=&\!\! \S^{\rm g.f.}_{\m \n}(p) 
+ \S^{\rm G }_{\m \n}(p)\;,
\lb{SIGMA_V} \\
\!\!\!\!\!\! \S^{\rm g.f.}_{\m \n}(p)\!\!&=&\!\! \frac{2\; g^2}{\xi} \; 
\left[ \quart \; \tr \; \tlsum{k} \sum_\a \D_{\a \a}^{V,(0)}(k)\;  
\khat^2_\a 
+ \quart \;\tr \; \phat^2_\m \; \tlsum{k} \sum_\a \D_{\a \a}^{V,(0)}(k) \; \cos^2 (k_\a/2)
\right. \nonumber \\
\!\!\!\!\!\! \!\!&\phantom{=}&\!\! \left. -\half \; \tlsum{k} \sum_\a 
\D_{\a \m}^{V,(0)}(k) \; \khat_\a \; \khat_\m \right] \; \d_{\m \n} 
\nonumber \\
\!\!\!\!\!\! \!\!&\phantom{=}&\!\! -\frac{g^2}{\xi} 
\; \left[ \tlsuma{k} \D_{\m \m}^{V,(0)}(k) 
-\quart \; \tr \; \tlsuma{k} \D_{\m \n}^{V,(0)}(k) \;
\khat_\m \; \khat_\n \right] \; \phat_\m \; \phat_\n\;, \lb{SIGMA_GF} \\
\!\!\!\!\!\! \S^{\rm G}_{\m \n}(p) \!\!&=&\!\! -\frac{g^2}{2} \; 
\left[ \sum_\a \phat^2_\a 
\tlsump{k} \left( \khat_\a^2 \; \D_{\m \m}^{V,(0)}(k)+   
      \khat^2_\m \; \D_{\a \a}^{V,(0)}(k) 
-2 \; \khat_\a \khat_\m \D_{\a \m}^{V,(0)}(k) \right) 
\right] \; \d_{\m \n} \nonumber \\
\!\!\!\!\!\! \!\!&\phantom{=}&\!\! -g^2
\; \left[ \tlsuma{k} \D_{\m \n}^{V,(0)}(k) \; \khat_\m \;  \khat_\n -
\tlsumpa{k} \khat_\m^2 \; \D_{\n \n}^{V,(0)}(k) \right] \; \phat_\m \; 
\phat_\n \;. 
\lb{SIGMA_G} 
\eea
$\S^{\rm g.f. }_{\m \n}(p) $ and 
$\S^{\rm G}_{\m \n}(p) $ are, respectively, the contributions from     
the gauge-fixing and plaquette actions.  The $Z$-factor, Eq.~(\eq{WAVEF}),
originates from the fact that we used the composite operator
$\mbox{Im}\;U_{\m x}$ instead of $A_{\m x}$ in the definition of
$\Delta_{\m\n}^V(p)$.  On a symmetric lattice ({\it i.e.} $L=T$) 
the self-energies are 
\bea
\S^{\rm g.f.}_{\m \n}(p)\!\!&=&\!\! \frac{2\; g^2}{\xi} \; 
\left[ \tr \; I_{11} -\half \; \left(I_{11}+3 \; I_{12} \right) 
+\phat_\m^2 \; \tr\; \left( K_{11}-\quart\; I_{11} 
\right) \right] \; \d_{\m \n} \nonumber \\
\!\!&\phantom{=}&\!\! -\frac{g^2}{\xi} \; \left[
K_{11} -\quart \; \tr \; I_{12} -\quart \; \tr \; \left( I_{11} -I_{12} \right)
\; \d_{\m \n} \right] \; \phat_\m \; \phat_\n \; ,
\lb{SIG_GF}\\
\S^{\rm G}_{\m \n}(p) \!\!&=&\!\! -g^2 \; (J_{12}-I_{12}) \; \left[ 
|\phat|^2\; \d_{\m \n} -\phat_\m \; \phat_\n \right] \lb{SIG_G}\; ,
\eea
(note that the last expression is transverse, as it should be)
with lattice integrals 
\bea
K_{\m \n}\!\!&=&\!\! \tlsuma{k} \D^{V,(0)}_{\m \n}(k)  \; , \lb{K}\\
I_{\m \n}\!\!&=&\!\! \tlsuma{k} \D^{V,(0)}_{\m \n}(k) 
\; \khat_\m  \; \khat_\n \; , \lb{I11}\\
J_{\m \n}\!\!&=&\!\! \tlsuma{k} \D^{V,(0)}_{\m \m }(k) \; \khat_\n^2  \;  .
\lb{J12}
\eea

The Slavnov--Taylor identity, Eq.~(\eq{BRST_I_TREE}), is satisfied 
to one-loop order for 
\be
\k=\k_{\rm FM-FMD} = -\frac{1}{\xi} \; \left[ 
                  \tr \; \left(\quart + X_{1} \; (\xi -1) \right) 
-\eighth \; \xi 
                                   \right] \;,
\lb{k_FMFMD}
\ee
\be
\l_2 = \frac{\; \tr}{\xi} \; \left[ 
\sixteenth -2 \; \left(1+\quart \; (\xi-1) \right) \; X_2 
+ \quart \; (\xi -1) \; (X_1+X_3) \right] \lb{lambda_2}
\ee
and 
\be
\l_1+\l_3 = - \tr \; \quart \;\left(1-\frac{1}{\xi}\right) \; X_3 \;, 
\lb{lambda_13}
\ee
with 
\bea
X_{1}\!\!&=&\!\! \tlsuma{k} \frac{\khat_1^4  }{|\khat|^4} = 0.0951(1)\; ,\\
X_{2}\!\!&=&\!\!\tlsuma{k} \frac{1 }{|\khat|^2} = 0.1549(1)\; , \lb{X2} \\
X_{3}\!\!&=&\!\!\tlsuma{k} \frac{\khat_1^2 \; \khat_2^2} 
{|\khat|^4} = \frac{1}{3}\left(\frac{1}{4}-X_{1}\right)
 = 0.0516(1)\; , \lb{X3} 
\eea
in the limits $L,T\to\infty$ and $m^2\to 0$.
Equation~(\eq{k_FMFMD}) provides us with an expression for the FM-FMD 
phase boundary, to be compared (in Sect.~\ref{NUM_PHAS}) 
with our Monte Carlo results.  One can also verify that in the limit 
$g \ra 0$ Eq.~(\eq{k_FMFMD}) turns into the corresponding 
result for the reduced model \cc{wmy_pd}, 
taking the limit $g \ra 0$ such that
$\tk$ in Eq.~(\eq{TKGXI}) is kept fixed, {\it i.e.} 
$\xi \ra \infty$. 
(Note that both $\l_2$ and the sum $\l_1+\l_3$
vanish in the limit $\tr \ra 0$.  This happens because of a
combination of two facts:
for $\tr=0$ a ghost action can be added such that the full
action has an exact BRST symmetry on the lattice \cc{hn_nogo}; however,
to one loop, the ghosts do not appear in the vacuum polarization
in the abelian case. The mass counterterm does not vanish for $\tr=0$
since BRST invariance allows for equal non-zero masses
for the U(1) gauge field and the Fadeev--Popov ghosts.)
We can fix $\l_1$ and $\l_3$ if we demand the
wavefunction renormalization constant to be equal to one at
the one-loop level, then
\be
\l_1= \left( \frac{1}{4}-\frac{3+\xi}{4}X_2 \right)\;, \lb{lambda_1} 
\ee
which then determines $\l_3$ using Eq.~(\eq{lambda_13}).
%
%
\begin{figure}[t]
\centerline{
\epsfxsize=12.0cm
\vspace*{0.5cm}
\epsfbox{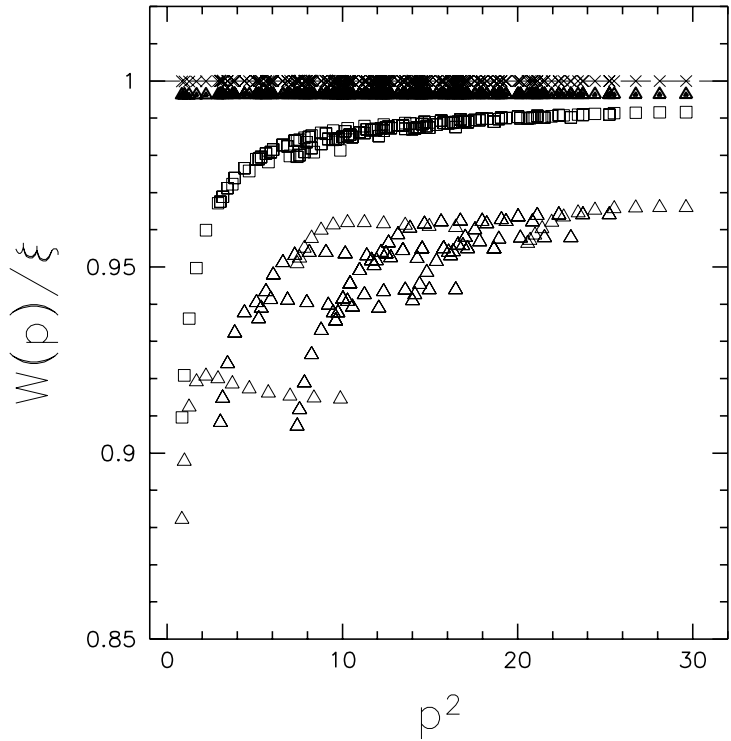}
}
\vspace*{-1.5cm}
\caption{ \noindent {\em The ratio $W(p) /\x$ as a 
function of $p^2$ on a $6^3 24$ lattice 
at the point $(g,\tk,\tr)=(0.6,0.8,1)$.  See text.
}}
\label{STID}
\end{figure}
%
%

As a check of our results, and to get a feeling for the effects 
of the various counterterms, 
we have calculated $W(p)$ in Eq.~(\eq{BRST_I_TREE_A}) to 
one-loop order, and plotted in Fig.~\ref{STID} the ratio $W(p)/\x$ 
as a function of $p^2$ for $0<p_\mu<\pi$, $\mu=1,\ldots,4$. 
After tuning the counterterm coefficients 
this ratio should approach one in the continuum limit $a \ra 0$. 
The inversion of the $4 \times 4$ 
matrix in Eq.~(\eq{PROP_ONE}) was done numerically.
The ratio $W(p)/\x$ was computed on a $6^3 24$ lattice at 
the point $(g,\tk,\tr)=(0.6,0.8,1)$. We used antiperiodic boundary
conditions
in the time direction to avoid the zero mode of the propagator.
The open triangles in Fig.~\ref{STID} 
correspond to setting $\k=0$, $\l_i=0$, $i=1,\ldots,5$. 
The plot shows that the ratio $W(p)/\x$ is clearly below one and, 
as shown by the irregularities, is not a continuous function of $p^2$. 
These irregularities are caused by the part of the 
self-energy with the structure of the $\l_2$ counterterm.   
Setting $\l_2$ equal to the value obtained from 
(\eq{lambda_2}), one gets the values represented 
in Fig.~\ref{STID} by the squares. All irregularities 
disappear and $W(p)/\x$ is a continuous 
function of $p^2$. The filled triangles in Fig.~\ref{STID} were 
obtained after setting $\k=\k_{\rm FM-FMD}$ (Eq.~(\eq{k_FMFMD}))
and the crosses after setting also $\l_1+\l_3$ to the value 
determined from Eq.~(\eq{lambda_13}). The graph shows the crosses to be 
very close to one for all momenta.
\subsection{Particle Spectrum at the FM-FMD Phase transition}
\lb{PROP_AN}
In this subsection we discuss the particle spectrum. We mentioned 
already in Sect.~\ref{MOD} that the particle 
spectrum in the FM phase of the pure gauge-Higgs model ($\tk=0$) contains 
a massive vector boson and a massive Higgs particle whose masses scale 
when $\k$ is tuned towards the FM-PM phase transition. 
On the other hand, at the FM-FMD phase transition 
the action (\eq{FULL_ACTION}) is supposed to provide a new 
lattice discretization of a theory of free photons, with nothing else. 
A Higgs particle associated with the longitudinal 
gauge degrees of freedom should be absent. 
Therefore, the particle spectrum has to change when 
crossing the FM-FMD-PM tricritical line 
in the FM phase: an (unstable) Higgs bound state should exist
near the PM-FM transition, and not near the FM-FMD transition.

The question of the existence of a Higgs bound state is 
a non-perturbative issue.  In this section, we present 
perturbative results for the various correlation functions
that will be used to probe the spectrum numerically in 
Sect.~\ref{NUM_PROP}.

The vector propagator defined in Eq.~(\eq{PROP_V})
was already calculated to one-loop order in the last 
section, and is given by Eq.~(\eq{PROP_ONE}).       
It is evident 
that one indeed obtains a free, canonically normalized vector propagator 
if the four counterterm coefficients $\k$, $\l_1$, $\l_2$ and 
$\l_3$ are tuned towards the values given in
Eqs.~(\eq{k_FMFMD}-\eq{lambda_13},\eq{lambda_1}). 

An operator containing the quantum numbers of the 
Higgs particle is $\mbox{Re } U_{\m x}$.
The corresponding Higgs two-point function on a cylindrical lattice is 
given by 
\be
\D^H_{\m \n} (p)=
\frac{1}{L^3 \; T} \left\lag \sum_{x,y}
\mbox{Re} \; U_{\m x} \; \mbox{Re} \; U_{\n y} \exp(i\;  p \; (x-y)  )
\right\rag \;, \lb{PROP_S} 
\ee
which has been used for the numerical determination of the Higgs mass 
in gauge-Higgs models \cc{spectrum}.

It is easy to verify that to one-loop order
\be
\D^{H, (1)}_{\m \n} (p)= \half \; g^4 \; \exp \left( i \; (p_\m -p_\n)/2 
\right)
\tlsuma{k} \D^{V,(0)}_{\m \n}(k)  \; \D^{V,(0)}_{\m \n}(p+k) \; .  
\lb{PROP_S_ONE}  \\
\ee
For small $p$, we can extract the non-analytic part by replacing the
integrand with its continuum expression, and we obtain (for $m^2=0$
and $L,T\ra \infty$)
\be
\sum_{\m\n} 
\left. \D^{H, (1)}_{\m \n} (p)\right|_{\rm non-analytic}= 
-(\half g^4)\frac{1}{(4\pi)^2}(3+\xi^2)\log{p^2}
\;.  \lb{PROP_SNA}
\ee
It is obviously not possible to conclude from this perturbative 
calculation alone that a Higgs particle does not exist.  However,
one may compare a nonperturbative evaluation of the same correlation
function with the perturbative result.  If they agree, this provides
evidence that a Higgs bound state does not occur in the theory, and
this is what we will investigate in Sect.~\ref{NUM_PROP}. 

Another equivalent way of looking at this is to consider 
the coordinate-space correlation function 
\be
G^{H}_{\m \n} (|x-y|)= \left\lag \mbox{Re} \; U_{\m x} \;
\mbox{Re} \; U_{\n y} \right\rag - 
\left\lag \mbox{Re} \; U_{\m x} \right\rag \; 
\left\lag \mbox{Re} \; U_{\n x} \right\rag
\;,  \lb{GS} 
\ee
which, if no bound state is present in the spectrum,
should factorize for $|x-y| \ra \infty$ as 
\be
G^{H}_{\m \n} (|x-y|) = C_{\m \n} \; \left[ G^{V}_{\m \n}(|x-y|) \right]^2
\;,  \lb{GS_FAC} 
\ee
where
\be
G^{V}_{\m \n} (|x-y|)= \left\lag \mbox{Im} \; U_{\m x} \;
\mbox{Im} \; U_{\n y} \right\rag 
  \lb{GV} 
\ee
is the vector correlation function, 
and $C_{\m \n}$ is a constant which can be determined in 
perturbation theory. 

\begin{figure}[t]
\centerline{
\epsfxsize=12.0cm
\vspace*{0.5cm}
\epsfbox{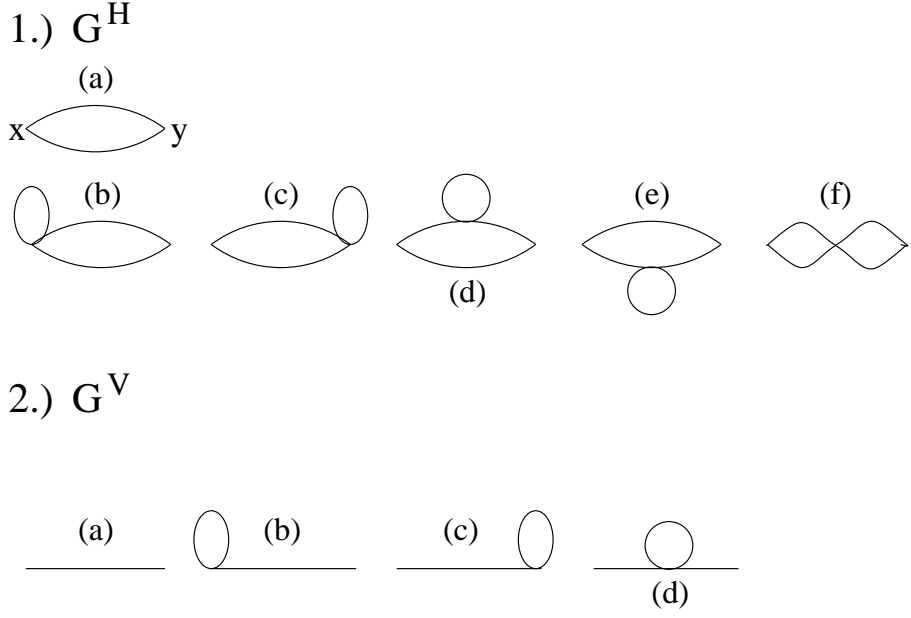}
}
\vspace*{0.0cm}
\caption{ \noindent {\em Feynman diagrams
for the coordinate-space correlation functions
$G^{H}_{\m \n} (|x-y|)$ and $G^{V}_{\m \n} (|x-y|)$.
}}
\label{FDIA1}
\end{figure}

To leading order in perturbation theory we find, 
\be
G^{H}_{\m \n} (|x-y|)= \frac{g^4}{4}\lag A^2_{\m x}\; A^2_{\n y} \rag_0
=\frac{g^4}{2} \D^{V,(0)\;2}_{\m x, \n y} \; \lb{FAC1}
\ee
and 
\be
G^{V}_{\m \n} (|x-y|)= g^2 \; \lag A_{\m x}\; A_{\n y} \rag_0
=g^2 \; \D^{V,(0)}_{\m x, \n y} \; , \lb{FAC2}
\ee
where $\lag \cdots \rag_0$ denotes the quantum
average with the part of the lattice action (\eq{FULL_ACTION}) 
that is quadratic in $A_\m$, and 
\be
\D^{V,(0)}_{\m x, \n y} = \tlsum{p} \exp \left(i\; p \; (x-y) \right)
\; \D_{\m \n}^{V,(0)} (p) \lb{DD}   \; ,
\ee
with $\D_{\m \n}^{V,(0)}(p)$ given in Eq.~(\eq{PROP_TREE}).
Substituting Eqs.~(\eq{FAC1}) and (\eq{FAC2}) into 
(\eq{GS_FAC}) leads to 
\be
C_{\m \n}=\frac{1}{2} \;. \lb{C1} 
\ee
The expressions in Eqs.~(\eq{FAC1}) and 
(\eq{FAC2}) are represented by 
Feynman diagrams 1a and 2a in Fig.~\ref{FDIA1}, 
respectively. Note that, to leading order in perturbation 
theory, factorization 
holds without any tuning of the six counterterm coefficients 
$\k$, $\l_1,\ldots,\l_5$. 

We now wish to verify explicitly that factorization holds also to 
next-to-leading order in $g^2$. To this order we obtain,   
\bea
\!\!\!\!\!
G^{H}_{\m \n} (|x-y|)\!\!\!&=& \!\!\!\frac{g^4}{4}\lag A^2_{\m x} A^2_{\n y} 
(1-S_{\rm I}^{(4)}) \rag_0
-\frac{g^6}{48} \lag A^2_{\m x} A^4_{\n y}+A^4_{\m x} A^2_{\n y}\rag_0
\nonumber \\
\!\!\!\!\!
\phantom{G^{H}_{\m \n} (|x-y|)} \!\!\!&=& \!\!\!\frac{g^4}{2} \left\{  \D^{V,(0)\;2}_{\m x, \n y} \left( 1-\frac{g^2}{2}
 \left[
 \D^{V,(0)}_{\m x,\m x}+\D^{V,(0)}_{\n y,\n y} \right] \right) 
 - {1\over 2}\lag A^2_{\m x}A^2_{\n y} 
 S_{\rm I}^{(4)} \rag_0 \right\} ,
\lb{GS3} 
\eea
where $S_{\rm I}^{(4)}$ designates all 
terms of the lattice action (\eq{FULL_ACTION})
which are quartic in the gauge potential $A_{\m x}$. 
Similarly,  we find for $G^V_{\m \n}(|x-y|)$
\bea
\!\!\!
\!\!\!
G^{V}_{\m \n} (|x-y|)\!\!&=&\!\! g^2\lag A_{\m x}A_{\n y} \; 
(1-S_{\rm I}^{(4)})\rag_0
-\frac{g^4}{6} \lag A_{\m x} A^3_{\n y}+A^3_{\m x} A_{\n y}\rag_0
\nonumber \\
\!\!\!
\!\!\!
\phantom{G^{V}_{\m \n} (|x-y|)}\!\!\!&=&\!\!g^2 \left\{ \D^{V,(0)}_{\m x, \n y} 
\left(1-\frac{g^2}{2}\left[ \D^{V,(0)}_{\m x,\m x} 
+\D^{V,(0)}_{\n y,\n y} 
\right]\right)-\lag A_{\m x}A_{\n y} 
\; S_{\rm I}^{(4)} \rag_0
\right\}  . 
\lb{GV2}
\eea
The various diagrams that contribute to $G^H$ and $G^V$ 
are displayed in Fig.~\ref{FDIA1}. 
Diagrams 1b, 1c, 2b  and 2c correspond 
to the terms in Eqs. (\eq{GS3}) and (\eq{GV2}) which are proportional to 
$[\Delta^{V,(0)}_{\m x, \m x}+\Delta^{V,(0)}_{\n y, \n y}]$,
and  give only a contribution to the wave-function renormalization constant.
The four-point vertices in diagrams 1d-1f (we will refer to diagram 1f
as the ``figure-eight diagram") and diagram 2d arise from $S_{\rm I}^{(4)}$. 
The integral expressions for those diagrams are given in Appendix B.

\begin{figure}[t]
\vspace*{-0.5cm}
\begin{tabular}{ll}
 \hspace*{-1.3cm} \epsfxsize=9.80cm
 \vspace*{-0.5cm}
\epsfbox{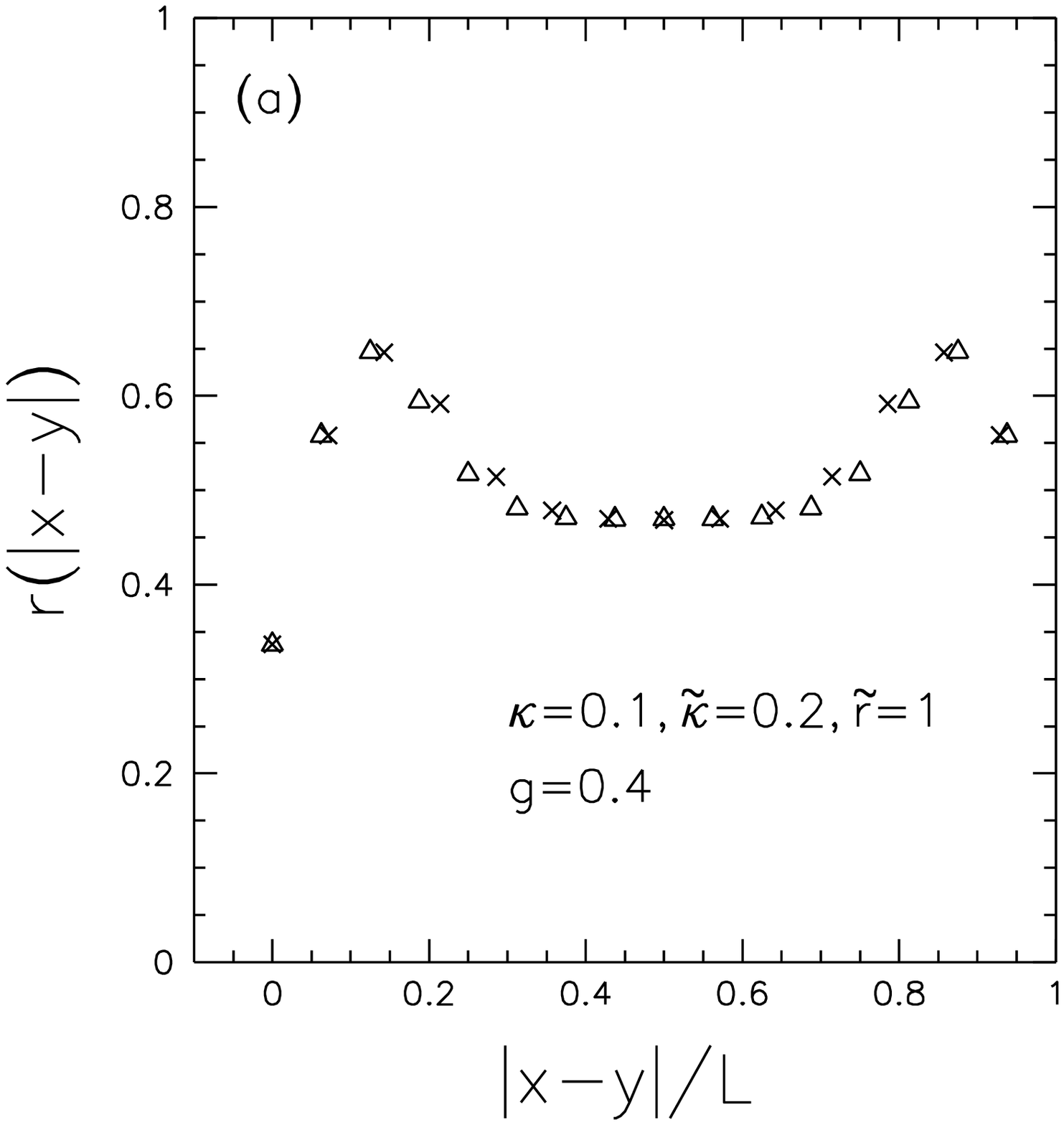}     &
 \hspace{-3.4cm} \epsfxsize=9.80cm
 \vspace*{0.0cm}
\epsfbox{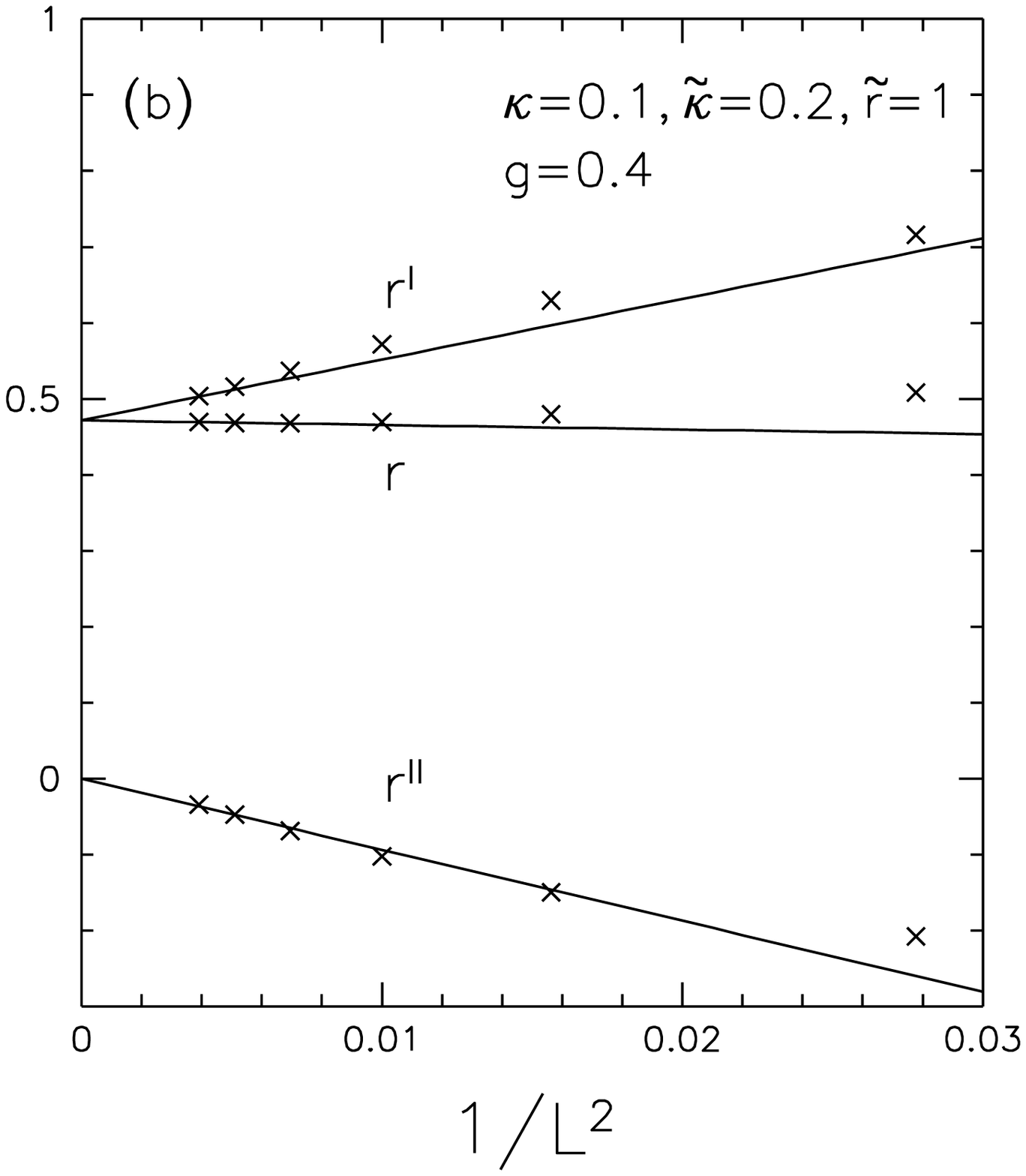}     \\
\end{tabular}
\vspace*{0.0cm}
\caption{ \noindent {\em  \mbox{}Fig.~a shows the ratio
$r(|x-y|)$ as a function of $|x-y|/L$ for
$L=14$ (crosses) and $16$ (triangles), with periodic boundary
conditions. In Fig.~b the
ratios $r(L/2)$, $r^{\rm I}(L/2)$ and $r^{\rm II}(L/2)$ are displayed as
a function of $1/L^2$. The straight lines in Fig.~b
are to guide the eye.
}}
\label{R1}
\end{figure}

In perturbation theory, one expects that $G^H_{\m\n}(|x-y|)\sim
(G^V_{\m\n}(|x-y|))^2$ for large $|x-y|$.  Here, we show this to be
true also at two loops.
It can easily be verified that, after squaring 
$G^{V}_{\m \n} (|x-y|)$, diagrams 2a, 2b and 2c combine 
into 1b and 1c and that similarly 2a and 2d combine into 
1d and 1e. The Higgs correlation function 
can then be written as 
\be
G^{H}_{\m \n} (|x-y|)=\left(\frac{1}{2} + g^2 \; C_{\m \n}^{(1)} \right) \; 
\left[ G^{V}_{\m \n} (|x-y|) \right]^2 
 - {g^4\over 4}\lag A^2_{\m x}A^2_{\n y} 
 S_{\rm I}^{(4)} \rag_0^{\rm 1f} +O(g^8)  \; , \lb{GS4}
\ee
where 
\be
C_{\m \n}^{(1)} =\quart \; \left( K_{\m \m} 
+K_{\n \n} \right)\;, 
\lb{CCC1}
\ee
({\it cf.} Eq.~(\eq{K}))
and $\lag A^2_{\m x}A^2_{\n y} S_{\rm I}^{(4)} \rag_0^{\rm 1f}$ 
is the contribution from the figure-eight diagram which is given 
by the complicated expression in Eq.~(\eq{D1f}) in Appendix B. 
The various terms contributing to the figure-eight diagram in Eq.~(\eq{D1f})  
can be divided into two classes according to the number of 
loop momentum factors in the loop integrals.  First, each of the terms
in Eq.~(\eq{D1f}) consists of the product of two one-loop integrals. 
The terms in Eq.~(\eq{D1f1}) contain in one of 
the two integrands no explicit momentum factors ($\phat$ {\it etc.}),
whereas the terms in Eq.~(\eq{D1f2}) contain 
in each of the integrands at least one such momentum factor. 
The momentum factors correspond in coordinate space to derivatives 
that act on that particular loop. 
A dimensional analysis then shows that 
the terms in Eq.~(\eq{D1f2}) do not give a contribution 
at large distances (they give only contact terms in momentum space). 
The only contribution at large separations comes 
from the terms in Eq.~(\eq{D1f1}). The loop integrals without any explicit
momentum factors 
behave at large separations $|x-y|$ as the leading order term 
$\lag A^2_{\m x} A^2_{\n y} \rag_0$ which, in the limit $m \ra 0$, 
is logarithmic divergent in momentum space.  In each term, 
the second integral
contains two momenta and approaches a constant in limit $k \ra 0$, thus
leading to a contribution to the constant $C_{\m \n}$.
Therefore, for $|x-y|\ra \infty$,
\be
-\frac{g^4}{4} \lag A^2_{\m x}A^2_{\n y} S_{\rm I}^{(4)} \rag_0^{\rm 1f}  
=g^2 \; C_{\m \n}^{(2)} \; \left[ G^{V}_{\m \n} (|x-y|) \right]^2 + O(g^8)
\;.
\ee
As a check (for the case $\mu=\nu$), 
and in order to determine the constant $C_{\m \m}^{(2)}$, 
we have numerically computed the three ratios 
\bea
r(|x-y|) \!\!&=&\!\!- \frac{ 
\lag A^2_{\m x}A^2_{\m y} 
 S_{\rm I}^{(4)} \rag_0^{\rm 1f} 
           }{
g^2 \D^{V,(0)\;2}_{\m x, \m y} 
           } \;, \lb{RR}  \\
r^{\rm I}(|x-y|) \!\!&=&\!\! -\frac{ 
\lag A^2_{\m x}A^2_{\m y} 
 S_{\rm I}^{(4)} \rag_0^{\rm 1f, I} 
           }{
g^2 \D^{V,(0)\;2}_{\m x, \m y} 
           } \;, \lb{RRI}  \\
r^{\rm II}(|x-y|) \!\!&=&\!\! -\frac{ 
\lag A^2_{\m x}A^2_{\m y} 
 S_{\rm I}^{(4)} \rag_0^{\rm 1f, II} 
           }{
g^2 \D^{V,(0)\;2}_{\m x, \m y} 
           }  \;,  \lb{RRII}  
\eea
where $\lag A^2_{\m x}A^2_{\m y} S_{\rm I}^{(4)} \rag_0^{\rm 1f}$,
$\lag A^2_{\m x}A^2_{\m y} S_{\rm I}^{(4)} \rag_0^{\rm 1f, I}$ 
and $\lag A^2_{\m x}A^2_{\m y} S_{\rm I}^{(4)} \rag_0^{\rm 1f, II}$ 
are given in Eqs.~(\eq{D1f}), (\eq{D1f1}) and (\eq{D1f2}) of Appendix B 
respectively. 
We have chosen $x$ and $y$ as two 
on-axis points. The ratios were computed on a symmetric 
lattice of size $L^4$ at the same point in the phase diagram 
($(g,\k,\tk,\tr)=(0.4,0.1,0.2,1)$)  
where the numerical simulations were performed ({\it cf.} Sect.~\ref{SFAC}.)
If factorization holds 
the two ratios $r(|x-y|)$ and $r^{\rm I}(|x-y|)$ should exhibit a plateau at 
large separations $|x-y|$. As an example we have plotted 
$r(|x-y|)$ in Fig.~\ref{R1}a 
as a function of $|x-y|/L$ for $L=14$ (crosses) and 
$16$ (triangles). 
%
%
   %
%
   %
%
   %
%
%
The plot shows that we obtain indeed  a plateau at large $|x-y|$. 
In Fig.~\ref{R1}b we have plotted the mid points, 
i.e. the values of $r(L/2)$, $r^{\rm I}(L/2)$ and $r^{\rm II}(L/2)$,
as a function of $1/L^2=(a/L^{\rm phys.})^2$ where  
$L^{\rm phys.}=a L$ is a fixed physical scale and $a$ is 
the lattice spacing. We see that $r^{\rm II}(L/2)$ indeed
vanishes in the limit $a \ra 0$. In contrast, the ratios 
$r(L/2)$ and $r^{\rm I}(L/2)$ approach a non-zero value in this 
limit. The constant $C_{\m \m }^{(2)}$ is then given by 
\be
C_{\m \m }^{(2)} =\quart \; \lim_{L \ra \infty } r(L/2) = 0.12(1) \;. 
\ee
For the constants $C_{\m \m }^{(1)}$ and $C_{\m \m }$ 
we find the values (for $g=0.4$)
\be
C_{\m \m }^{(1)} =0.3277(1) \;\;\;\mbox{and} \;\;\; 
C_{\m \m }=\frac{1}{2}+g^2 C_{\m \m }^{(1)} +g^2 C_{\m \m }^{(2)}
=0.5716(16) \;.  \lb{C2}
\ee
The above arguments do not lead to a constraint on the 
counterterm coefficients $\l_1$, $\l_2$, $\l_3$ and $\k$ at this
order. It is however clear that our above arguments 
are true only if $\l_4=\l_5=0$, which is consistent to this order
in $g^2$. At higher orders, factorization holds 
only when the counterterms $\l_4$ and $\l_5$ are  
tuned  appropriately.  (Note that then for any $\l_1$, $\l_2$ and $\l_3$
the theory is free in the $a\to 0$ limit.)
%
%
%
%
%
\section{Numerical results}
\secteq{4}
\lb{NUM}
In all our Monte Carlo simulations we set 
$\l_1 =\l_2 =\l_3 =\l_4 =\l_5=0$. The action depends then only 
on the four parameters $g$, $\k$, $\tk$ and $\tr$.
We have seen in the previous section that, for the quantities
we will consider, $\l_4=\l_5=0$ in perturbation theory to the
order we have taken into account.  This means that we can set
them equal to zero also in our numerical computations, as long
as the latter agree well with perturbation theory (within our
precision) with the same choice of values.  
In addition, we will be mainly concerned
with factorization ({\it cf.} previous section), which should
work for any choice of $\l_1$, $\l_2$ and $\l_3$. 
\subsection{Phase diagram}
\lb{NUM_PHAS}
For the determination of the $(\k,\tk,\tr)$-phase diagram we have to  
construct observables which allow us to locate the various 
phase transitions. The observables we used are three internal 
energies 
\bea
E_p \!\!&=&\!\! -\frac{1}{6 L^4}\; \frac{\del}{\del g^{-2}} \ln Z\;,  \lb{EP}\\
E_\k \!\!&=&\!\! -\frac{1}{8 L^4} \; \frac{\del}{\del \k} \ln Z\;,  \lb{EK}\\
E_{\rm g.f.} \!\!&=&\!\! -\frac{1}{64 L^4} \; \frac{\del}{\del \tk} \ln Z\;,
  \lb{ETK}
\eea
which in the vector picture are given by the expressions 
\bea
E_p \!\!&=&\!\! \frac{1}{6 L^4} \; \left\lag 
\sum_{x, \m < \n} \mbox{Re } U_{\m \n x} \right\rag \;.  \lb{VEP}\\
E_\k \!\!&=&\!\!  \frac{1}{4 L^4} \left\lag 
\sum_{x \m} \mbox{Re } U_{\m x } \right\rag \;, \lb{VEK} \\
E_{\rm g.f.} \!\!&=&\!\! \frac{1}{64 L^4} \;  \left\lag \sum_x \left\{
\quart \; \left( C_x -C_x^{\dagger} \right)^2 - \tr \;
\left[ \quart\;\left(C_x +C_x^{\dagger} \right)^2 -B_x^2 \right] 
\right\} \right\rag \;. \lb{VETK}
\eea
These quantities are not order parameters, and hence 
do not vanish in any of the various phases, but they signal 
phase transitions by an abrupt change. In the case of 
a second order phase transition we expect to find, 
in the infinite-volume limit,     
an ``S"-like curve with an infinite slope at the phase transition.   
At a first-order phase transition 
the internal energies exhibit a jump. 
On a finite lattice, however,             
it is difficult to distinguish between first- and second-order phase 
transitions, and it is usually necessary 
to perform a careful finite-size scaling analysis to settle 
the question of the order. 
In the FMD phase the hypercubic rotation invariance 
is broken by the non-vanishing vector condensate, $\lag A_\m \rag \neq 0$.
A true order parameter which 
allows to distinguish the FMD phase 
from the other phases can be defined on the lattice by the expression
\be
V= \left\lag \sqrt{\quart \; \sum_\m \left[ \frac{1}{L^4} \;
\sum_x \mbox{Im } U_{\m x} \right]^2 } \right\rag \; , \lb{ORDER}
\ee
which reduces in the constant-field approximation to 
$[\quart \sum_\m \sin^2 (g A_\m)]^{1/2}$. On a small lattice, 
the system tunnels from one of the $16$ discrete 
minima in Eq.~(\eq{CFMD}) to the others. This is the reason 
why, in Eq.~(\eq{ORDER}), we took the modulus of 
$\frac{1}{L^4} \sum_x \mbox{Im } U_{\m x}$. The summation over $x$ 
is to project onto zero momentum. 

The Monte Carlo simulations were done with a standard 5-hit 
Metropolis algorithm, and were   
performed either in the vector or in the Higgs picture. We wrote 
two codes and checked that the results obtained in the two 
pictures are consistent. The vector-picture 
simulations require less CPU time   
since only the gauge fields have to be updated. 
However, the autocorrelation time for gauge non-invariant 
quantities turns out to be slightly larger for the vector-picture 
simulations. We furthermore find
that the Higgs-picture simulations perform slightly better in regions 
of the phase diagram where metastabilities occur 
(the region near the FM-PM 
phase transition at $\tr=0$ and large $\tk$).
Most of our simulations were carried out in the vector picture.
\begin{figure}
\vspace*{-0.5cm}
\begin{tabular}{ll}
 \hspace*{-2.0cm} \epsfxsize=9.40cm
 \vspace*{-0.5cm}
\epsfbox{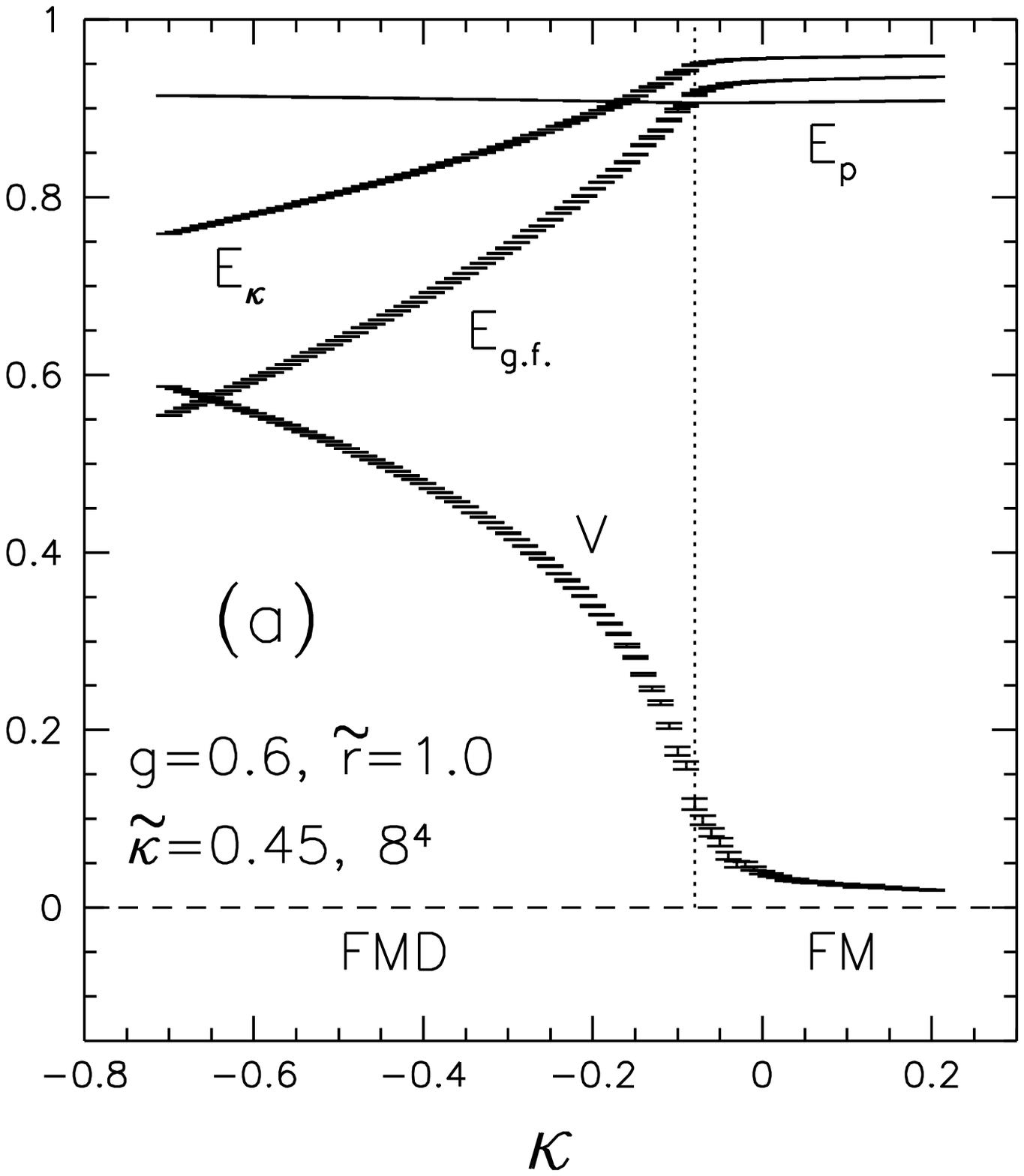}     &
 \hspace{-2.2cm} \epsfxsize=9.40cm
 \vspace*{-1.0cm}
\epsfbox{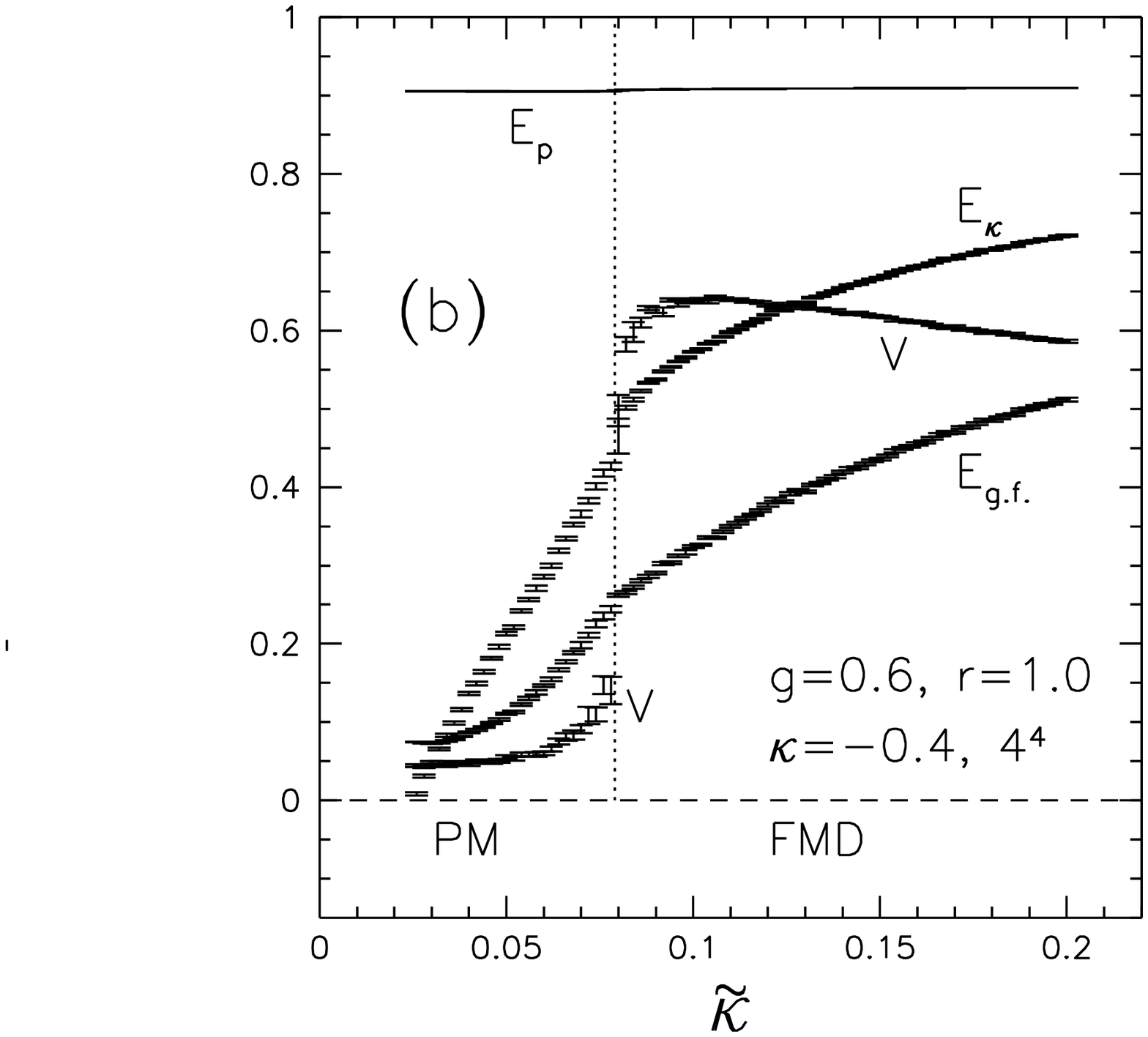}     \\
\multicolumn{2}{l}{
 \hspace*{2.0cm} \epsfxsize=9.40cm
 \vspace*{-0.5cm}
\epsfbox{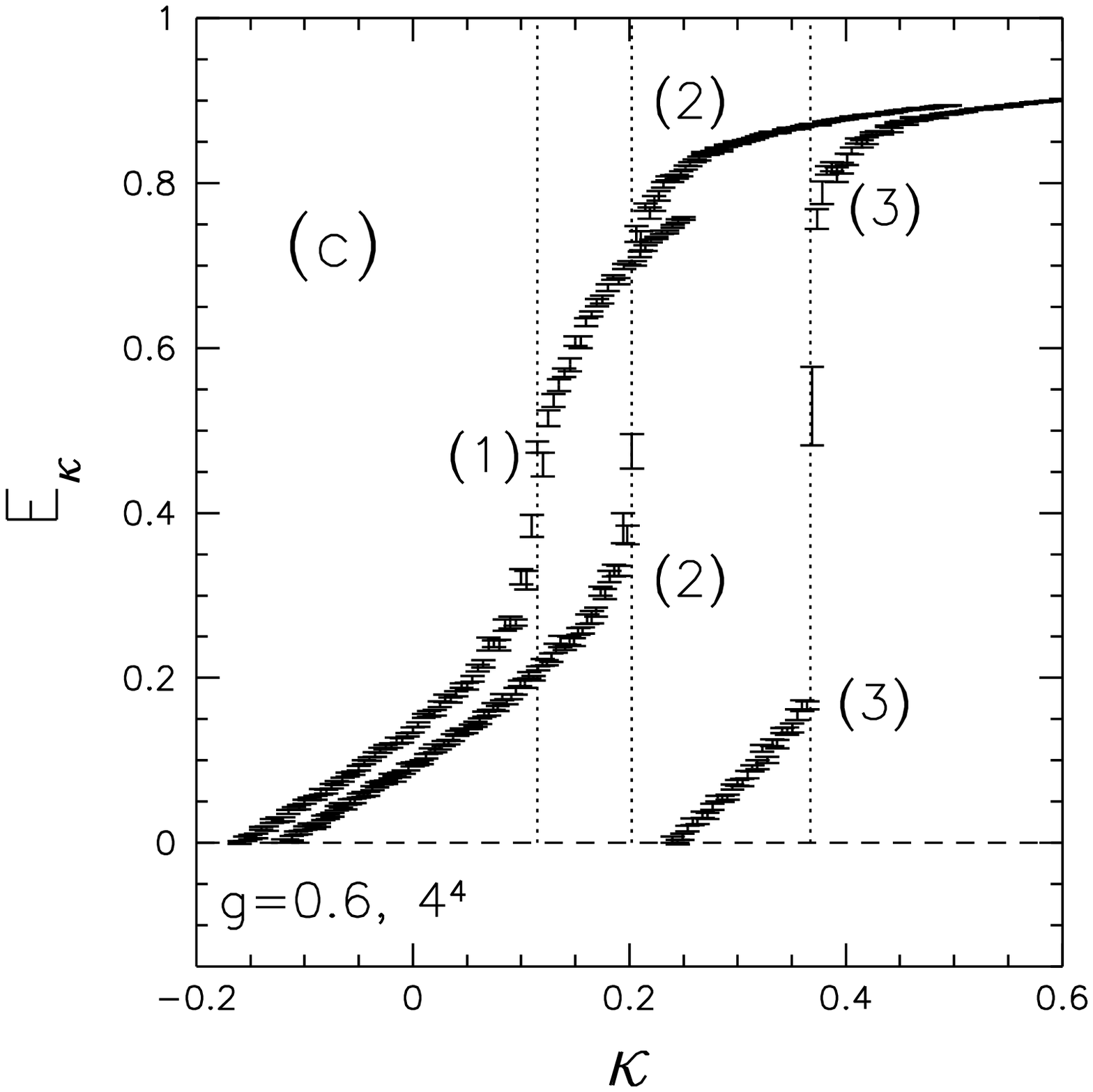}  }  \\  
\end{tabular}
\vspace*{-0.0cm}
\caption{ \noindent {\em  Scans across the FM-FMD (Fig.~a), 
FMD-PM (Fig.~b) and 
FM-PM (Fig.~c) phase transitions. The three curves in Fig.~c which are 
labeled by the numbers 
$(1)$, $(2)$, $(3)$ were obtained at 
$(\tk,\tr)=(0.01,1)$, $(\tk,\tr)=(0.05,0.15)$ and  
$(\tk,\tr)=(0.05,-0.30)$. The estimated positions of 
the phase transitions are marked in all 
graphs by the vertical dotted lines. 
}}
\label{LOCAL}
\end{figure}
 
We have explored the phase diagram again at $g=0.6$, 
and, as in the mean-field analysis, 
we kept either the value of $\tk$ 
or of $\tr$ fixed and scanned the two-dimensional 
$(\k, \tr)$ or $(\k, \tk)$ plane. 
At each point of the scan we accumulated $5000$ Metropolis sweeps, 
which were preceded by $3000$ equilibration sweeps. The observables in 
Eqs.~(\eq{VEP}-\eq{ORDER}) were measured after each sweep. We corrected 
for autocorrelation time effects by multiplying the statistical 
error bars with $\sqrt{2 \tau}$ where $\tau$ is an estimate for 
the integrated autocorrelation 
time, which for the local observables (\eq{VEP}-\eq{ORDER}) 
is in the range between $2$ and $5$.
Most of the phase diagram scans were performed on a $4^4$ lattice. 
A few runs at $\tr=1$ were also done on an $8^4$ lattice. 

In Figs.~\ref{LOCAL}a-c we have displayed the various observables 
(\eq{VEP}-\eq{ORDER}) for some exemplary scans across 
the FM-FMD (Fig.~\ref{LOCAL}a), the FMD-PM (Fig.~\ref{LOCAL}b) 
and the FM-PM phase transition (Fig.~\ref{LOCAL}c). 
Figure~\ref{LOCAL}a shows that the order parameter $V$ is very small in 
the FM phase and rises sharply at the FM-FMD transition. The fact 
that $V$ is also non-zero in the FM phase is a finite-size effect. The 
internal energies $E_\k$ and $E_{\rm g.f.}$ show a sharp kink at the 
phase transition, whereas $E_p$ changes only very little. The transition 
seems to be of second order, in accordance with our 
mean field results and with perturbation theory.

The position of a 
phase transition on a finite lattice can be defined 
in different ways ({\it e.g.} the position of the 
maximum of the specific heat, or the real part 
of the partition function zero with the smallest imaginary 
part), 
and these transition points will all differ slightly from each other by 
an amount which vanishes in the infinite-volume limit. 
In our case we have identified the FMD-FM phase transition 
from the point where the slope of $V$ is maximal, but one should keep 
in mind that the position of the phase transition in the infinite-volume 
limit will slightly deviate from that value.  

The observables for a scan 
across the FMD-PM phase transition at $\tr=1$ 
are shown in Fig.~\ref{LOCAL}b. 
We believe that this phase transition is of second order 
which agrees with the findings from 
our mean-field calculation.  The point where the slope 
of $V$ is maximal always coincides nicely with the point where 
the slopes of $E_\k$ and $E_{\rm g.f.}$ are maximal. 
Finally, we have displayed in Fig.~\ref{LOCAL}c the internal energy $E_\k$
for three scans across the FM-PM transition 
((1): $\tr=1$, $\tk=0.01$; (2): $\tr=0.15$, $\tk=0.05$; 
(3): $\tr=-0.3$, $\tk=0.05$) as a function 
of $\k$. The plot indicates that the order of the 
FM-PM transition changes from second to first order 
when $\tr$ is lowered from $1$ to $-0.3$ at $\tk>0$. This 
agrees with our mean-field calculation. In contrast to the 
mean-field calculation, our Monte Carlo simulations seem to indicate that 
the FM-FMD transition is of second order at small $\tr$.
  
We have compiled our results for the various 
phase transitions into $(\k,\tk)$-or 
$(\k,\tr)$-phase-diagram plots, which are displayed on the right 
in Figs.~\ref{PHASED1}-\ref{PHASED4}. We 
have again determined the phase diagram only 
above the symmetry lines (dash-dotted lines in 
Figs.~\ref{PHASED1}-\ref{PHASED4}), but we 
checked with a few scans that the phase diagram
is indeed symmetric with respect to those lines.
A schematic three-dimensional phase diagrams of the $\tr>0$ region 
is shown in Fig.~\ref{SCHEM}. 
Qualitatively, the Monte Carlo phase diagrams 
comply nicely with the mean-field results, which 
are displayed on the left in Figs.~\ref{PHASED1}-\ref{PHASED4}, 
and which were discussed already in Sect.~\ref{MEAN}.

As mentioned already in Sect.~\ref{MEAN},
a difference between the mean-field and Monte Carlo phase diagrams 
is observed at $\tr=0$ (Figs.~\ref{PHASED2}c and d).  
The mean-field approximation predicts 
an FMD phase at $\tk \apgt 0.25$, $|\k | \aplt  0.125$.
This is not confirmed by the Monte Carlo simulations which give 
strong evidence that this region is filled by a PM phase. 
The Monte Carlo simulations indicate that the PM phase extends   
to very large values of $\tk$ but, on the basis of 
the simulations, it is of course impossible to 
decide if it ends at a finite value of $\tk$, or if it 
extends to $\tk = \infty$. \mbox{}Figure~\ref{LP} shows
that the observed difference at $\tr=0$ is connected to the fact 
that in case of the mean-field calculation 
the FM-FMD-PM tricritical line (in 
Fig.~\ref{LP} we displayed a projection of that line to a 
constant-$\k$
plane) penetrates through the $\tr=0$ plane and at $\tk \approx 0.3$ 
merges with the FM-AM-PM line into a FM-FMD-AM line, whose projection 
approaches asymptotically the $\tr=0$  axis at large $\tk$.          
The FMD phase extends therefore slightly into the $\tr < 0$ half space. 
In contrast, in the case 
of the Monte Carlo simulation, we find that the FM-FMD-PM line 
stays in the $\tr >0$ half space, and approaches the $\tr=0$ plane from above. 
The FMD phase resides only in the $\tr>0$ half space.
The solid curves in Fig.~\ref{LP}b
are to guide the eye, and were obtained 
by fitting the empirical {\it ansatz} $\tr=\pm a/(\tk+b) $ to the data 
with $a$ and $b$ constants. It turns out that the constants 
$a$ and $b$ are very similar for the FM-FMD-PM and FM-AM-PM lines. 

The continuum limit relevant for the gauge-fixing approach 
should be performed by approaching the 
FM-FMD phase transition from the FM side
(away from the tricritical line at which the
FM-FMD transition surface ends, for instance at $\tr\approx 1$).  
Our Monte Carlo simulations
indicate that this transition is of second order.
In the next section,
we will show that the particle spectrum in this continuum 
limit contains only a massless vector particle, the photon. 
The FM-PM phase transition at small $\tr$ near the FM-FMD-PM line 
and the whole FM-AM phase transition seem to be 
of first order, and no continuum theory can be defined at those
transitions. 

Finally, we compare our one-loop result for $\k_{\rm FM-FMD}$ 
(Eq.~(\eq{k_FMFMD})) with our Monte Carlo results. The one-loop 
results for $\k_{\rm FM-FMD}$ are represented in the graphs on the 
right in Figs.~\ref{PHASED1}-\ref{PHASED4} 
by the solid lines. They agree reasonably well with our Monte Carlo 
data. In the region of the FM-FMD phase transition we find that 
the departure of the perturbative results from the Monte Carlo data 
is in all cases smaller than two standard deviations.   
In some cases (see for instance
the phase diagram at $\tr=1$ in Fig.~\ref{PHASED1}b), 
however, the one-loop curve is systematically above the numerical results. 
To see if this deviation is due 
to finite-size effects we repeated some of the scans across 
the FM-FMD phase transition on an $8^4$ lattice. The obtained 
transition points are marked by the open circles in Fig.~\ref{PHASED1}b. 
They do not significantly deviate from our data on the $4^4$ lattice. We 
also evaluated the lattice integral $X_1$ in Eq.~(\eq{k_FMFMD}) 
on a $4^4$ and $8^4$ lattice 
which again did not result in a significant change 
of the situation encountered in Fig.~\ref{PHASED1}.
We therefore believe that the observed deviations are due to higher loop 
corrections.  They become smaller for smaller $\tr$.
This is not unreasonable: as long as perturbation theory
applies, some couplings are proportional to $\tr$, so that higher loop
terms maybe less important for smaller $\tr$.  (Of course, when $\tr$
is too small, perturbation theory breaks down all together, as is 
clear from the case $\tr=0$, where we do not even find an FM-FMD
transition, numerically.) 
\subsection{Vector and Higgs two-point functions}
\lb{NUM_PROP}
To see whether the spectrum at the FM-FMD phase transition indeed contains 
only a massless photon we have computed the vector and Higgs two-point 
functions numerically.
\subsubsection{Vector two-point function }
In our Monte Carlo simulations we computed the correlation function 
$\D_{\m \n}^V(p)$ in Eq.~(\eq{PROP_V}). 
We have set $\m=\n$ and $p_\m=0$. 
The two-point functions $\D^{V}_{\m \m} (p)$, $\m=1,2,3$
were determined for a table of lattice momenta $p$ which 
lead to different values of 
$\phat^2=\sum_{\n \neq \m} \phat_\n^2$. Finally, we 
took the average of the three two-point functions 
$\D^{V}_{\m \m} (p)$, 
$\m=1,2,3$. All Monte Carlo simulations were 
performed at $\l_i=0$, $i=1,\dots,5$.

\begin{figure}[t]
\vspace*{-0.5cm}
\begin{tabular}{ll}
 \hspace*{-1.3cm} \epsfxsize=9.80cm
 \vspace*{-0.5cm}
\epsfbox{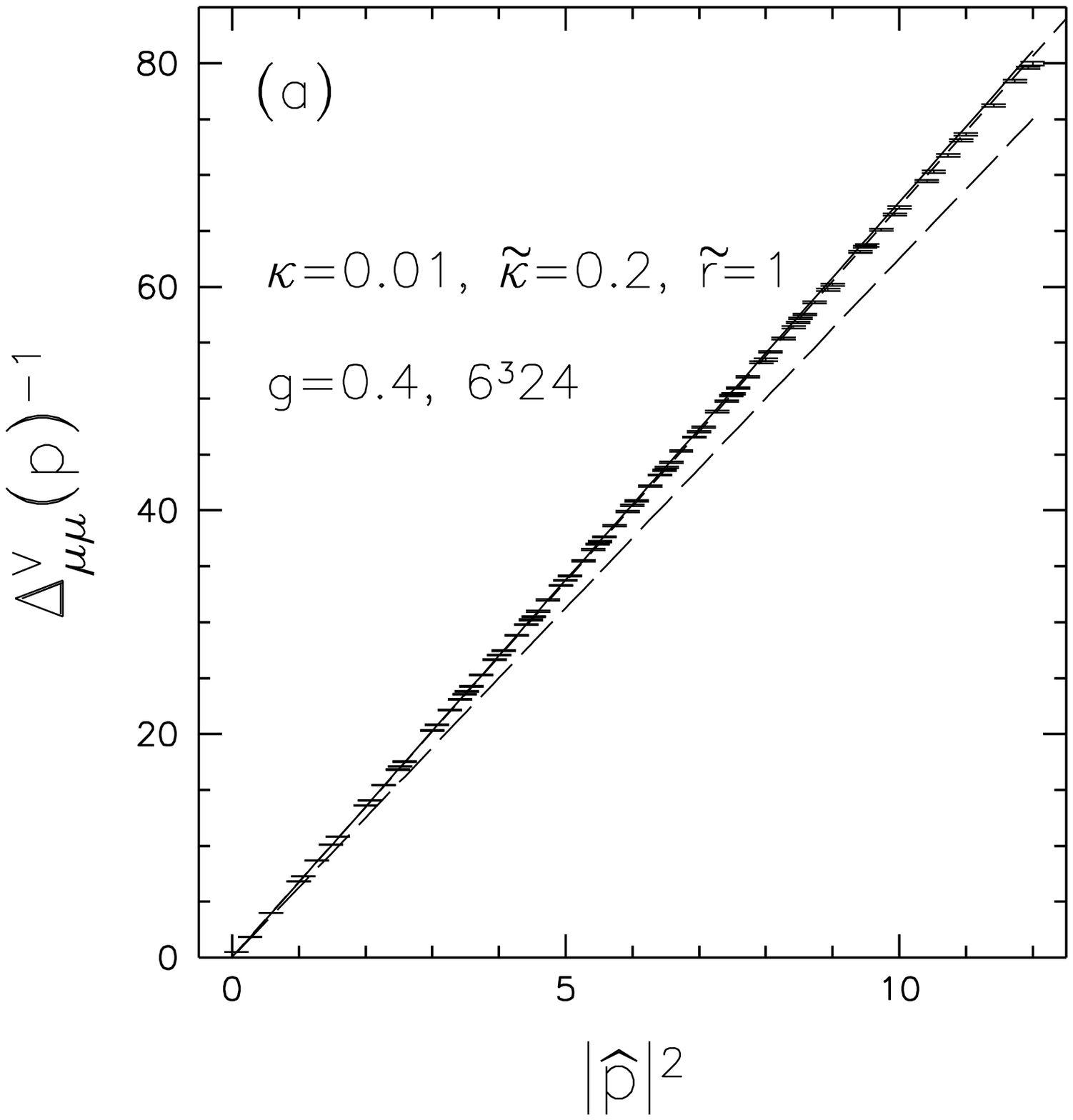}     &
 \hspace{-3.4cm} \epsfxsize=9.80cm
 \vspace*{0.0cm}
\epsfbox{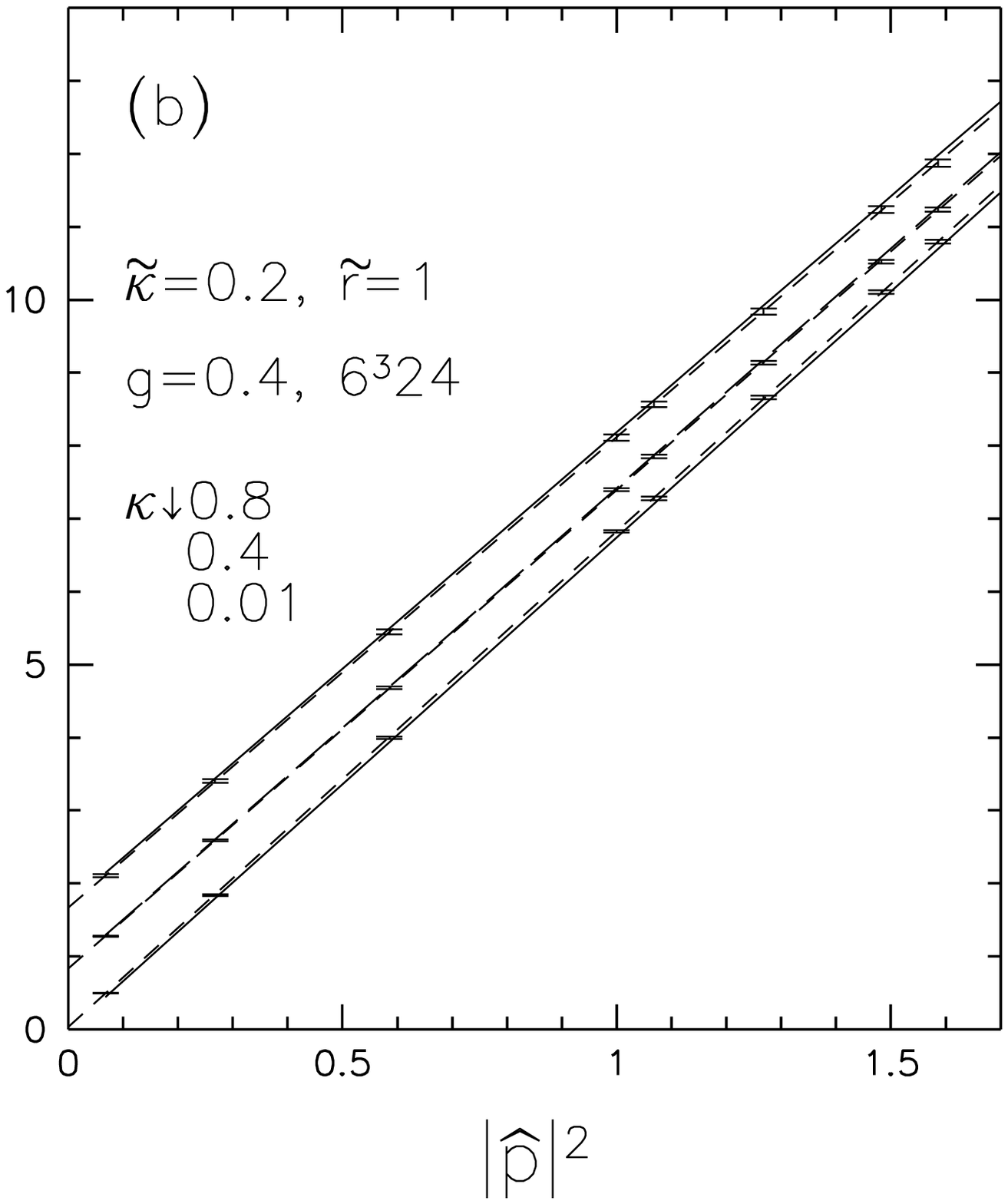}     \\
\end{tabular}
\vspace*{0.0cm}
\caption{ \noindent {\em The inverse vector two-point function (points
with error bars).  The long-dashed line is the tree-level result, and the
solid line the one-loop result.  The latter is almost on top of a linear
fit (short dashes)
to the data.  Fig.~a shows results for $\k=0.01$, while Fig.~b shows
a blow-up of the small momentum region for $\k=$0.8, 0.4 and 0.01.
}}
\label{RATIOV}
\end{figure}
 
For the above choices, we first derive a simplified expression for the 
vector two-point function to one-loop order, which can then
easily be compared with the Monte Carlo results. After setting 
$\m=\n=1$ and $p_1=0$, the one-loop  
vector two-point function in Eq.~(\eq{PROP_ONE}) can be written as,
\bea
\D^{V, (1)}_{11} (p)&=&\left[ \frac{Z_{11}(g^2)}{\D^{V,(0)}(
p)^{-1} +\S(p) }
\right]_{11, \; p_{1}=0}\lb{SV_ONE_A} \nonumber \\
&=& Z_{11}(g^2)\left[ \D^{V,(0)}(p)-\D^{V,(0)}(p)\; 
\S(p) \; \D^{V,(0)}(p) +\cdots \right]_{11, \; p_{1}=0}
\;, 
\lb{SV_ONE_A_expand} 
\eea
with 
\be 
\S_{\m\n}(p)=\left( \l_1 g^2|\phat^2|+\l_2g^2\phat_\m^2\right)
\d_{\m\n}+\l_3g^2\phat_\m\phat_\n+\S^{\rm bare}_{\m\n}(p)\;,
\ee
where 
$Z_{1 1}(g^2)$ is the wave-function renormalization constant 
in Eq.~(\eq{WAVEF}) and $\S^{\rm bare}(p)$ 
is the self-energy in Eq.~(\eq{SIGMA_V}) ($m^2=2g^2\k$ is already
included in $\Delta^{V,(0)}$, {\it cf.} Eq.~(\eq{PROP_TREE})). 
After writing Eq.~(\eq{SIGMA_V}) 
as $\S^{\rm bare}_{\m \n}(p)=a_{\m}(p) \d_{\m \n}+ b_{\m \n}(p)
\phat_\m \phat_\n$ and Eq.~(\eq{PROP_TREE}) as  
$\D^{V,(0)}_{\m \n}(p)={\cal A}(p)\d_{\m \n} 
+{\cal B}(p)\phat_\m \phat_\n$, and using the fact that 
$p_1=0$, we obtain 
\bea
\D^{V, (1)}_{11} (p)&=&Z_{11}(g^2)\left[ 
{\cal A}(p)+{\cal A}(p)\; a_{1}(p)\; {\cal A}(p) 
+\cdots\right]_{p_{1}=0}
\nonumber  \\
&=& \frac{Z_{11}(g^2)}{ m^2+(1+\l_1) \; 
\sum_{\n \neq 1} \phat_\n^2 +a_1(p) 
}\; .
\lb{SV_ONE_B} 
\eea
Note that the expression in Eq.~(\eq{SV_ONE_B}) 
does not depend on $\l_2$ and $\l_3 $ as we set 
$p_1=0$. This also implies that we 
will not see any effects of the Lorentz-symmetry breaking part 
of the one-loop self-energy in our data. 
In the following we set also $\l_1$ equal to zero 
since in our Monte Carlo simulations we have also 
not included this counterterm. On a lattice which is
asymmetric in space and time $a_1(p)$
is given by the expression
\bea
\left. a_1(p)\right|_{p_{1}=0}&\!\!=&\!\!\frac{g^2 \; \tr}{2\xi} \; 
\left[ 3 I_{11}+I_{44} \right] -\frac{g^2}{\xi} \; 
\left[ 2 I_{12}+I_{14}+I_{11} \right] \nonumber \\
&&\!\!-  \half \; g^2 \; \left[ 2\; (J_{12}-I_{12} )  
\; (\phat_2^2 + \phat_3^2) 
+\phat_4^2 \; \left(J_{14}+J_{41}
-2 \; I_{14}) \right)\right] \,, 
\lb{A_MU} 
\eea
where the integrals $I_{\m \n}$ and  $J_{\m \n}$ are given 
in Eqs.~(\eq{I11}) and (\eq{J12}). 
%
%

In Fig.~\ref{RATIOV} we plotted the results of our numerical 
computation of the inverse vector propagator in momentum space,
as a function of $|\phat|^2$.
The long-dashed and solid lines represent tree-level and one-loop
perturbation theory evaluations of the same quantity, at values
of the parameters equal to those used in the numerical computation.
In Fig.~\ref{RATIOV}a, these are $(\k,\tk,\tr)=(0.01,0.2,1)$, while
in Fig.~\ref{RATIOV}b we show an enlargement of the small-momentum
region for $(\tk,\tr)=(0.2,1)$ and three different values of $\k$,
0.8, 0.4 and 0.01, the latter being very close to the FM-FMD
transition point.  A linear fit (short dashes) 
to the numerical results can, within
the resolution of the plots, hardly be distinguished from the one-loop
curve.  
The gauge coupling $g$ is $0.4$ and the lattice size is $6^3 24$. 
The perturbative expressions for (the inverses of)
$\D^{V, {\rm (0)}}_{\m \m} (p)$
and $\D^{V, {\rm (1)}}_{\m \m} (p)$ were evaluated  
on a lattice of the same size.  We have measured the vector two-point 
function $\D^{V}_{\m \m} (p)$ 
on $2 \times 10^5$ configurations which were generated 
by our 5-hit Metropolis program. The error bars were again computed by 
multiplying the standard deviation with $\sqrt{2 \tau}$ where $\tau$ 
is an estimate of the integrated autocorrelation time obtained from 
the autocorrelation function.

{}From these results we draw two conclusions: the fact that a linear fit
works very well confirms that the theory is a theory of free photons
near the FM-FMD transition, and the good agreement with perturbation
theory implies that this can be understood in perturbation theory,
as explained in Sect.~\ref{ANA}.
\begin{figure}[t]
\centerline{
\epsfxsize=11.0cm
\vspace*{-.5cm}
\epsfbox{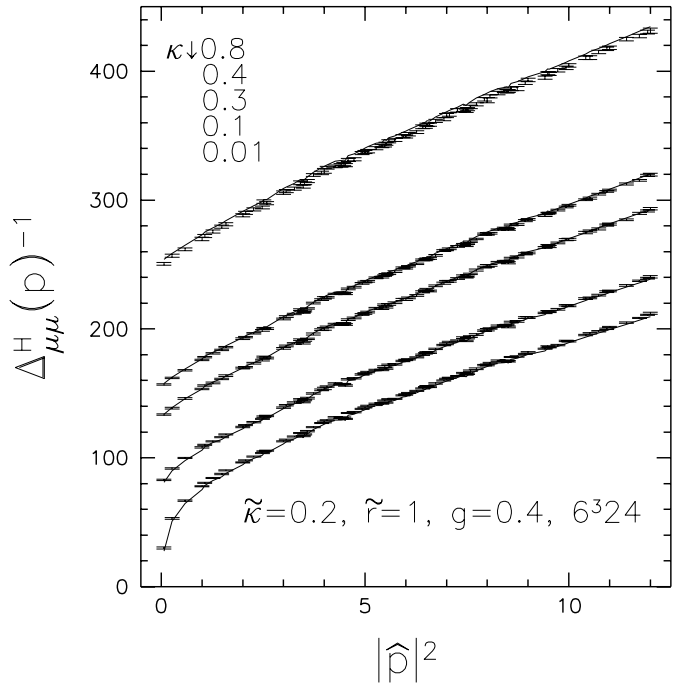}
}
\vspace*{0.0cm}
\caption{ \noindent {\em The inverse Higgs two-point function, 
${\D_{\m \m}^H(p)}^{-1}$, 
plotted as a function of $|\phat|^2$  
for a series of points near the FM-FMD phase transition 
at $g=0.4$, $\tk=0.2$ and $\tr=1$. 
The error bars mark the Monte Carlo results. 
The lattice size is again $6^3 24$. 
The perturbative results (Eq.~({\protect \eq{PROP_S_ONE}}))
are represented by the solid lines.  
}}
\label{HIGGSPA}
\end{figure}

For comparison, we have repeated this analysis at a series of points near the
FM-PM phase transition, at $\tk>0$ and $\tr=1$.  We found that the perturbative
results do not
converge (one loop is not close to tree level), and also do not describe
the numerical data in this case (with deviations well over 100\%).
We also determined 
the vector boson mass $m_V$ by fitting an {\it ansatz} 
$Z_V/(\phat^2 +m^2_V)$ to 
the numerical data of the vector two-point function 
at small $|\phat|^2$. We find that the obtained 
vector boson mass $m_V$ shows qualitatively the 
same $\k$ dependence as reported in Ref.~\cc{spectrum} for the 
U(1) gauge-Higgs model. The vector boson 
mass decreases when $\k$ is lowered from the FM side towards the FM-PM 
transition. However, it does not vanish at the phase transition which 
is probably, like in the U(1) gauge-Higgs model, a finite-size effect. 
\subsubsection{Higgs two-point function} 
As mentioned before, 
the spectrum at the FM-FMD phase transition should contain only
a massless photon, and no Higgs particle.                    
In this section we present the Monte Carlo results for the 
Higgs two-point function, and compare them with the analytic formulas derived 
in Sect.~\ref{PROP_AN}. 
\begin{figure}[t]
\centerline{
\epsfxsize=11.0cm
\vspace*{-.9cm}
\epsfbox{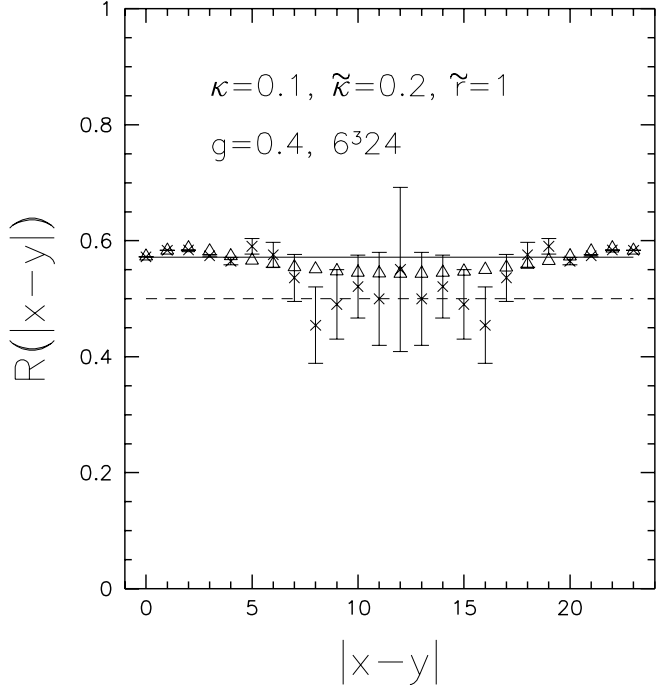}
}
\vspace*{-.0cm}
\caption{ \noindent {\em  The ratio $R(|x-y|)$ 
as a function of $|x-y|$ at a point near the FM-FMD phase transition 
($(g,\k,\tk,\tr)=(0.4,0.1,0.1,1)$). The lattice size is $6^3 24$. 
The horizontal dashed and solid lines were obtained by evaluating the 
ratio in Eq.~({\protect \eq{RATIO}}) in the infinite volume limit
to leading and next-to-leading order in perturbation theory 
(Eqs.~({\protect \eq{C1}}) and ({\protect \eq{C2}}).)
The triangles were obtained by evaluating 
$G^{H}_{\m \m} (|x-y|)$ and $G^{V}_{\m \m} (|x-y|)$
in Eq.~({\protect \eq{RATIO}}) to next-to-leading order 
on the same lattice as used in the simulations. 
}}
\label{FAC_A}
\end{figure}

We have computed the momentum space 
Higgs two-point function in Eq.~(\eq{PROP_S}). As in the case 
of the vector two-point function, we have set $\n=\m$ and $p_\m=0$.
The Higgs correlation function was determined for the same 
lattice-momenta as the vector two-point function.  

To see if the two-point function leads to a pole we have plotted in
Fig.~\ref{HIGGSPA} ${\D_{\m \m}^H(p)}^{-1}$ as a 
function of $|\phat|^2=\sum_{\n\neq \m} \phat_\n^2 $ 
for several $\k$ values near the FM-FMD phase transition.
The simulations were performed at 
$g=0.4$, and at $\tk=0.2$ and $\tr=1$.         
The numerical data are represented in Fig.~\ref{HIGGSPA} 
by the error bars. The five data sets from the 
top to bottom correspond to $\k=0.8$, $0.4$, $0.3$, $0.1$ and $0.01$.
At each $\k$ we have measured ${\D_{\m \m}^H(p)}$ on $2\times 10^5$ 
equilibrium configurations which were again generated with the 5-hit 
Monte Carlo algorithm. As in the case of the vector two-point 
function, we corrected for the autocorrelation time effects by 
multiplying the standard deviation with a factor $\sqrt{2 \tau}$.             
Our Monte Carlo simulations indicate that the FM-FMD phase 
transition is situated at $\k \approx 0$.

If the pole scenario would be correct, 
the ${\D_{\m \m}^H(p)}^{-1}$ data at sufficiently small momenta 
should fall on a straight line, for $\k\searrow\k_{\rm FM-FMD}$.
What we find, however, is that,
when lowering $\k$ towards the FM-FMD phase transition, 
a cusp emerges at small momenta. 
Evidently, the data at small $|\phat|^2$ do not fall 
on a straight line. The cusp is due to the logarithm in 
Eq.~(\eq{PROP_SNA}). 
The solid lines in Fig.~\ref{HIGGSPA} were obtained by    
evaluating Eq.~(\eq{PROP_S_ONE}) on a lattice of the same size and for 
the same parameter values as used in the simulations. We find 
that the perturbative formula (\eq{PROP_S_ONE}) describes the data 
very well. 
(A similar behavior was discovered before in the reduced limit
of the model at $\tr=1$ for the left-handed neutral and right-handed
charged fermion propagators, which also exhibit such a logarithmic
singularity
at small momenta, and do not exist as bound states, see 
Refs.~\cc{wmy_prl,wmy_edin}.)

Again for comparison, we looked also at the Higgs two-point function  
near the FM-PM phase transition at $\tk=0$ and $g=0.6$. 
We find that $\k_{\rm FM-PM} \approx 0.18$. 
We find that in this case, in accordance with expectations, 
the spectrum at the 
FM-PM phase transition contains a massive Higgs particle, 
giving rise to a pole in  ${\D_{\m \m}^H(p)}$.        
To extract the Higgs mass we fitted the 
$\D^H(p)^{-1}$ data at small momenta to the {\it ansatz} 
$(m_H^2 +|\phat|^2)/Z_H$. 
The Higgs mass decreases when $\k$ 
is lowered toward the FM-PM phase transition but, 
because of finite size effects, does not vanish at the phase transition 
(again, as in Ref.~\cc{spectrum}).
We also find that, 
as in the case of the vector two-point function,
perturbation theory does not describe 
the PM side of the FM-FMD-PM tricritical line.
\begin{figure}[t]
\centerline{
\epsfxsize=11.0cm
\vspace*{-.9cm}
\epsfbox{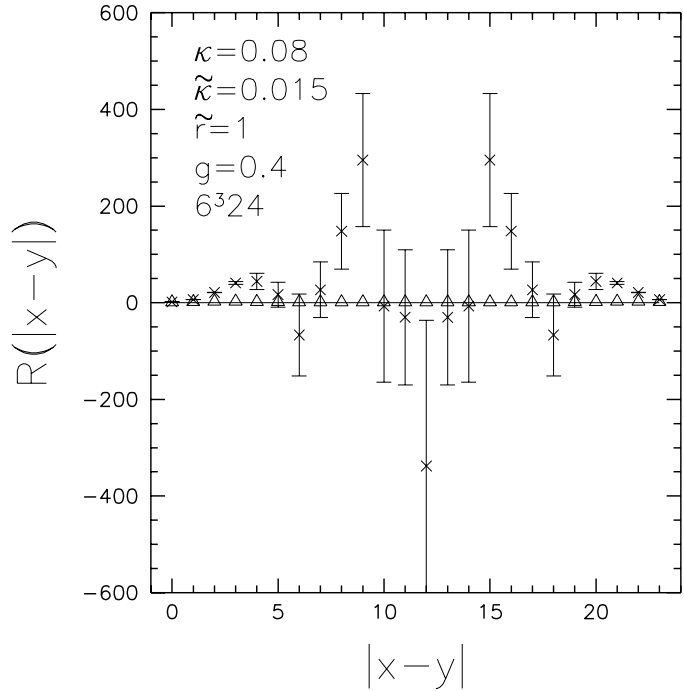}
}
\vspace*{0.0cm}
\caption{ \noindent {\em  The ratio $R(|x-y|)$ 
as a function of $|x-y|$ 
at  a point near the FM-PM phase transition 
($(g,\k,\tk,\tr)=(0.4,0.08,0.15,1)$). The lattice size is $6^3 24$. 
The horizontal solid line and the triangles were obtained by evaluating the 
ratio in Eq.~({\protect \eq{RATIO}})
to leading and next-to-leading order in perturbation theory. 
}}
\label{FAC_B}
\end{figure}
%
%
\subsubsection{Factorization} 
\lb{SFAC}
The results of the previous section already give
strong evidence that the Higgs two-point function in the continuum limit 
at the FM-FMD phase transition does not have a 
pole, and that consequently 
a Higgs particle does not exist in the spectrum. In Sect.~\ref{PROP_AN}
we have shown that, to 
next-to-leading order in perturbation theory, 
the Higgs two-point function in coordinate 
space factorizes into the product of two vector two-point functions 
(Eq.~(\eq{GS_FAC})). For $\l_4=\l_5=0$, 
factorization holds to this order for arbitrary 
values of the other counterterm coefficients. 

In this section we will 
investigate whether factorization also
holds  non-per\-tur\-ba\-ti\-ve\-ly, which would imply that a Higgs bound 
state does not exist.                                     
To this end, we have computed the Higgs and vector two-point correlation 
functions $G^{H}_{\m \m} (|x-y|)$ and $G^{V}_{\m \m} (|x-y|)$
in Eqs.~(\eq{GS}) and (\eq{GV}) in our Monte Carlo simulations as function 
of $|x-y|$ where $x$ and $y$ were chosen to be
two on-axis points. The simulations were 
carried out at the point $(g,\k,\tk,\tr)=(0.4,0.1, 0.2,1)$,
which is located in the FM phase near the FM-FMD phase 
transition. As in our other Monte Carlo simulations, we have set 
$\l_1= \ldots =\l_5=0$. (Strictly speaking, we expect factorization only
to hold for the properly tuned values of $\l_4$ and $\l_5$, but to the
extent that our simulation results agree with our perturbative results,
we do not expect to see the difference,
{\it cf.} end of Sect.~\ref{ANA}.)  The lattice size is again $6^3 24$. 
We find that enormous statistics 
is required to obtain a stable signal  for
$G^{H}_{\m \m} (|x-y|)$. This is partly due to the subtraction
in Eq.~(\eq{GS}). We used translation invariance on the 
lattice to improve the signal. We have 
accumulated in total about $27 \times 10^6$ Metropolis sweeps. 
The Higgs- and vector-correlation functions 
were measured after each sweep. 

In Fig.~\ref{FAC_A} we plotted the ratio
\be
R(|x-y|) =\frac{G^{H}_{\m \m} (|x-y|)} 
               {\left[ G^{V}_{\m \m} (|x-y|) \right]^2 } 
\;  \lb{RATIO} 
\ee
as a function of $|x-y|$. The error bars of the ratio 
were calculated by a blocking procedure. 
Figure~\ref{FAC_A} shows that the Monte Carlo data for $R(|x-y|)$ (crosses)
are, within error bars, independent of $|x-y|$, indicating
that factorization holds also non-perturbatively.  In fact, 
factorization sets in almost immediately (both perturbatively
and numerically).

\mbox{}From our calculation in Sect.~\ref{PROP_AN} it follows 
that this ratio 
is equal to $1/2$ to leading order in perturbation theory.
Figure~\ref{FAC_A} shows that the Monte Carlo results are indeed very 
close to this value, marked by the dashed line. We find however  
that the data at small distances where the error bars are smallest
are systematically above this value by a small amount. To understand this small 
discrepancy, we have computed the ratio also to next-to-leading 
order in perturbation theory. The next-to-leading order result 
for $C_{\m \m}$, which was computed in Sect.~\ref{PROP_AN} in the 
infinite volume limit, is represented in Fig.~\ref{FAC_A} by 
the horizontal solid line, which is indeed much closer to the numerical 
data. In addition, we have also evaluated all 
Feynman diagrams in Fig.~\ref{FDIA1} numerically on the same 
lattice and for the same parameter values as used in the Monte Carlo 
simulations. Figure~\ref{FAC_A} shows that 
the next-to-leading order result for $R(|x-y|)$ (triangles)
agree within two standard deviations for all separations $|x-y|$ with the 
Monte Carlo data. 
We consider the $|x-y|$-independence of our Monte Carlo results for $R(|x-y|)$ 
and the good agreement with the perturbative result as 
strong indication that factorization holds, 
and that consequently the spectrum  
near the FM-FMD phase transition does not contain a Higgs particle.         
We repeated the same calculation at a point in the FM phase  
near the FM-PM phase transition where a Higgs particle is known to exist.
The Monte Carlo simulations were performed at 
$(g,\k,\tk,\tr)=(0.4,0.08,0.015,1)$ with less statistics, 
about $5 \times 10^6$ Metropolis sweeps. Figure~\ref{FAC_B} shows 
clearly that, in contrast to Fig.~\ref{FAC_A}, 
the ratio $R(|x-y|)$ depends 
strongly on $|x-y|$ (notice also the difference in ordinate scale between
Figs.~\ref{FAC_A} and \ref{FAC_B}!), 
and seems to oscillate when the separation 
$|x-y|$ is increased. The solid lines and the triangles represent again the 
leading and next-to-leading order perturbative results for 
the ratio, both far off from the Monte Carlo data. 
It is obvious 
that factorization does not hold on the PM side of the FM-FMD-PM 
tricritical line.                                                 
\section{Conclusion and Outlook}
\secteq{5}
\lb{OUT}
In this paper we investigated the gauge sector of the 
gauge-fixing approach for the case of a U(1) 
gauge group. This approach provides a completely new non-perturbative 
formulation of a U(1) gauge theory on the lattice, which is more complicated 
than Wilson's manifestly gauge-invariant compact formulation,
but closer in spirit to the continuum formulation,
for which gauge-fixing is indispensable. 
Gauge fixing allows us to control the longitudinal 
degrees of freedom which otherwise, as we mentioned in Sect.~\ref{INTRO}, 
form a central obstruction to the construction of lattice chiral gauge 
theories. 

The action is rather complicated in comparison 
with the Wilson plaquette action. Apart from the  
Wilson plaquette term and the 
gauge-fixing term, it also includes six counterterms whose coefficients 
have to be adjusted such that Slavnov--Taylor identities are restored in the 
continuum limit. We have demonstrated in this paper that 
this new lattice formulation reproduces 
the desired properties of the continuum theory at a new type of continuous
phase transition, the FM-FMD transition. 
The spectrum contains at this phase transition
only a massless photon, while 
a Higgs-like excitation, associated with propagating longitudinal 
gauge degrees of freedom, does not exist. 

This new phase transition occurs in a region of the phase diagram
accessible to weak-coupling lattice perturbation theory, as one would
expect from the close relation to the continuum theory.  
In perturbation theory, by construction, the
theory (without fermions) at the phase transition is a theory of free
massless photons.  The very good agreement between one-loop perturbation
theory and numerical results makes us confident that this is also true
non-perturbatively.

As mentioned above, in order to make this work, 
six counterterms need to be adjusted.
However, only one of those (a gauge-field mass term)
has dimension smaller than four.  One expects that the others can
be reliably calculated in perturbation theory.  Our results indicate
that this is indeed true: within our numerical precision, we find that
we can do with just the tree-level values of all dimension four
counterterms.  Adding fermions to the theory will require a few
additional counterterms, but (using a fermion formulation with
shift symmetry \cc{gp}) all of those are dimension four, and we expect
that they can be calculated in perturbation theory as well.  This
means that the fact that counterterms are required does not make
this formulation of lattice chiral gauge theories particularly
expensive.

It is clear that the precise form of the lattice action should be
chosen such that lattice perturbation theory applies. 
It is therefore important to construct 
the lattice gauge-fixing action such 
that the dense set of lattice Gribov copies of the perturbative
vacuum ($U_{\m x}=1$), which occurs for a naive discretization
of the gauge-fixing action, 
is removed by adding higher-dimensional operators (with dimension
six or higher). For comparison, we studied also the limit where those 
higher-dimensional terms are omitted (by setting $\tr=0$), such that 
these lattice Gribov copies are present. Our numerical 
results for this naive choice show that, at small $\tk$, there is an FM-PM type
phase transition, in a universality class different from the FM-FMD
transition (there is even some evidence that the FM-PM 
phase transition is of first order, implying that 
a continuum limit cannot be 
performed at all). The situation at large $\tk$ is 
very unclear since the FM-PM phase 
transition is ``wedged" (in the $(\k,\tk,\tr)$ phase diagram) 
between two tricritical lines, resulting in a 
very complicated phase structure, where four phases get 
very close to each other. We furthermore find that the numerical simulations 
in that region are hampered by strong metastabilities. These
findings strongly suggest that for $\tr=0$ ({\it i.e.} naive
gauge fixing) no phase transition in the desired universality class
of the continuous FM-FMD transition (found at $\tr\approx 1$ 
and large enough
$\tk$) occurs.  This implies that the naive, $\tr=0$ action does not lead
to a theory of free photons, and is unsuitable for the construction
of lattice chiral gauge theories.  It is likely that this is related
to the dense set of lattice Gribov copies which occurs at $\tr=0$,
since they represent unsuppressed rough fluctuations of the longitudinal
gauge field.  In addition, our mean-field 
results indicate that for small nonzero
$\tr$ the FM-FMD transition may become first order, which would imply
that small values of $\tr$ should be avoided altogether. 

Our previous results on the fermion spectrum in the reduced model
\cc{wmy_prl}, combined with
the results of this paper, provide what we consider to be convincing
evidence that the gauge-fixing approach does indeed lead to a viable 
non-perturbative lattice formulation of chiral gauge theories 
for abelian gauge groups.  We would like to emphasize that a key
element of this approach lies in the fact that lattice perturbation
theory provides a valid approximation to our lattice theory (including
fermions \cc{wmy_pert,wmy_prl}).

We believe that, in addition, our gauge-fixed lattice formulation
has matured to the point where it may be used as an alternative
to other gauge-fixing methods for abelian theories.
In the traditional approach, one first performs a Monte-Carlo 
update using only the gauge action (Eq.~(\ref{SG})).
Then, a sequence of gauge transformations is performed, 
aiming to find ``the best'' configuration on the same orbit.
For example, in the Landau-gauge method one attempts to maximize 
${\rm Re} \sum_{x,\m} U_{x,\m}$. A well-known problem is 
that local algorithms cannot (and do not) always find the global maximum.
Some specific obstructions of global nature 
have been described in the literature (see {\it e.g.} Ref.~\cite{LG}
and references therein).

In contrast, in our approach one always performs the Monte-Carlo update
on the entire gauge-field space, with one single Boltzmann weight. 
While the action~(\ref{FULL_ACTION}) is more complicated,
this approach may nevertheless have several advantages:
1) one has all systematic errors completely under control;
2) we believe that global features ({\it e.g.} Double Dirac Sheets~\cite{LG})
do not cause any special difficulty in our approach,
and that this is related to the fact that at no stage is 
our Monte-Carlo update constrained to stay only on a single orbit;
3) finally, as mentioned earlier, 
we find that our photon propagator is in excellent agreement 
with theoretical expectation.
(In order to maintain the good agreement of the photon propagator
with the continuum theory for all four-momenta
we expect that, of the five marginal counterterms,
at least the one-loop $\l_2$ counterterm should be included
in the Monte-Carlo update, see Sect.~\ref{LAMBDAS}, in particular
the discussion of Fig.~\ref{STID}.)

Coming back to the program of constructing chiral lattice gauge theories,
as a next step the gauge-fixing approach should be generalized 
to non-abelian gauge groups. This, even without fermions, is 
a difficult task.   The main obstacle is the existence of ``continuum"
Gribov copies \cc{grib}, 
which occur in non-abelian theories, in addition to the
lattice Gribov copies discussed in this paper.
It is well-known that 
the determinant of the Fadeev--Popov operator can be negative for some 
of these continuum Gribov copies, giving rise to a non-positive 
integration measure. 
It is therefore very likely that the weighting of gauge configurations 
in the path integral will deteriorate due to the presence of
these continuum Gribov copies. In the worst case, contributions from different 
Gribov copies may even 
cancel each other (this is in fact not unlikely, in view of 
Neuberger's theorem \cc{hn_nogo}). This obstacle may be circumvented by 
adopting a gauge-fixing procedure as proposed in 
Refs.~\cc{parrinello,zwanziger}. An advantage of this procedure is that 
the (gauge-field) integration measure is guaranteed to be positive. 
A possible disadvantage of this method is the fact that the 
counterpart of the Fadeev--Popov action is a highly non-local 
functional of the gauge fields. This makes a perturbative analysis,
and, in particular,
the construction of the counterterm action non-trivial.  Work on this
is in progress. 

Another project for future investigation concerns fermion-number 
violation. Most lattice chiral fermion actions (including
that of Ref.~\cc{wmy_prl}) can be          
written in the form $\sum_{x,y} \psb_x D_{x,y}(U) \psi_y$
with $D_{x,y}(U)$ the lattice Dirac operator.  Obviously, 
this action (and also the fermion measure) are invariant 
under an exact global U(1) 
symmetry which, at first glance, seems to be in contradiction
with fermion-number violation \cc{banks}. However, Ref.~\cc{fviol} 
demonstrated, in a two-dimensional 
toy model, that fermion-number violation can still occur despite this exact 
symmetry. The central observation is that fermionic states 
are excitations relative to the vacuum. The global 
U(1) symmetry prohibits a given state to change fermion 
number, but nothing prevents the ground state 
to change when an external field is applied.
We expect that a similar phenomenon may explain how fermion-number
violating processes take place in our four-dimensional dynamical 
theory.
\subsubsection*{Acknowledgements}

We would like to thank Giancarlo Rossi and Massimo Testa for
discussions. 
M.G. would like to thank the Physics Departments of the 
University of Rome II ``Tor Vergata", the Universitat Autonoma
of Barcelona, and the University of 
Washington for hospitality.  M.G. is supported in part
by the US Department of Energy, and Y.S. is supported in part by 
the Israel Academy of Science. The numerical calculations
were performed on the workstations of the Computer Center 
of the University of Siegen and numerous workstations
and PCs at the Physics Departments of Washington
University, St. Louis and the University of Siegen. 
\appendix
\section*{Appendix A}
\secteq{A}
We look for a translation-invariant solution, choosing
the mean-field {\it ansatz} 
\be
\ph_x= \f \; , \lb{MFF}
\ee
\be
U_{\m x} = u\; \exp \left(i \;  A_\m   \right) \;.\lb{MFU}
\ee
The corresponding magnetic fields were replaced by
\be
h_x= h_\f \; , \lb{MFFH}
\ee
and
\be
H_{\m x}= h_u \; \exp \left(i \;  A_\m   \right)\;. \lb{MFUH}
\ee
The $4+d$ mean fields $\f$, $u$, $h_\f$, $h_u$ and $A_\m$,
$\m=1,\ldots,d$
are space-time independent.
Using these expressions we obtained for the free energy 
\bea
&& \!\!\!\!\!\!\!\!\! {\cal F}(\f,h_\f,u,h_u,A;\k,\tk,\tr )=
L^d \; \left\{ \phantom{\sum_{i=1}^4} \!\!\!\!\!\! 2 \; \f \; h_\f +2\;
d\; u\;
h_u
- \log \; I_0 (2h_\f)- d\; \log \; I_0 (2h_u) \right. \nonumber \\
&& \!\!\!\!\!\!\!\!\!  \left. -\half \; d \; (d-1) \; \frac{1}{g^2} \;
\left( u^
4 -1\right)
+\sum_{i=1}^4 \; \f^{2i} \; f^{(i)} (u,A;\k,\tk,\tr) \right\} \;,
\lb{FREE}
\eea
where $L$ is the extent of the $d$-dimensional lattice
in each direction, and
\bea
 \!\!\!\!\!\! f^{(1)}(u,A;\k,\tk,\tr)\!\!\!&=&\!\!\!
                           -2 \; (4\; d\; \tk \;\tr+\k) \; u\; F(A)
                          + \; \tk \; (1+\tr)\; u^2\;(2 F(A)^2 -d)
\nonumber \\
                          \!\!\!&\phantom{=}&\!\!\! + \frac{1}{16}
\;\tk\; \tr
                          \; u^2\; F(2A) \; (2d+1)  \;,
                          \lb{F_1} \\
 \!\!\!\!\!\! f^{(2)}(u,A;\k,\tk,\tr)\!\!\!&=&\!\!\! 2\; \tk \; u^2\;
F(A)^2 \;
                         (\tr-1)
                         - \frac{1}{64} \; \tk\; \tr \; u^2\; \left[
                           6 \; u^2 F(2A)^2 \right. \nonumber \\
 \!\!\!&\phantom{=}&\!\!\!  -4 \; (2 \; (d+1) +(d-1)\; u^2 )
                           \; F(2A) \nonumber \\
\!\!\!&\phantom{=}&\!\!\! \left. +4\; d\; (1+2d)
                            + d\; (2d -5)\; u^2  \right] \;,
                          \lb{F_2} \\
 \!\!\!\!\!\! f^{(3)}(u,A;\k,\tk,\tr)\!\!\!&=&\!\!\!
                         - \frac{1}{16} \; \tk \; \tr \; u^4 \; \left(
                           2 \; F(2A)^2 -2 \; (d-1) \; F(2A)
                   -d  \right) \;,
                          \lb{F_3} \\
 \!\!\!\!\!\! f^{(4)}(u,A;\k,\tk,\tr)\!\!\!&=&\!\!\!
                          - \frac{1}{32} \; \tk \; \tr u^4 \;
                           ( F(2A) -d )^2   \;.
                          \lb{F_4}
\eea
The two quantities $F(A)$ and $I_0(h)$ are given by
\be
F(A)=\sum_\m \cos A_\m  \lb{FQ}
\ee
and
\be
I_0(h)=\frac{1}{\p}\int_0^\p d \a \; \exp( \pm h \; \cos \a ) \;.
\ee
It can be checked that the above expression for ${\cal F}$ reduces
in the limit $u \ra 1$ and $\tr \ra 1$
to the free energy which we obtained before
in the reduced model \cc{wmy_pd}.

The values of the $4+d$ mean fields which are realized at
a given point in the ($\k$, $\tk$, $\tr$) parameter space
are determined from the absolute minimum of the free energy.
The local extrema of the free energy
are obtained by solving the $4+d$ saddle-point equations
\bea
\frac{\del {\cal F} }{ \del h_\f}     \!\!\!&=&\!\!\! 2\;  L^d \;
\left\{ \f -\frac{I_1(2h_\f) }{I_0(2h_\f)} \right\}=0  \;, \lb{SPEH1}
\\
\frac{\del {\cal F} }{ \del h_u}     \!\!\!&=&\!\!\! 2\;  L^d \; d\;
\left\{ u -\frac{I_1(2h_u) }{I_0(2h_u)} \right\}=0  \;, \lb{SPEH2} \\
\frac{\del {\cal F} }{ \del A_\m } \!\!\!&=&\!\!\!  -L^d \left\{
C(\f,u;\k, \tk,
 \tr)
+B(\f,u;\k, \tk, \tr) \; \cos A_\m \right\} \; \sin A_\m =0 \; ,
\lb{QMU} \\
\frac{\del {\cal F} }{ \del \f }     \!\!\!&=&\!\!\! 0 \;,
\;\;\;\;
\;\;\;\;
\frac{\del {\cal F} }{ \del u }     =  0\;,\lb{SPEFU}
\eea
where
\be
I_1(h)=\frac{d I_0(h)}{d h}
\ee
and
\bea
C(\f,u;\k, \tk, \tr)\!\!\!&=&\!\!\! -2 \; \k \; u\; \f^2
+ 4 \; \tk \; (1+\tr) \; u^2 \; \f^2 \;  F(A)  \nonumber  \\
\!\!\!&\phantom{=}&\!\!\! + 4 \; \tk \; (\tr-1) \; u^2 \; \f^4 \;  F(A)
-8d \; \tk \; \tr \; u \; \f^2 \; , \lb{AA} \\
B(\f,u;\k, \tk, \tr)\!\!\!&=&\!\!\! +\quart \; \tk \; \tr \;
(1+2d)\; u^2 \; \f^2 -\tk \; \tr \; \left\{
\quart \; u^2 \; \f^4 \; (3 \; u^2 \; F(2A)   \right. \nonumber  \\
\!\!\!&\phantom{=}&\!\!\!  -(2\; (d+1) + (d-1)\; u^2))
+\half \; (2\; F(2A) -(d-1) ) \; u^4 \; \f^6 \nonumber  \\
\!\!\!&\phantom{=}&\!\!\! \left. +\quart \; (F(2A)-d) \; u^4 \; \f^8
\right\}
 \;. \lb{BB}
\eea
Equations (\eq{QMU}-\eq{SPEFU}) can have
multiple solutions and it is therefore important to insert the various
solutions back into the free energy to
find out which of them corresponds to the absolute minimum.
Since it is difficult to find a closed expression for the
solutions of the saddle-point equations (\eq{SPEH1}-\eq{SPEFU}),
we minimized the free energy numerically. We set
$d=4$.

This was done in the following steps:
The two saddle point equations (\eq{SPEH1}) and (\eq{SPEH2})
were used to express the magnetic fields
$h_\f$ and $h_u$ in terms of $\f$ and $u$.
The free energy depends then only
on $\f$, $u$ and $A_\m$. The saddle-point equation
(\eq{QMU}) has the solutions
\be
A_\m=(z,z,z,z), (\pm A,z,z,z), \ldots, 
(\pm A,\pm A,\pm A,\pm A)  \;,  \lb{QSOL}
\ee
where $z=0,\pi$ and $\cos A=-C/B$ (provided $|C/B|<1$). 
We have inserted each of these solutions back into the
free energy  ${\cal F}$ in Eq.~(\eq{FREE}),
which is now only a function of $\f$ and $u$.
Note that in the PM phase, where
$\f=0$, $A_\m$ remains undetermined because
all the dependence of the free energy on $A_\m$ 
({\it cf.} Eq.~(\eq{FREE})) disappears.
The minimization
with respect to $\f$ and $u$ was done numerically by discretizing
the $(\f,u)$-space by a fine grid and calculating the free energy
on each site of that grid. Finally we have
picked the absolute minimum among the various $A_\m$
solutions in Eq.~(\eq{QSOL}).
The only solutions for $A_\m$ which lead to an absolute minimum of the free
energy turn out to be $A_\m=(0,0,0,0)$, $(\p, \p, \p, \p)$ and 
$(\pm A,\pm A, \pm A, \pm A)$
(with $A \neq 0, \p$),
which correspond respectively to FM, AM and FMD phases.

After this procedure we obtain at a given point in the
phase diagram a numerical value for each of the eight mean
fields $\f$, $u$, $h_\f$, $h_u$ and $A_\m$, $\m=1,\ldots,4$.
The phase transitions, finally, were located by monitoring
$u$, $\f$ and $A$ as a function of the four coupling constants
$g$,  $\k$, $\tk$ and $\tr$. For locating the FM-PM and
FMD-PM boundaries, we used the fact that $\f=0$ on the PM side
and $\f > 0$ on the FM and FMD side of the transition.
Similarly, the FM-FMD (AM-FMD) phase transition was located using
the fact that $A=0$ ($A=\p$)  in the FM (AM)
phase and $A\neq 0,\p$ in the FMD phase.
\section*{Appendix B}
\secteq{B}
In this appendix we give the explicit expressions for
$\langle A_{\m x} A_{\m y} S_{\rm I}^{(4)} \rangle_0$
and $\langle A^2_{\m x} A^2_{\m y} S_{\rm I}^{(4)} \rangle_0$. 
We find 
\be
\langle A_{\m x}A_{\n y} S_{\rm I}^{(4)} \rangle_0 =g^2 \tlsumpa{k} 
e^{ik\left(x-y \right)} \; e^{i \; (k_\m -k_\n)/2 } \;
\sum_{\rho \sigma} \dmn{\mu}{\rho}{k} 
\Sigma^{\rm bare}_{\rho \sigma} (k) \dmn{\sigma}{\nu}{k} \;
\lb{B1}\ee
and
\be
\langle A^2_{\m x} A^2_{\n y} S_{\rm I}^{(4)} \rangle_0 
=
\langle A^2_{\m x} A^2_{\n y} S_{\rm I}^{(4)} \rangle_0^{\rm 1d} 
+\langle A^2_{\m x} A^2_{\n y} S_{\rm I}^{(4)} \rangle_0^{\rm 1e} 
+\langle A^2_{\m x} A^2_{\n y} S_{\rm I}^{(4)} \rangle_0^{\rm 1f} 
\;,\ee
where 
$\langle A^2_{\m x} A^2_{\n y} S_{\rm I}^{(4)} \rangle_0^{\rm 1d}$, 
$\langle A^2_{\m x} A^2_{\n y} S_{\rm I}^{(4)} \rangle_0^{\rm 1e}$ 
and $\langle A^2_{\m x} A^2_{\n y} S_{\rm I}^{(4)} \rangle_0^{\rm 1f}$ 
are the contributions that correspond to the 
Feynman diagrams 1d, 1e and 1f in Fig.~\ref{FDIA1}, 
\bea
&& \!\!\!\!\!\!\!\!\!\!\!\!
\langle A^2_{\m x} A^2_{\n y} S_{\rm I}^{(4)} \rangle_0^{\rm 1d} =
\langle A^2_{\m x} A^2_{\n y} S_{\rm I}^{(4)} \rangle_0^{\rm 1e} 
=g^2 \tlsump{k}
\left[ e^{ i \; k \;(x-y )} \; \; e^{i \; (k_\m -k_\n)/2 } \times 
\phantom{\tlsum{p} } \right.
\nonumber  \\
&& \!\!\!\!
\left. \left(
\tlsuma{p} \dmn{\mu}{\nu}{p+k} \times \sum_{\rho \sigma}
\dmn{\mu}{\rho}{p} \Sigma_{\rho \sigma} (p)  
\dmn{\sigma}{\nu}{p} \right) \right] \; , \lb{D1ed}  \\
&& \!\!\!\!\!\!\!\!\!\!\!\!
\langle A^2_{\m x} A^2_{\n y} S_{\rm I}^{(4)} \rangle_0^{\rm 1f} =
\langle A^2_{\m x} A^2_{\n y} S_{\rm I}^{(4)} \rangle_0^{\rm 1f,I} +
\langle A^2_{\m x} A^2_{\n y} S_{\rm I}^{(4)} \rangle_0^{\rm 1f,II} 
\;, \lb{D1f} \\
&& \!\!\!\!\!\!\!\!\!\!\!\!
\langle A^2_{\m x} A^2_{\n y} S_{\rm I}^{(4)} \rangle_0^{\rm 1f,I}=
g^2 \tlsump{k}
\left[ e^{ i \; k \;(x-y )} \; \; e^{i \; (k_\m -k_\n)/2 } \times 
\phantom{\tlsum{p} } \right.
\nonumber  \\
&& \!\!\!\!
\left( {\tr \over \xi} \sum_{\rho} \tlsuma{p} \dmn{\mu}{\rho}{p} 
\dmn{\mu}{\rho}{p+k}
\mma{\hp}{\rho}\mmb{\hkp}{\rho}  \right. \nonumber \\
&& \!\!\!\!
\times
\sum_{\lambda} \tlsuma{q} \dmn{\nu}{\lambda}{q} \dmn{\nu}{\lambda}{q+k}
\cma{q}{\lambda}\cmb{q+k}{\lambda}\nonumber \\
&& \!\!\!\!
-{2 \over \xi}\sum_{\rho \lambda} \left\{
\tlsuma{p} \dmn{\mu}{\rho}{p} \dmn{\mu}{\lambda}{p+k}
\mma{\hp}{\rho} \mma{\hp}{\lambda} \right. \nonumber \\
&& \!\!\!\!
\left. \left. \left. \times
\tlsuma{q} \dmn{\nu}{\rho}{q} \dmn{\nu}{\rho}{q+k} \right\} \right) 
+ \left( \phantom{\tlsum{q}}
\m \leftrightarrow \n \phantom{\tlsum{q}} \right) \right] \; , \lb{D1f1} \\
&& \!\!\!\!\!\!\!\!\!\!\!\!
\langle A^2_{\m x} A^2_{\n y} S_{\rm I}^{(4)} \rangle_0^{\rm 1f,II}=
g^2 \tlsump{k}
\left[ e^{ i \; k \;(x-y )} \; \; e^{i \; (k_\m -k_\n)/2 } \times 
\phantom{\tlsum{p} } \right.
\nonumber  \\
&& \!\!\!\!
\left( {4 \tr \over \xi} \left\{ {1\over 32}
\sum_{\rho} \tlsuma{p} \dmn{\mu}{\rho}{p} \dmn{\mu}{\rho}{p+k}
\mma{\hp}{\rho} \mmb{\hkp}{\rho} \right. \right. \nonumber \\
&& \!\!\!\!
\times \sum_{\l} \tlsuma{q} \dmn{\n}{\l}{q} \dmn{\n}{\l}{q+k}
\mma{\hq}{\l} \mmb{\hkq}{\l} \nonumber \\
&& \!\!\!\!
+{1\over 16}
\sum_{\rho \lambda}  \; \tlsuma{p} \dmn{\mu}{\rho}{p}
\dmn{\mu}{\lambda}{p+k}
\mma{\hp}{\rho} \mmb{\hkp}{\lambda}  \nonumber \\
&& \!\!\!\!
\times
\tlsuma{q} \dmn{\nu}{\rho}{q} \dmn{\nu}{\lambda}{q+k}
\mma{\hq}{\rho} \mmb{\hkq}{\lambda} \nonumber \\
&& \!\!\!\!
+{1\over 2} \sum_{\rho \lambda} \; \tlsuma{p} \dmn{\mu}{\rho}{p}
\dmn{\mu}{\lambda}{p+k}
\mma{\hp}{\rho} \cmb{p+k}{\lambda}   \nonumber \\
 && \!\!\!\!
\left.  \times
\tlsuma{q} \dmn{\nu}{\rho}{q} \dmn{\nu}{\lambda}{q+k}
\mma{\hq}{\rho} \cmb{q+k}{\lambda}\right\}\nonumber \\
&& \!\!\!\!
-\left\{ {1 \over 2} \sum_{\rho \lambda} \;  
\tlsuma{p} \dmn{\mu}{\rho}{p} \dmn{\mu}{\rho}{p+k}
\mma{\hp}{\lambda} \mmb{\hkp}{\lambda}  \right. \nonumber \\
&& \!\!\!\!
 \times
\tlsuma{q} \dmn{\nu}{\rho}{q} \dmn{\nu}{\rho}{q+k}
\mma{\hq}{\lambda} \mmb{\hkq}{\lambda}  \nonumber \\
&& \!\!\!\!
+{1 \over 2} \sum_{\rho \lambda} \;  \tlsuma{p} \dmn{\mu}{\rho}{p}
\dmn{\mu}{\rho}{p+k}
\mma{\hp}{\lambda}\mmb{\hkp}{\lambda}   \nonumber \\
&& \!\!\!\!
\times
\tlsuma{q} \dmn{\nu}{\lambda}{q} \dmn{\nu}{\lambda}{q+k}
\mma{\hq}{\rho}\mmb{\hkq}{\rho} \nonumber \\
&& \!\!\!\!
+\sum_{\rho \lambda}\;  \tlsuma{p} \dmn{\mu}{\rho}{p}
\dmn{\mu}{\lambda}{p+k}
\mma{\hp}{\lambda}\mmb{\hkp}{\rho} \nonumber \\
&& \!\!\!\!
\times
\tlsuma{q} \dmn{\nu}{\lambda}{q} \dmn{\nu}{\rho}{q+k}
\mma{\hq}{\rho}\mmb{\hkq}{\lambda} \nonumber \\
&& \!\!\!\!
-2\sum_{\rho \lambda} \; \tlsuma{p} \dmn{\mu}{\rho}{p}
\dmn{\mu}{\rho}{p+k}
\mma{\hp}{\lambda}\mmb{\hkp}{\lambda}  \nonumber \\
&& \!\!\!\!
\left. \left. \left.  \times
\tlsuma{q} \dmn{\nu}{\rho}{q} \dmn{\nu}{\lambda}{q+k}
\mma{\hq}{\lambda}\mmb{\hkq}{\rho}
\right\} \right) 
+ \left( \phantom{\tlsum{q}}
\m \leftrightarrow \n \phantom{\tlsum{q}} \right) \right] \; , \lb{D1f2} 
\eea
where the self-energy $\Sigma^{\rm bare}_{\m \n} (p)$ in Eqs.~(\eq{B1})
and (\eq{D1ed}) is given in Eq.~(\eq{SIGMA_V}).
The terms in Eqs.~(\eq{D1f1}) and (\eq{D1f2}) 
which are proportional to $1/\x$ are 
the contribution from the gauge-fixing action and all other terms 
arise from the $F_{\m \n}^4$ part of the plaquette action.

\end{document}